\documentclass[preprint,floatfix,nofootinbib,showpacs,preprintnumbers,amsmath,amssymb,Here,prd]{revtex4}

\usepackage{graphicx}
\usepackage{epsfig}
\usepackage{dcolumn}
\usepackage{bm}
\usepackage{subfigure}
\usepackage{graphicx}
\usepackage{feynmp}
\usepackage[usenames]{color}
\usepackage{textfit}

\begin{document}

\newcommand{\hsp}{\hspace{0.08in}}
\newcommand{\be}{\begin{eqnarray}}
\newcommand{\ee}{\end{eqnarray}}
\newcommand{\sbe}{\begin{eqnarray*}}
\newcommand{\see}{\end{eqnarray*}}
\newcommand{\kb}{\overline\xi}
\newcommand{\ii}{\it i}
\newcommand{\fb}{\overline\Psi}
\newcommand{{\de}}{{\delta}}
\newcommand{{\D}}{{\Delta}}
\newcommand{\co}{\frac{{\ii}\pi^2}{2}}
\newcommand{\lef}{\left}
\newcommand{\rig}{\right}
\newcommand{\mul}{k \cdot p}
\newcommand{\ld}{\ell\cdot q}
\newcommand{\nn}{\nonumber}
\newcommand{\ep}{\epsilon}
\newcommand{\kslash}{\not\!k}
\newcommand{\p}{\not\!p}
\newcommand{\w}{\not\!w}
\newcommand{\q}{\not\!q}
\newcommand{\gm}{\gamma_{\mu}}
\newcommand{\g}{\gamma}
\newcommand{\cof}{\frac{\alpha_0 \xi}{4\pi}}
\newcommand{\pdl}{\frac{p^2}{\Lambda^2}}
\newcommand{\qdl}{\frac{q^2}{\Lambda^2}}
\newcommand{\ql}{\frac{q^2}{\Lambda^2}}
\newcommand{\llm}{\frac{\ell^2}{\Lambda^2}}
\newcommand{\plp}{\frac{q^2}{\Lambda^{\prime 2}}}
\newcommand{\kdl}{\frac{k^2}{\Lambda^2}}
\newcommand{\sr}{\scriptscriptstyle R}
\newcommand{\al}{\alpha}
\newcommand{\ar}{\alpha_{\scriptscriptstyle R}}
\newcommand{\lp}{l^{\scriptscriptstyle P}}
\newcommand{\lk}{l^{\scriptscriptstyle K}}
\newcommand{\tp}{t^{\scriptscriptstyle P}}
\newcommand{\tk}{t^{\scriptscriptstyle K}}
\newcommand{\lpp}{l'^{\scriptscriptstyle P}}
\newcommand{\lkp}{l'^{\scriptscriptstyle K}}
\newcommand{\tpp}{t'^{\scriptscriptstyle P}}
\newcommand{\tkp}{t'^{\scriptscriptstyle K}}
\newcommand{\z}{{\it z}}
\newcommand{\dgk}{\displaystyle{G(k^2)}}
\newcommand{\dgp}{\displaystyle{G(p^2)}}
\newcommand{\dgq}{\displaystyle{G(q^2)}}
\newcommand{\dfk}{\displaystyle{F(k^2)}}
\newcommand{\dfp}{\displaystyle{F(p^2)}}
\newcommand{\dfq}{\displaystyle{F(q^2)}}
\newcommand{\at}{(\ell+q/2)}
\newcommand{\eek}{(\ell-q/2)}
\newcommand{\e}{\ell}
\newcommand{\es}{\not\!\ell}
\newcommand{\ea}{\ell_{+}}
\newcommand{\ei}{\ell_{-}}
\newcommand{\f}{\it f}
\newcommand{\ek}{\ell^{\scriptscriptstyle K}}
\newcommand{\epp}{\ell^{\scriptscriptstyle P}}
\newcommand{\el}{\ell^{\scriptscriptstyle L}}
\newcommand{\tl}{t^{\scriptscriptstyle L}}
\newcommand{\eb}{\overline\eta}
\newcommand{\Tr}{{\rm Tr}}

\begin{titlepage}
\preprint{ IPPP/09/28, DCPT/09/56, ADP-09-06/T686 }
 {\centerline{\bf { BUILDING THE FULL FERMION-PHOTON VERTEX OF QED}}}
\vskip 2mm
{\centerline{\bf {BY IMPOSING MULTIPLICATIVE RENORMALIZABILITY}}}
\vskip 2mm
{\centerline{\bf { OF THE SCHWINGER-DYSON EQUATIONS }}}
\vskip 2mm
{\centerline{\bf { FOR THE FERMION AND PHOTON PROPAGATORS}}}
\vskip 0.8cm
\baselineskip=7mm
{\centerline{\bf{A. K{\i}z{\i}lers\"{u}{$^{1,2}$} and M.R. Pennington$^2$}}}
\vskip 4mm
{\centerline{{$^1$} Special Research Centre for the Subatomic Structure of Matter,}}
 {\centerline{Department of Physics,}}
 {\centerline{Adelaide University}}
 {\centerline{Adelaide, 5005, Australia}}
 \vskip 4mm
{\centerline{{$^2$}Institute for Particle Physics Phenomenology,}}
{\centerline{ Physics Department}}
{\centerline{Durham University}}
{\centerline{Durham DH1 3LE, U.K.}}
\vskip 0.4cm
{\centerline {ABSTRACT}}
\vspace{4mm}
\baselineskip=6mm
\noindent
    In principle, calculation of a full Green's function in any field
    theory requires knowledge of the infinite set of multi-point
     Green's functions,
    unless one can find some way of truncating the corresponding
    Schwinger-Dyson equations.  For the fermion and boson propagators in QED
    this requires an {\it ansatz} for the full three point vertex.
    Here we illustrate how the properties of gauge invariance, gauge
 covariance
     and multiplicative renormalizability impose severe constraints on
        this fermion-boson interaction, allowing a consistent truncation of
    the propagator equations.   We demonstrate how these conditions imply
    that the 3-point vertex {\bf in the propagator equations} is largely
    determined by the behaviour of the fermion propagator itself and not by
knowledge of the many higher point functions. We give an
explicit form for the fermion-photon vertex, which in the fermion and photon
propagator fulfills these constraints to
all orders in leading logarithms for massless QED, and accords with the weak coupling limit in perturbation theory at ${\cal O}(\alpha)$.
This provides the first attempt to deduce non-perturbative Feynman rules
    for strong physics calculations of propagators in massless QED that ensures
    a more consistent truncation of the 2-point Schwinger-Dyson equations.
 The generalisation to next-to-leading order
and masses will be described in a longer publication.
\vskip  3mm
\footnotesize
\baselineskip=4.8mm
\pacs{11.15.-q, 11.15.Tk, 12.20.-m, 12.20.Ds, 12.38.-Lg}
\begin{flushleft}
Electronic address: m.r.pennington@durham.ac.uk\\
Electronic address: akiziler@physics.adelaide.edu.au
\end{flushleft}
\end{titlepage}
\vfil\eject
\vskip 1cm
\section{Introduction}
\baselineskip=6mm
\parskip=2.5mm
\normalsize
\baselineskip=6mm
\noindent
Solution of the Schwinger-Dyson equations (SDE) for any field theory would
constitute the complete determination of that theory and every possible
measurable quantity would be known. Even though it is nearly 60 years
since these field equations were first derived
\cite{Schwinger:1951ex,Schwinger:1951hq,Dyson:1949bp,Dyson:1949ha,Green:inteqn}, we are far from obtaining their
solution even for a relatively simple theory like QED. Progress has been
hampered by the very structure that makes field theory interesting, namely
 that the Schwinger-Dyson equations form an infinite
nested set. Each $n$-point function must be multiplicatively renormalizable
and, in a gauge theory, respects gauge invariance. To achieve this,
the solution even for the
2-point functions (the propagators) appears to require knowledge of all the
other $n$-point functions. Consequently, studies in gauge theories have
resorted foremostly to a perturbative approximation, in which each
Green's function is expanded to a given order in the coupling squared.
Or as an approximation to non-perturbative physics, simple (even simplistic)
 ansatz have been used for the 3-point function to allow the fermion
propagator to be investigated. In return dynamical mass generation has been
studied in the {\it rainbow approximation} \cite{Kondo:1990ig,Fomin:1984tv,Miransky:1984ef,Miransky:1986ib,Miransky:1986xp,Curtis:1993py,Hawes:1994ce} and some level of understanding of when
chiral symmetry breaking can occur has been reached. While valuable
for gaining intuition, this is no substitute for a genuine non-perturbative
study. While formal results on gauge invariance and multiplicative
renormalizability (MR) have long been known using the gauge technique of
Salam, Delbourgo \cite{Delbourgo:1978bu,Salam:1963sa,Zumino:1959wt} and others, this method has not proved useful for
providing equations that can be readily solved either analytically or
numerically. Here, an alternative approach, an attempt to develop
non-perturbative Feynman rules, has proved more fruitful. The aim is
to write down explicit representations for the effective $n$-point
functions, in particular, for the 3-point function, which ensures that
the solutions of the Schwinger-Dyson equations for the 2-point functions
respect gauge invariance and are multiplicatively renormalizable \cite{collins,zuber}.

\noindent
What has previously impeded the practical study of the Schwinger-Dyson
equations has been the need to handle overlapping divergences that
dramatically complicate the renormalization of the equations. The
present approach overcomes this difficulty by requiring that the
2-point functions must be multiplicatively renormalizable  and
no overlapping divergences can thereby occur. This procedure is genuinely
non-perturbative and is not readily relatable to attempts at summing
subsets of Feynman graphs with these same properties \cite{Curtis:1990zs,Curtis:1993py,Kizilersu:2001pd,Kizilersu:2001,Curtis:1991fb,Burden:1993gy,Hawes:1991qr}.

\baselineskip=6mm
\noindent The first of such non-perturbative studies has been in the case
of quenched
QED \cite{Dong:1994jr,Haeri,Haeri:1990ty,Fischer:2004ym,Atkinson:1989fp,Roberts:1994dr,Atkinson:1993mz,Bloch:1994if,Hawes:1991qr,Hawes:1994ce,Hawes:1996ig,Hawes:1996mw,Hawes:1996pe,Alkofer:2000wg}
 --- that is QED in which the explicit factor of $N_F$ multiplying
the fermion loop corrections to the photon propagator is set equal to zero.
 Then a form for the fermion-boson vertex that satisfies the Ward identity,
 the Ward-Green-Takahashi identity \cite{Ball:1980ay} and renders the fermion propagator
multiplicatively renormalizable, has been written down explicitly \cite{Curtis:1990zs,Bashir:1994az}.
While the form is non-perturbative, the fact that it must agree
with perturbation theory in the weak coupling regime is a key pointer to the
ultraviolet structure, expressed in terms of logarithms of momenta. The
purpose of the present paper is to extend this study by developing
the constraints that have to be fulfilled in the case of {\bf massless unquenched}
QED to ensure {\bf both} the fermion and photon propagators are
multiplicatively renormalizable (at least as far as leading logarithms are concerned).

\noindent In general, the full fermion-boson vertex has 12 components, all of which are in principle independent,
though one is forced to be zero by gauge invariance. The fermion and photon propagators do not require complete
knowledge of the full complexity of this structure, but just 2 projections that arise in the Schwinger-Dyson equations
for these 2-point functions. We present a simple solution to the constraints from multiplicative renormalizability.
While the general structure of the full vertex is not complete, the projections within the SDEs for the 2-point functions have no
freedom.

\noindent While it is clear that the full 3-point function must involve knowledge of the 4-point kernel and
higher-point functions, as far as its role in the equations for the propagators is concerned, this is not the case.
Thus it can be that the {\it effective} 3-point function involves only the full 2-point functions. A clue to this is
provided by the Ward-Green-Takahashi \cite{Ward,Green:1953te,Takahashi:takahashi}
 identity, which tells us that part of the 3-point vertex (often called the
longitudinal part) is precisely fixed by the fermion propagator alone.
Moreover, a hint that the remaining transverse part may be similarly
constrained is the fact that the vertex and fermion wavefunction
renormalization have common renormalization factors $(Z_1=Z_2)$ as a
 consequence of gauge invariance. Thus the transverse part must know
about the fermion propagator functions too. How, this works in full QED is
what we investigate in this paper.

\noindent
In Sect.~2 we consider the structure of the fermion-boson vertex
and its ultraviolet behaviour. In Sect.~3 we compute the Schwinger-Dyson
equations for the fermion and boson propagators. In Sect.~4 we deduce
the ultraviolet
structure imposed by multiplicative renormalizability.
Sect.~5 gives the constraints on the vertex imposed by MR conditions. The pattern of constraints indicates a general analytic form for the
transverse part of the vertex structure.
In Sect.~6 we deduce a solution to these constraints involving the full fermion wavefunction renormalization. The vertex in
the weak coupling limit is studied
in Sect.~7 and the restrictions it imposes derived.  In Sect.~8 we conclude
 and outline a programme for
future work.
Since this procedure is rather complicated, we show in Fig.~1 a flow diagram
of this calculation.

\baselineskip=7mm
%
%
\begin{figure}[htbp]
\refstepcounter{figure} \addtocounter{figure}{-1} \setlength{\unitlength}{1mm}
\begin{center}
\vspace*{-2.5cm}\hspace*{-1cm}
\begin{picture}(160,220) \thicklines
%
%
%
\put(40,190){\framebox(80,10){\parbox{8cm}{\begin{center} {\bf Non-perturbative massless QED} \end{center}}}}
\put(80,190){\vector(0,-1){5}} \put(50,175){\framebox(60,10){\parbox{6cm}{\begin{center} {\bf Schwinger-Dyson
equations} \end{center}}}} \put(80,175){\vector(0,-1){4.5}} \put(40,160){\framebox(80,10){\parbox{8cm}{\begin{center}
{\small{ Truncation is needed\\[-3mm] Make an ansatz for 3-point vertex}} \end{center}}}} \put(80,160){\vector(0,-1){4.55}}
\put(0,50){\framebox(160,105){}}
\put(5,135){\framebox(50,18){\parbox{5cm}{\begin{center} {\footnotesize{The ansatz must satisfy  criteria,\\[-3mm] which the
full vertex itself\\[-3mm] satisfies}}\end{center}}}} \put(55,144){\vector(1,0){35}}
\put(90,135){\framebox(50,18){\parbox{5cm}{\begin{center} {\footnotesize{
Full vertex is divided into \\[-4mm]longitudinal and transverse \\[-3mm]parts\\[-3mm]
$\Gamma_F^{\mu}=\Gamma_L^{\mu}+\Gamma_T^{\mu}$}}\end{center}}}}
\put(115,135){\vector(0,-1){8}} \put(5,110){\framebox(50,18){\parbox{5cm}{\begin{center} {\footnotesize{ As a result of
{\bf Gauge}\\[-3mm]{\bf Invariance}
 Ward Identities\\[-3mm] must be fulfilled}}\end{center}}}}
\put(55,120){\line(1,0){20}} \put(75,130){\line(0,-1){10}} \put(75,130){\vector(1,0){40}}
\put(90,111.5){\framebox(50,15){\parbox{5cm}{\begin{center} {\footnotesize{
{\bf Longitudinal vertex is fixed}\\[-3mm]
by Ball-Chiu \cite{Ball:1980ay}\\[-3mm] $\Gamma_L^{\mu}=\Gamma_{BC}^{\mu}$}}\end{center}}}}
\put(115,111.5){\vector(0,-1){5}} \put(90,86){\framebox(50,20){\parbox{5cm}{\begin{center} {\footnotesize{ {\bf
Transverse part} left to be\\[-3mm] determined.
The vector \\[-3mm]structure of the vertex leads to\\[-3mm]
$\Gamma_T^{\mu}=\sum_{i=1}^8\tau_{\,i}(k^2,p^2,q^2)\,T_i^{\mu}$}}\end{center}}}}
%
%
\put(10,90){\framebox(40,10){\parbox{4cm}{\begin{center} {\footnotesize{ $T_{\,i}^{\mu}$ basis tensors are\\[-3mm] given by
Ball-Chiu }}\end{center}}}} \put(50,95){\line(1,0){20}} \put(70,95){\line(0,-1){12.5}} \put(70,82.5){\vector(1,0){45}}
%
%
\put(115,86){\vector(0,-1){7}} \put(90,69){\framebox(50,10){\parbox{5cm}{\begin{center} {\footnotesize{
$\tau_{\,i}$, the coefficient functions\\[-3mm]
 are the only unknowns}}\end{center}}}}
\put(115,69){\vector(0,-1){7}} \put(90,52){\framebox(50,10){\parbox{5cm}{\begin{center} {\footnotesize{
Substitute this vertex into \\[-3mm]
the coupled SD-equation}}\end{center}}}}
%
%
\put(5,55){\framebox(50,20){\parbox{5cm}{\begin{center} {\footnotesize{
Using charge conjugation, \\[-3mm]other information and,\\[-3mm]
perturbative expansion of \\[-3mm]
$\tau_{\,i}$ are suggested }}\end{center}}}} \put(55,65){\vector(1,0){60}}
%
\put(115,50){\vector(0,-1){5}} \put(80,30){\framebox(70,15){\parbox{7cm}{\begin{center} {\footnotesize{ Solve SD-eqn.
for the $1/F$ and $1/G$ \\[-3mm]in terms of the constants of \\[-3mm]the perturbative expansion of $\tau_{\,i}$}}\end{center}}}}
\put(5,30){\framebox(50,15){\parbox{5cm}{\begin{center} {\footnotesize{
Calculate the general form of \\[-3mm]
{\bf multiplicatively}\\[-3mm]{\bf renormalisable $F$ and $G$}}}\end{center}}}} \put(30,30){\line(0,-1){2}}
\put(30,28){\vector(1,0){85}}
\put(115,30){\vector(0,-1){4.5}} \put(85,7.5){\framebox(60,18){\parbox{6cm}{\begin{center} {\footnotesize{ Find the {\bf{constraints}} on the vertex \\[-3mm]
function imposed by \\[-3mm]
{\bf multiplicative} {\bf renormalisation }\\[-3mm]
and generalize these constraints }}\end{center}}}} \put(115,7.5){\vector(0,-1){4}}
\put(5,12.5){\framebox(60,10){\parbox{6cm}{\begin{center} {\footnotesize{
Perturbative calculation and \\[-3mm]
all other vertex information}}\end{center}}}} \put(35,12.5){\line(0,-1){7}} \put(35,5.5){\vector(1,0){80}}

\put(75,-2.5){\framebox(80,6){\parbox{8cm}{\begin{center} {\footnotesize{ {\bf Construct non-perturbative vertex
ansatz} }}\end{center}}}}
\end{picture}
\end{center}
\caption{Flow diagram of the Schwinger-Dyson
 calculation presented here.}
\label{Fig:flownon}
\end{figure}
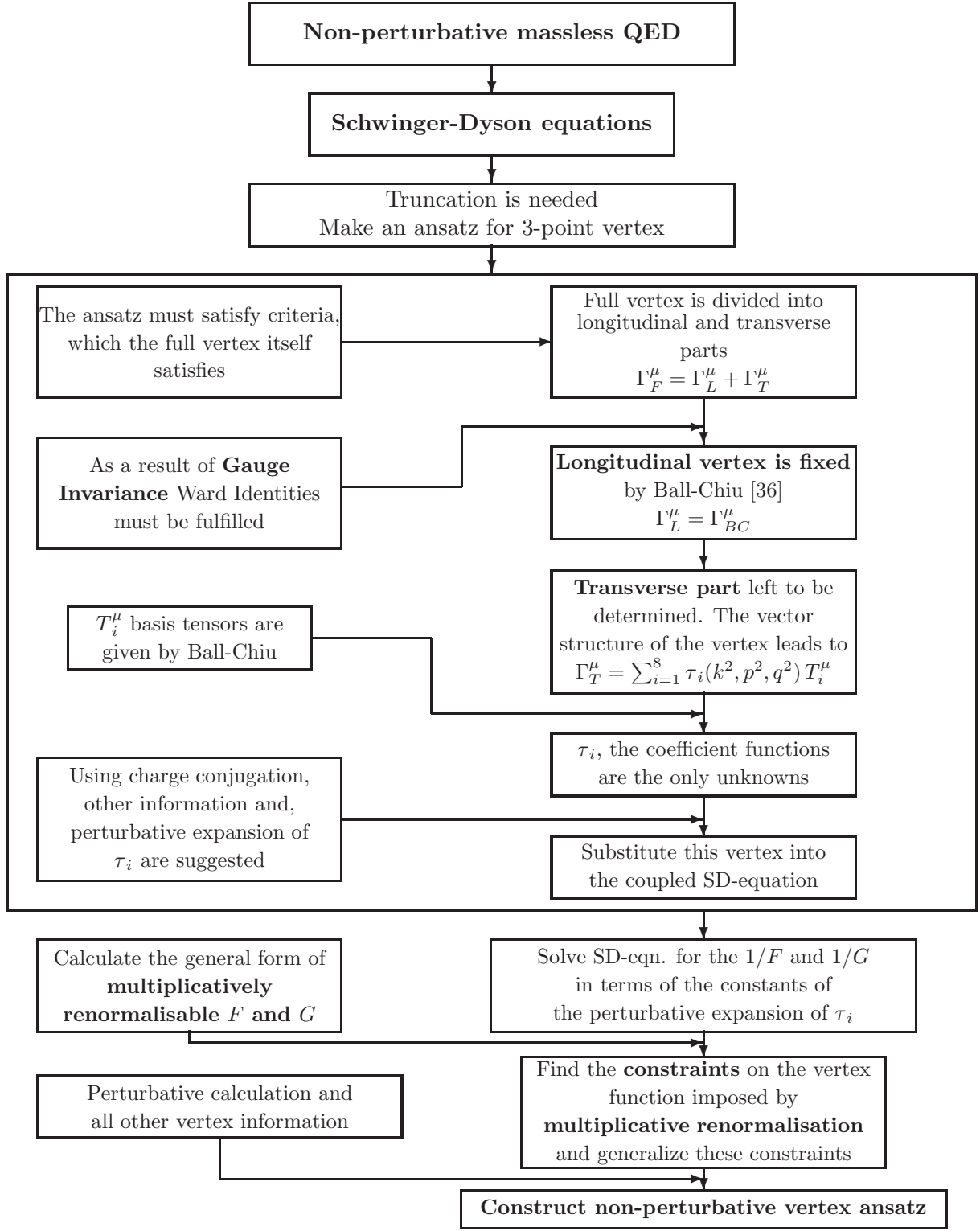

\newpage
\section{\label{sec:vertex}Vertex and Propagators and their Renormalizations\protect\\  Defined}

\noindent The two key constraints on the fermion-boson vertex are provided by the gauge
invariance of the theory and by multiplicative renormalizability.
Here we begin with the
first of these and describe the importance of the Ward-Green-Takahashi
identity~\cite{Ward,Green:1953te,Takahashi:takahashi}.  Though this is well known, it forms the essential background
allowing us to establish our notation.


\begin{figure}[h]
\begin{center}
\refstepcounter{figure}
\addtocounter{figure}{-1}
~\epsfig{file=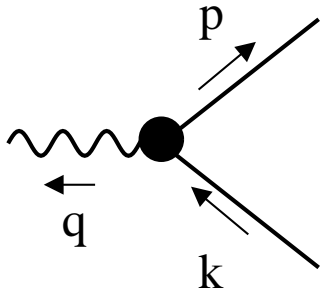,width=130pt}
\end{center}
\vspace{-1cm}
\hspace{4.5cm}
\caption{Fermion-boson vertex.}
\label{fig:vertex}
\end{figure}

\noindent
The vertex, displayed in Fig.~\ref{fig:vertex}, is a function of the two independent momenta flowing through the vertex. We
take these to be the fermion momenta, $k$ and $p$. The vertex function is $\Gamma^{\mu} (k,p\,;\,q)$ with $q\,=\,k\,-\,p$\,. It is
well-known that the coupling of two spin-$1/2$ particles to a spin-1 boson involves 12 independent vectors, of these
eight are transverse to the boson momentum $q$. The structure of the four (longitudinal) components are constrained by the
Ward-Green-Takahashi identity (WGTI)
\be
q^{\mu}\,\Gamma_{\mu}(k,p ~ ;~q)=S_F^{-1}(k)-S_F^{-1}(p\,)\,\qquad ,
\label{eq:WGT}
\ee
where $S_F(p\,)$ is the full fermion propagator carrying momentum $p$.
In general
\be
{\ii}S_F(p\,)=\;{\ii}\,\frac{F(p^2)}{\p-{\cal M}(p^2)}=\;{\ii}\,\frac{1}{A(p^2) \p-{\cal B}(p^2)}\quad,
\label{eq:fermionprop}
\ee
where $F(p^2)$ (or $A(p^2)= 1/F(p^2)$) is the fermion wavefunction renormalization and
 ${\cal M}(p^2)$ (or $ B(p^2)= M(p^2)/F(p^2)$) is its mass function. The bare fermion propagator is just
 $S_F^0(p\,)= 1/(\p -m_0) $.
From the form of this propagator, the Ward-Green-Takahashi identity, Eq.~(\ref{eq:WGT}),
contains terms with both one and no gamma matrices, so that the vertex component involving two through
$\sigma^{\mu\nu}\equiv \frac{1}{2}[\gamma^{\mu},\gamma^{\nu}]$ must be zero. Thus in a gauge theory there are in fact
11 independent non-zero vectors in terms of which to decompose $\Gamma^{\mu}(p,k;q)$. Of these, six occur if the fermions are
massless as we consider here, {\it i.e.} ${\cal M}(p^2) = 0$. Eq.~({\ref{eq:WGT}}) has a well known zero photon momentum limit;
the Ward identity~:
\be
\Gamma_{\mu}(p,p\,;0)\,=\lim_{k \longrightarrow p} \Gamma_{\mu}(p,k \,;q)\,=\, \frac{\partial S_F^{-1}(p)}{\partial p_{\mu}}\quad .
\label{eq:ward}
\ee
The full vertex can be divided into
longitudinal and transverse components
\be
\Gamma^{\mu}(p,k\, ;q)=\Gamma_L^{\mu}(p,k\, ;q)+\,\Gamma_T^{\mu}(p,k \,;q)\quad,
\ee
where
\be
q_{\mu}\Gamma^{\mu}_T(p,k\, ;q)=0\qquad.
\ee
We demand that the longitudinal part {\bf alone} is responsible for the vertex
satisfying both
Eqs.~(\ref{eq:WGT},\ref{eq:ward}). This means that each component must be separately free of
kinematic singularities, so that
\be
\Gamma^{\mu}_T(p,p\, ;0)=0\quad.
\ee
The longitudinal part is then defined, following Ball-Chiu~\cite{Ball:1980ay}, to be
\be
\Gamma_{L}^{\mu}(p,k,q)&\equiv&\Gamma^{\mu}_{BC}(p,k,q)\,,\nn\\
&=&\sum_{i=1}^4\,\lambda_i(p^2, k^2, q^2)\,L_i^{\mu}(p,k;q)\quad ,
\label{eq:longitudinal}
\ee
where
\begin{alignat}{2}
\lambda_1(p^2,k^2,q^2)=&\,\frac{1}{2}\lef(\frac{1}{F(k^2)}+\frac{1}{F(p^2)}\rig)\,,                      \qquad\quad         &L_1^{\mu}(p,k;q)=& \,\g^{\mu},\nn\\
\lambda_2(p^2,k^2,q^2)=&\,\frac{1}{2}\,\frac{1}{(k^2-p^2)}\lef(\frac{1}{F(k^2)}-\frac{1}{F(p^2)}\rig)\,, \qquad \quad        &L_2^{\mu}(p,k;q)=&\,\lef(k^{\mu}+p^{\mu}\rig)\lef(\kslash+\p\rig), \nn\\
\lambda_3(p^2,k^2,q^2)=&\,-\frac{1}{(k^2-p^2)}\lef(\frac{M(k^2)}{F(k^2)}-\frac{M(p^2)}{F(p^2)}\rig)\,,   \qquad \quad        &L_3^{\mu}(p,k;q)=&\,(k^{\mu}+p^{\mu}), \nn\\
\lambda_4(p^2,k^2,q^2)=& \,0                                                                        \,,  \qquad \quad        &L_4^{\mu}(p,k;q)=&\,\sigma^{\mu\nu}\,\lef(k_{\nu}+p_{\nu}\rig).
\label{eq:longitcoeff}
\end{alignat}

\noindent Crucially because of gauge invariance, this longitudinal component
of the vertex is wholly determined by the fermion propagator. Moreover, it
is this longitudinal component that gives the dominant ultraviolet behaviour
of the vertex~\cite{Kizilersu:1995iz}.

\noindent
Quite generally, the
transverse vertex can  be decomposed in the {\underline{massless}} fermion case
in terms of the remaining four basis vectors as~:
\be
\hspace{4.5cm}
\Gamma_T^{\mu}(p,k;q)
=\sum_{i=2,3,6,8}\,\tau_{\ii}(p^2,k^2,q^2)\,T^{\mu}_{\ii}(p,k;q)\,,
\qquad\qquad\qquad\qquad\quad
\label{eq:transverse}
\ee
where the $\tau_i$ are coefficient functions depending on momenta $k^2, p^2$
and $q^2$, which are as yet undetermined, and
the $T_i$ are the basis tensors defined by Ball and Chiu~\cite{Ball:1980ay}
--- the modification of this basis by K{\i}z{\i}lers\"{u} et al. \cite{Kizilersu:1995iz}
 does
not affect
these four vectors~:
\be
&T^{\mu}_{2}(p,k;q)&=\left
(\,p^{\mu}(k\cdot q)-k^{\mu}(p\cdot q)\,\right)({\not\! k}
+{\not\! p})\;,\nonumber\\
&T^{\mu}_{3}(p,k;q)&=q^2\gamma^{\mu}-q^{\mu}{\not \! q}\;,\nonumber\\
&T^{\mu}_{6}(p,k;q)&=\gamma^{\mu}(p^2-k^2)+(p+k)^{\mu}{\not \! q}\;,\nonumber\\
&T^{\mu}_{8}(p,k;q)&=-\gamma^{\mu}k^{\lambda}p^{\nu}\,{\sigma_{\lambda\nu}}
+k^{\mu}{\not \! p}-p^{\mu}{\not \! k}\;.
\label{eq:Ts}
\ee
With these basis vectors, the $\tau_i (i=2,3,6,8)$ are individually free of
kinematic singularities at ${\cal O}(\alpha)$ in perturbation theory in any
covariant gauge as shown in Ref.~\cite{Kizilersu:1995iz}.
It is  these $\tau_i$'s that are constrained by multiplicative
renormalizability~\cite{Curtis:1990zs}.
It is our key presumption that this will {\bf force} these
transverse components (or at least their projections in the Schwinger-Dyson
equations for the 2-point functions) to depend only on
propagator functions just like the longitudinal part of Eqs.~(\ref{eq:longitudinal},~\ref{eq:longitcoeff}).

\noindent {\underline{\textbf{What we can say about these coefficients?}}

\noindent
Here we discuss the fundamental constraints on the transverse vertex that follow from ({\it{i}})~ dimensional analysis, ({\it{ii}})~symmetry properties,
({\it{iii}})~order of perturbation theory, ({\it{iv}})~gauge dependence and ({\it{v}})~renormalization~:

\noindent
({\it{i}})~The transverse vertex is
dimensionless. Knowing the dimensions of the basis vectors
from Eq.~(\ref{eq:Ts}) tells us the dimensions of the $\tau_i$'s. With $d \equiv \hbox{ momentum}^2$, then:
\be
\mbox{dim. of} \;T_2^{\mu}\;&:&\;d^2\;\longrightarrow\qquad
\mbox{dim. of} \;\tau_2\;:\;\frac{1}{d^2}\;,\nn\\
\mbox{dim. of} \;T_3^{\mu}\;&:&\;d\;\;\,\longrightarrow\qquad
\mbox{dim. of} \;\tau_3\;:\;\frac{1}{d}\;,\nn\\
\mbox{dim. of} \;T_6^{\mu}\;&:&\;d\;\;\,\longrightarrow\qquad
\mbox{dim. of} \;\tau_6\;:\;\frac{1}{d}\;,\nn\\
\mbox{dim. of} \;T_8^{\mu}\;&:&\;d\;\;\,\longrightarrow\qquad
\mbox{dim. of} \;\tau_8\;:\;\frac{1}{d}\;.
\label{eq:dim}
\ee
({\it{ii}})~The $C$-parity operation~\cite{drell, Dong:1994jr} on
Eqs.~(\ref{eq:longitudinal},~\ref{eq:transverse}) requires
\begin{alignat}{2}
\tau_2(k^2,p^2,q^2)&=
\,\;\;\;\;\tau_2(p^2,k^2,q^2)                 \qquad, \qquad   &\lambda_1(k^2,p^2,q^2)&=\,\;\;\;\;\lambda_1(p^2,k^2,q^2)\,,\nn\\
\tau_3(k^2,p^2,q^2)&=
\,\;\;\;\;\tau_3(p^2,k^2,q^2)                 \qquad,\qquad   &\lambda_2(k^2,p^2,q^2)&=\,\;\;\;\;\lambda_2(p^2,k^2,q^2)\,, \nn\\
\tau_6(k^2,p^2,q^2)&=\;\,-\,\tau_6(p^2,k^2,q^2)\qquad,\qquad   &\lambda_3(k^2,p^2,q^2)&=\,\;\;\;\;\lambda_3(p^2,k^2,q^2)\,,\nn\\
\tau_8(k^2,p^2,q^2)&=
\,\;\;\;\;\tau_8(p^2,k^2,q^2)                 \qquad,\qquad        &\lambda_4(k^2,p^2,q^2)&=\;\,-\,\lambda_4(p^2,k^2,q^2)\,.
\label{eq:sym}
\end{alignat}

\noindent
({\it{iii}})~At zeroth order in perturbation
theory
the full vertex is $\gamma^{\mu}$.  Since at this order $F =1$, we see
from Eqs.~(\ref{eq:longitudinal},~\ref{eq:longitcoeff})
that $\Gamma_L^{\mu}\,=\,\gamma^{\mu}$,  consequently, $\Gamma_T^{\mu}\,=\,0$.
Thus the $\tau_i\,=\,{\cal O}(\alpha)$ in perturbation theory.

\noindent
({\it{iv}})~The propagator for the photon carrying
momentum $q$ is
\be
{\ii}\D_{\mu\nu}(q)&=&
-{\ii}\,\left[\,\frac{G(q^2)}{q^2}\lef(g_{\mu\nu}-\frac{q_{\mu}q_{\nu}}{q^2}\rig)
\,+\,\xi\,\frac{q_{\mu}q_{\nu}}{q^4}\,\right]\quad ,\nn\\[3mm]
&=&-{\ii}\,\left[\,\D_{\mu\nu}^T+\,\xi\,\frac{q_{\mu}q_{\nu}}{q^4}\,\right]\,,
\qquad\qquad\qquad\qquad\qquad\qquad
\label{eq:photonprop}
\ee
where $G(q^2)$ is the photon renormalization function, $\xi$ is the
covariant gauge parameter and the $\Delta^T_{\mu \nu}$ is the transverse
part of the photon propagator. The bare photon propagator, $\Delta_{\mu\nu}^0$, has $G(q^2)\equiv 1$ in Eq.~(\ref{eq:photonprop}).

\noindent
Gauge covariance is expressed through the Landau-Khalatnikov-Fradkin (LKF)
transformations~\cite{LK,Fradkin:Fradkin}.
These mean that once a Green's function is known in some gauge, then its form
in all other gauges is determined. In general, this is, of course, only
useful if we know the relevant Green's function precisely in some gauge.
Nevertheless, the LKF transformations provides two key results we shall use.
The first concerns the fermion wavefunction renormalization, $F(p^2)$,
which can only depend on the covariant gauge through a unique factor of
$\xi$ in its anomalous dimension. The second fact
is that the photon wavefunction renormalization, $G(q^2)$, must be gauge
independent. Both of these requirements place restrictions on the form of
the non-perturbative interactions.

\noindent
({\it{v}})~In QED the full-propagators and the vertex function are all
divergent.
However, as is well known ~\cite{collins,zuber,ryder}, one can define
finite (renormalized) propagators
and vertex function by absorbing these divergences
into functions, $Z_i$ $ (i=1,2,3)$. As usual we introduce field
renormalizations~:
\be \Psi_{\sr}\,=\,{\it Z}_2^{-1/2}\,\Psi_0\;,\qquad
A_{\sr}^{\mu}\,=\,{\it Z}_3^{-1/2}\,A^{\mu}_0\;,\qquad
%
%
\ee
where the subscripts
${R}$ and ${\small 0}$ denote renormalized and bare quantities, respectively.
The latter are conveniently made finite by introducing an ultraviolet
momentum cutoff
$\Lambda$ and the former renormalized quantities depend on the
momentum scale $\mu$ at which we choose to renormalize.
The divergence of the fermion propagator is absorbed into $Z_2$, the fermion renormalization function, by~:
\be S_{\sr}(p,\mu)&=&{\it Z}_2^{-1}(\mu,\Lambda)\,S_0(p,\Lambda)\quad, \label{eq:mrfermion} \ee and similarly for the
photon~: \be \D_{\mu\nu}^{\sr}(p, \mu)&=&{\it Z}_3^{-1}(\mu,\Lambda)\,\D_{\mu\nu}^0(p,\Lambda)\;.
\label{eq:mrphoton} \ee
The gauge covariance of the photon propagator requires that the covariant
gauge parameter is similarly renormalized~:
\be \xi_{\sr}\,=\,{\it Z}_3^{-1}\,\xi\;. \label{eq:mrgauge} \ee
The divergence of the vertex function is cancelled
by the factor $Z_1$~:
\be \Gamma_{\mu}^{\sr}(p,\mu)=\,Z_1(\mu,\Lambda)\,\Gamma_{\mu}^{\,0}\,(p,\Lambda)\;, \label{eq:mrvertex} \ee
with the above definitions, the coupling constant is renormalized
 according to,
\be e_{\sr}\,=\,\frac{{\it Z}_2}{{\it Z}_1}\,\sqrt{{\it Z}_3}\,e\;. \label{eq:mrcoupling} \ee
Making use of the Ward-Green-Takahashi  identity~\cite{Ward,Green:1953te,Takahashi:takahashi}~: \be {\it Z}_1={\it Z}_2\;, \ee the coupling
constant renormalization becomes \be e_{\sr}={\it Z}_3^{1/2}\,e\;. \label{eq:coupling} \ee
As usual, we define $\alpha = e^2/(4\pi)$, where $\alpha_0 , \alpha_{\sr}$ denote the bare and renormalized couplings
related to  $e$ and $e_{\sr}$, respectively.

\noindent What we want to determine are the constraints these
 renormalizations of the fermion and photon propagators impose
on the transverse part of the fermion-boson vertex.
The renormalization of the 3-point vertex is
proportional to fermion renormalization constant $Z_2^{-1}$.
This can be seen already in the longitudinal vertex from the WGTI \cite{Ball:1980ay}. Consequently, the
 non-perturbative structure of the transverse component, and hence
the $\tau_i$'s, must
be proportional to the inverse of the fermion wavefunction renormalization,
{\it i.e.} ${\tau}_i\,(F,G) \sim 1/F$, just as the longitudinal $\lambda_i$'s of Eq.~(8) are.

\noindent To go further, the basic idea is easily explained by
considering the fermion propagator in quenched massless QED.
The non-perturbative
quantity is the fermion wavefunction renormalization $F(p^2, \Lambda^2)$.
Let us imagine expanding this perturbatively and just keeping leading
logarithms, so that we have
\be
F(p^2, \Lambda^2)\,=\, 1\,+\,\alpha_0\, A_1\, \ln\frac{p^2}{\Lambda^2}\,+
\,\alpha_0^2\, A_2\, \ln^2\frac{p^2}{\Lambda^2}\,
+\,\alpha_0^3\, A_3\, \ln^3\frac{p^2}{\Lambda^2}\,+\,\cdot\cdot\cdot \quad,
\ee
then inserting such a form in the loop integral of Fig.~\ref{fig:fermion}.  For this to be
a solution of the Schwinger-Dyson equation, the equation has to deliver
$F(p^2, \Lambda^2)$ with the same
perturbative expansion as output. However, to be
multiplicatively renormalizable, the coefficients $A_n$ cannot be
independent, but related by
$A_2 = A_1^{\,2}/2$, $A_3 = A_1^{\,3}/6$ and finally $A_n = A_1^{\,n}/n!$.
This requirement places a severe constraint on the fermion-boson
vertex. Since its longitudinal part is known, it is
its transverse components that are constrained.  The aim of this paper is
to determine these conditions on the $\tau_i$ of Eq.~(\ref{eq:transverse})
for full QED.
In general, these $\tau_i(p^2,k^2,q^2)$ functions can be written as a sum of terms,
each with the correct dimensions, Eq.~(\ref{eq:dim}), symmetry properties, Eq.~(\ref{eq:sym}), and renormalization requirements, as~:
\be
\tau_i(p^2,k^2,q^2)=\,\sum_j\,f_{ij}\,(p^2,k^2,q^2)\,{\overline{\tau}}_i^{(j)}\,(F,G)\,.
\label{eq:kinetic}
\ee
Each of these $\tau_i's$ has been divided into two parts ~:
{\it a kinematic part} encoded in $f_{ij}$, giving the right dimensions, Eq.~(\ref{eq:dim}), which
depends on momenta squared, and {\it a functional part}, $\overline{\tau}_i^{(j)}$,
that is {\bf assumed}
only to know about the fermion and photon renormalization
functions $F$ and $G$ at $k^2$, $p^2$ or $q^2$. Such a form
would provide a genuine non-perturbative construction,
\be
\tau_i^{sym}(p^2,k^2,q^2)&=&\sum_j\lef[f_{ij}^{anti}\,(p^2,k^2,q^2)\,{\tau^{anti}_i}^{(j)}\,(F,G)
                           +\, f_{ij}^{sym}\,(p^2,k^2,q^2)\,{\tau^{sym}_i}^{(j)}\,(F,G)\rig]\,,\nn\\[3mm]
\tau_i^{anti}(p^2,k^2,q^2)&=&\sum_j\lef[f_{ij}^{sym}\,(p^2,k^2,q^2)\,{\tau^{anti}_i}^{(j)}\,(F,G)
                           +\, f_{ij}^{anti}\,(p^2,k^2,q^2)\,{\tau^{sym}_i}^{(j)}\,(F,G)\rig]\,.\nn\\
\ee

\noindent
The forms of the $\tau_i$'s
are structured such that the integrals are soluble.
First we deal with the kinematic factors for each $\tau_i$'s, which are included in the following way:
\vspace{4mm}
\be
\nn\tau_2^M(p^2,k^2,q^2)&=\displaystyle{\frac{2}{(k^4-p^4)}}&
\lef[\beta_2+\gamma_2\,\frac{2\, k\cdot p}{k^2+p^2}\rig]
                    \; \tau_2^{anti}(p^2, k^2,q^2)\\[2.mm]
\nn &\,\,+\displaystyle{\frac{2}{(k^2+p^2)^2}}&\lef[\delta_2+\epsilon_2 \frac{2\,k\cdot p}{k^2+p^2}\,\,\rig] \;\tau_2^{sym}(p^2, k^2, q^2) \quad,
\\ [4mm]
\nn \tau_3^M(p^2,k^2,q^2)&=
    \displaystyle{\frac{1}{(k^2-p^2)}}&
\lef[\beta_3+\gamma_3\,\frac{2\, k\cdot p}{k^2+p^2}\rig]\;
                  \tau_3^{anti}(p^2,k^2,q^2) \\ [2.mm]
\nn &+\displaystyle{\frac{1}{(k^2+p^2)}}&\lef[\delta_3+\epsilon_3\frac{2\, k\cdot p}{k^2+p^2}\,\,\rig] \;
                  \tau_3^{sym}(p^2, k^2, q^2) \quad,
\\[4mm]
\nn
\tau_6^M(p^2,k^2,q^2)&=
    \displaystyle{\frac{1}{(k^2+p^2)}}&
\lef[\beta_6+\gamma_6\,\frac{2\, k\cdot p}{k^2+p^2}\rig] \;
        \tau_6^{anti}(p^2, k^2,q^2) \\[2.mm]
\nn &+\displaystyle{\frac{(k^2-p^2)}{(k^2+p^2)^2}}&\lef[\delta_6+\epsilon_6\frac{2\, k\cdot p}{k^2+p^2}\,\,\rig]\;
        \tau_6^{sym}(p^2, k^2,q^2) \quad,
\\[4mm]
\nn
\tau_8^M(p^2,k^2,q^2)&=
    \displaystyle{\frac{1}{(k^2-p^2)}}&
\lef[\beta_8+\gamma_8\,\frac{2\, k\cdot p}{k^2+p^2}\rig]  \;
        \tau_8^{anti}(p^2, k^2,q^2) \\[2.mm]
\nn &+\displaystyle{\frac{1}{(k^2+p^2)}}&\lef[\delta_8+\epsilon_8\frac{2\, k\cdot p}{k^2+p^2}\,\,\rig]\;
        \tau_8^{sym}(p^2, k^2, q^2)\quad.
\\[-8.5mm]
\label{eq:tauform}
\ee

\noindent
The factor 2 in the numerator of $\tau_2$ is merely for later convenience and superscript \lq\lq$M$" stands for Minkowski space.
The kinematic factors, $f_{ij}^{sym,~anti}$ play two roles:
first to ensure that each of
$\tau_i^{sym,~anti}$ is dimensionless, and to define the appropriate symmetry of these functions under the interchange of $k,p$.
 To make the problem tractable we do {\bf not} include $q^2$ dependence in the denominator factors. However, the dimensions and symmetry of the $\tau_i^{sym,~anti}$ is,
 of course, maintained by multiplying by a factor of $q^2/(k^2+p^2)$. Such a factor can be rewritten as $1 - 2 k \cdot p/(k^2+p^2)$,
 and this is the origin of the inclusion of the $\beta_i,\gamma_i,\delta_i, \epsilon_i$ terms in Eq.~(\ref{eq:tauform}).

\noindent
The $\tau_i^{anti}$ and $\tau_i^{sym}$ are antisymmetric and symmetric under
$k^2 \leftrightarrow p^2$, respectively.
The $\tau_i^{sym,~anti}$ are assumed to be solely functions of the fermion and boson renormalization functions $F$ and $G$, with consequently simplified dependence
 on $k^2$, $p^2$ and $q^2$. Since here we expand these functions in terms of leading logarithms, it is helpful to note that combinations like
$\log (k^2/p^2)$ are antisymmetric, while $\log(q^4/k^2p^2)$ is clearly symmetric under the interchange of $k$ and $p$, with each power of a \lq\lq log" being multiplied by
 a factor of $\alpha_0$. Such forms are the basis for the leading logarithmic expansion of the $\tau_i^{sym,~anti}$.  Before renormalization,
 these will depend on the ultraviolet cutoff $\Lambda$, and we can represent the $\tau_i^{sym,~anti}$ by:

\begin{alignat}{2}
\tau_i^{anti}(p^2,k^2,q^2)&=\sum_{m=1}^{\infty}\sum_{n,r=0}^{\infty}&{\cal{A}}_{mnrr}
          \lef[\left(\alpha_0 \ln \kdl \right)^m - \left(\alpha_0 \ln \pdl \right)^m \rig]
               \left(\alpha_0 \ln \qdl \right)^n \left(\alpha_0^2 \ln \kdl \ln \pdl \right)^r,
               \label{eq:nonperttauanti}\\[2mm]
\tau_i^{sym}(p^2,k^2,q^2)&=\sum_{m=0}^{\infty}\sum_{n,r=0}^{\infty}&{\cal{S}}_{mnrr}
          \lef[\left(\alpha_0 \ln \kdl \right)^m + \left(\alpha_0 \ln \pdl \right)^m \rig]
               \left(\alpha_0 \ln \qdl \right)^n  \left(\alpha_0^2 \ln \kdl \ln \pdl \right)^r.
\label{eq:nonperttausym}
\end{alignat}
\noindent
The fact mentioned earlier that the zeroth order vertex contribution comes from the longitudinal component, $\gamma^\mu$,  imposes the condition that there can be no leading order
 term in any transverse component. Consequently ${\cal S}_{0000}=0$.
It is important to note that the coefficients ${\cal{A}}$ and ${\cal{S}}$ are
constants in the above expressions and these depend on indices $m,n,r$. These are labelled by $mnrr$
to make it easy to read off that such a term contributes at
${\cal{O}}(\alpha_0^{m+n+r+r})$. Expanding Eqs.~(\ref{eq:nonperttauanti},~\ref{eq:nonperttausym}) to ${\cal{O}}(\alpha_0^3)$:
\small
\begin{alignat}{5}
\tau_i^{anti}(p^2,k^2,q^2)&= \alpha_0            &{\cal{A}}_{1000}&\lef(\ln\kdl  - \ln\pdl \rig)&&\nn \\
                          &+ \alpha_0^2 \Bigg\{  &{\cal{A}}_{2000} & \lef(\ln^2\kdl - \ln^2\pdl\rig)
                                               &+{\cal{A}}_{1100}& \lef(\ln\kdl-\ln\pdl\rig) \ln\qdl\Bigg\}
                                              \nn\\
                          &+ \alpha_0^3 \Bigg\{& {\cal{A}}_{3000} &\lef(\ln^3\kdl-\ln^3\pdl\rig)
                                             &+{\cal{A}}_{2100}& \lef(\ln^2\kdl-\ln^2\pdl\rig)\ln\qdl \nn\\
                          &                  &+{\cal{A}}_{1200} &\lef(\ln\kdl-\ln\pdl\rig)\ln^2\qdl\,
                                             &+\,{\cal{A}}_{1011}& \lef(\ln\kdl-\ln\pdl\rig)\ln\kdl\,\ln\pdl\Bigg\}+{\cal{O}}(\alpha^4)\,,\nn\\
\label{eq:perttauanti}
\end{alignat}
\vspace*{-12mm}
\begin{alignat}{6}
\tau_i^{sym}(p^2,k^2,q^2)&= \alpha_0\,\,\Bigg\{   &&{\cal{S}}_{1000} \Bigg(\ln\kdl     &+& \ln\pdl  & \Bigg)&                &+& 2\,&{\cal{S}}_{0100}&\ln\qdl  \,\,\Bigg\}  \nn \\
                         &+ \alpha_0^2  \Bigg\{   &&{\cal{S}}_{2000} \Bigg(\ln^2\kdl  &+& \ln^2\pdl & \Bigg)&                &+& 2\,&{\cal{S}}_{0200}&\ln^2\qdl       \nn\\
                         &                      &+&{\cal{S}}_{1100} \Bigg(\ln\kdl    &+&\ln\pdl   & \Bigg)&\ln\qdl         &+& 2\,&{\cal{S}}_{0011}&\ln\kdl \ln\pdl\,\,\Bigg\} \nn\\
                         &+ \alpha_0^3 \Bigg\{   & &{\cal{S}}_{3000} \Bigg(\ln^3\kdl  &+&\ln^3\pdl  & \Bigg)&               &+& 2\,&{\cal{S}}_{0300}&\ln^3\qdl  \nn\\
                         &                     &+&{\cal{S}}_{2100} \Bigg(\ln^2\kdl  &+&\ln^2\pdl  & \Bigg)&\ln\qdl        &+&    &{\cal{S}}_{1200}&\lef(\ln\kdl  + \ln\pdl \rig)\,\ln^2\qdl  \nn\\
                         &                     &+&{\cal{S}}_{1011} \Bigg(\ln\kdl    &+&\ln\pdl    & \Bigg)&\ln\kdl\ln\pdl &+& 2\,&{\cal{S}}_{0111}&\ln\kdl\,\ln\pdl\,\ln\qdl \,\,\Bigg\}
                         +{\cal{O}}(\alpha_0^4)\,. \nn\\
\label{eq:perttausym}
\end{alignat}
\normalsize
One should keep in mind in the rest of this
section that the sum of ${m,n,r,r}$ adds up to the order of the expansion. Thus, for example at ${\cal{O}}(\alpha_0^2)$ one only has coefficients
(${\cal{A}}_{2000}$,\, ${\cal{A}}_{1100}$) in $\tau_i^{anti}$ and
(${\cal{S}}_{2000}$,\, ${\cal{S}}_{0200}$,\,${\cal{S}}_{1100}$,\,${\cal{S}}_{0011}$)  in $\tau_i^{sym}$.
In turn, the dependence of ${\cal A}_{mnrr}$ and ${\cal S}_{mnrr}$ on $\xi$ and $N_F$
can only happen such that the maximum power of each of them is $m+n+2r$, {\it i.e.} the order of $\alpha_0$ too.


\noindent
As mentioned earlier the dominant ultraviolet behaviour of the vertex to ${\cal O}(\al_0)$
is given by the longitudinal component~\cite{Kizilersu:1995iz},
Eq.~(\ref{eq:longitudinal}), the transverse vertex has no leading
logarithms, {\it i.e.} $ (\alpha_0^n \ln^n \Lambda^2)$ terms must vanish.
Consequently, in Eqs.~(\ref{eq:nonperttausym},~\ref{eq:perttausym}) the relation~:

\noindent
at ${\cal{O}}(\alpha_0\,\ln\Lambda^2)$~:
\be
{\cal{S}}_{1000}^i +\,{\cal{S}}_{0100}^i=0\quad .
\label{eq:doubleprime}
\ee
\noindent
at ${\cal{O}}(\alpha^2_0\,\ln^2\Lambda^2)$~:
\be {\cal{S}}_{2000}^i+{\cal{S}}_{0200}^i+{\cal{S}}_{0011}^i+{\cal{S}}_{1100}^i=0\quad .
\label{eq:secondorderS} \ee
\noindent
at ${\cal{O}}(\alpha^3_0\,\ln^3\Lambda^2)$~:
\be
{\cal{S}}_{2100}^i+{\cal{S}}_{3000}^i+{\cal{S}}_{0300}^i+{\cal{S}}_{1011}^i+{\cal{S}}_{0111}^i+{\cal{S}}_{1200}^i=0\quad.
\label{eq:thirdorderS}
\ee

\noindent
and in general at ${\cal{O}}(\alpha^u_0\,\ln^u\Lambda^2)$~:
\be
\sum_{nr=0}^{u}\,{\cal{S}}_{m=[u-n-2r]nrr}^i=0\quad,
\ee
must hold.

\noindent
Our aim is to determine the conditions on the constants
${\cal{A}}^i_{mnrr}$ and ${\cal{S}}^i_{mnrr}$ for $i=2,3,6,8$
imposed by the fact that the
fermion and photon propagators satisfy the appropriate Schwinger-Dyson
equations
and that these must be multiplicatively renormalizable. These constraints must,
of course, be fulfilled by the full 3-point vertex.
In the weak coupling limit, perturbative calculation of the
relevant Feynman graphs will give explicit values for these constants.
However, the $\tau_i$'s
that enter here determine not the full vertex, but projections defined by the
Schwinger-Dyson equations of the next section.
\section{Unquenched Schwinger-Dyson Calculations}
\label{sec:unquenched}
\subsection{Fermion propagator }
\label{subsec:unfermion}
\begin{figure}[h]
\begin{center}
\refstepcounter{figure}
\addtocounter{figure}{-1}
\hspace{0.5cm} ~\epsfig{file=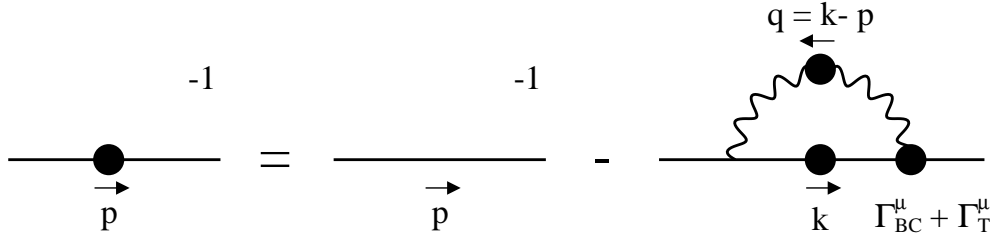,width=410pt}
\end{center}
\vspace{-0.5cm}
\hspace{1cm}
\caption{Unquenched Schwinger-Dyson equation for fermion propagator.}
\label{fig:fermion}
\end{figure}
%
%
\noindent
The Schwinger-Dyson equation for the fermion propagator displayed in
Fig.~\ref{fig:fermion}
 can be written as~:
\be
-{\ii}S_F^{-1}(p)=-{\ii}S_F^{{0}^{-1}}(p)
-\int_M\frac{d^4k}{(2\pi)^4}(-{\ii}e\Gamma^{\mu}(p,k;q))\,{\ii}S_F(k)\,
(-{\ii}e\gamma^{\nu})\,{\ii}\D_{\mu\nu}(q)\;.
\label{eq:ferone}
\ee
Substituting the form of the longitudinal part of the photon propagator from
Eq.~(\ref{eq:photonprop}) and using the Ward-Green-Takahashi identity
of Eq.~(\ref{eq:WGT}),
we can rewrite Eq.~(\ref{eq:ferone}) as~:
\be
{\ii}S_F^{-1}(p)&=&{\ii}S_F^{{0}^{-1}}(p)
-e^2\int_M\frac{d^4k}{(2\pi)^4}\Bigg\{\;\;\;\;
\Gamma^{\mu}(p,k;q)\,S_F(k)\,\gamma^{\nu}\,\D_{\mu\nu}^T(q)\nn\\
&& \hspace{43mm}
+\,\xi\,\lef(S_F^{-1}(k)-S_F^{-1}(p)\rig)\,S_F(k)\,\frac{\not\!q}{q^4}
\Bigg\}\;,\nn\\\nn\\
&=&{\ii}S_F^{{0}^{-1}}(p)
-e^2\int_M\frac{d^4k}{(2\pi)^4}\Bigg\{\;\;\;\;
\Gamma^{\mu}(p,k;q)\,S_F(k)\,\gamma^{\nu}\,\D_{\mu\nu}^T(q)\nn\\
&& \hspace{43mm}
+\,\xi\,\lef(\frac{\q}{q^4}-\,S_F^{-1}(p)\,S_F(k)\,\frac{\not\!q}{q^4}\rig)
\Bigg\}\,.
\label{eq:fertwo}
\ee
The second term in the integrand being an odd integral gives zero~:
\be
\int\frac{d^4k}{(2\,\pi)^4}\,\frac{\not\!q}{q^4}=0\, ,
\ee
if a translation invariant regularization is employed ~\cite{Dong:1994jr}.
After substituting the fermion and photon propagators,
Eqs.~(\ref{eq:fermionprop},~\ref{eq:photonprop}),
explicitly in Eq.~(\ref{eq:fertwo}), we obtain~:
\vspace{3mm} \be \frac{\p}{F(p^2,\Lambda^2)}&=&\p\,+\frac{{\ii}e^2}{(2\pi)^4}\,\int_M d^4k\,\Bigg\{
\;\;\;\;\Gamma^{\mu}(p,k;q)\,\frac{F(k^2)}{\kslash}\,\gamma^{\nu}\,
\frac{G(q^2)}{q^2}\,\lef(g_{\mu\nu}-\frac{q_{\mu}q_{\nu}}{q^2}\rig)\nn\\
&& \hspace{37mm}
-\,\xi\,\frac{\p}{F(p^2)}\,\frac{F(k^2)}{\kslash}\,\frac{\q}{q^4}\;\;\Bigg\}\,.
\ee
Multiplying this equation by $ \not\!\!p /4$, taking its trace and rearranging, we arrive at the following equation
for the fermion wavefunction renormalization~:
\vspace{5mm} \be \frac{1}{F(p^2,\Lambda^2)}&=&1+\frac{{\ii}e^2}{4p^2(2\pi)^4}\int_M \frac{d^4k}{k^2
q^2}\,\Tr\,\p\,\Bigg\{\;\;\;\;\,\Gamma^{\mu}(p,k;q)\kslash\,\gamma^{\nu}\,F(k^2)\,
G(q^2)\lef(g_{\mu\nu}-\frac{q_{\mu}q_{\nu}}{q^2}\rig)\nn\\
&& \hspace{55mm} -\,\frac{\xi}{q^2},\not\!p\,\kslash\,\not\!q\,\frac{F(k^2)}{F(p^2)}\,
\;\;\Bigg\}\,.
\label{eq:thfertwo}
\ee
We see this equation involves a particular projection
of the full vertex $\Gamma^{\mu}$.  To make this explicit
we substitute into this equation the general form given by
the Ball-Chiu longitudinal part, Eq.~(\ref{eq:longitudinal}),
and the transverse component, Eq.~(\ref{eq:transverse})~:
\vspace{4mm}
%
%
\be
\frac{1}{F(p^2,\Lambda^2)}&=&1+\frac{{\ii}e^2}{4p^2{(2\pi)}^4}\,
\int_M\frac{d^4k}{k^2q^2}\,
F(k^2)\nn\\\nn\\
&&
\times\,\Bigg\{
-\frac{\xi}{q^2}\,\frac{1}{F(p^2)}\,\Tr(\p\p\kslash\q)\nn\\
&& \hspace{13mm} +\,\lambda_1(p^2,k^2,q^2)\,G(q^2)\,\Tr
\Bigg[\p\,L_1^{\mu}(p,k,q)\kslash\,\g^{\nu}\lef(g_{\mu\nu}-\frac{q_{\mu}q_{\nu}}{q^2}\rig)
\Bigg]\nn\\
&& \hspace{13mm} +\,\lambda_2(p^2,k^2,q^2)\,G(q^2)\,\Tr\Bigg[\p\,L_2^{\mu}(p,k,q)\,\kslash\,\gamma^{\nu}
\lef(g_{\mu\nu}-\frac{q_{\mu}q_{\nu}}{q^2}\rig)\Bigg]\nn\\
&& \hspace{13mm} +\hspace{24mm}G(q^2)\,\Tr\Bigg[\p\,\Gamma^{\mu}_T(p,k,q)\,\kslash\,\gamma^{\nu}
\lef(g_{\mu\nu}-\frac{q_{\mu}q_{\nu}}{q^2}\rig)\Bigg]\Bigg\}\,, \label{eq:ferthree} \ee
where $d^4k=2\,\pi\,k^2\,dk^2\,\sin^2\Psi\,d\Psi$ and
 $\Psi$ is the angle between the 4-vectors $k$ and $p$.
To perform these integrals, we  move to Euclidean space
using the Wick rotation ($k_0 \rightarrow i k_0$, $k_i \rightarrow k_i$). After performing an explicit trace algebra in Eq.~(\ref{eq:ferthree}) and inserting
the transverse vertex, $\Gamma^{\mu}_T$, Eqs.~(\ref{eq:transverse},\ref{eq:Ts}), with its undetermined $\tau_i$'s, we obtain~:

\be
\frac{1}{F(p^2,\Lambda^2)}&=&1-\frac{e^2}{(2\pi)^3\,p^2}\,\int_E\,k^2\,dk^2\,\int_0^{\pi}\,\sin^2\Psi\,d\Psi\,
\frac{1}{k^2\,q^2}\nn\\[2.4mm]
&\times&
\Bigg\{
\,-\,\xi\,\frac{F(k^2)}{F(p^2)}\,\,\frac{p^2}{q^2}\,(k^2- k \cdot p)\nn\\
&& \hspace{4mm}
+\,\,F(k^2)\,G(q^2)
\Bigg[
\,\,\lambda_1^E(p^2,k^2,q^2) \lef\{ \frac{1}{q^2} \lef[-2\Delta^2-3q^2\, k \cdot p \,\rig] \rig\} \nn\\
&& \hspace{30mm}
+  \lambda_2^E(p^2,k^2,q^2)   \lef\{ \frac{1}{q^2}   \lef[ 2\,(k^2+p^2)\,\Delta^2 \rig] \rig\}
\nn\\[3mm]
&& \hspace{30mm}
+\,\tau_2^E(p^2,k^2,q^2)  \lef\{-(k^2+p^2)\,{\Delta}^2\right\} \nn\\
&& \hspace{30mm}
+\,\tau_3^E(p^2,k^2,q^2)\lef\{
2\Delta^2+\,3q^2 k \cdot p \rig\}\nn\\
&& \hspace{30mm}
+\,\tau_6^E(p^2,k^2,q^2)\lef\{-3\,(k^2-p^2)\,k \cdot p\rig\}\nn\\
&& \hspace{30mm}
+\,\tau_8^E(p^2,k^2,q^2)\lef\{-2\,\Delta^2\rig\}\;\;\Bigg]\Bigg\}\quad,
\label{eq:ferfour}
\ee
where ${\Delta}^2=(k \cdot p)^2-k^2p^2$.

\noindent Since multiplicative renormalizabilty  is closely related to the
ultraviolet
behaviour of the Green's functions, we make a general perturbative expansion
of
the non-perturbative fermion and photon wavefunction renormalizations in powers of leading logarithms as follows~:
\be
F(p^2,\Lambda^2)&=&\sum_{u=0}^{\infty} \,\alpha_0^u \,A_u \,\ln^u \pdl\,,
\label{eq:Fseries}\\
G(q^2,\Lambda^2)&=&\sum_{u=0}^{\infty} \,\alpha_0^u \,B_u \,\ln^u \qdl\,.
\label{eq:Gseries}
\ee
In this paper we will consider leading logarithms only in order to present the ideas and techniques and
postpone to a future paper the more involved next-to-leading order.
Of course, in perturbation theory the coefficients $A_{u}, B_{u}$ have
definite values. However, it is the general structure that multiplicative
renormalizability determines. We substitute these expansions into
Eq.~(\ref{eq:ferfour}) in order to calculate this. The photon wavefunction renormalization $G(q^2)$ depend on the momentum $q^2=k^2+p^2-2\,k \cdot p\,$
 therefore it has both a   radial and an  angular
component. However, the angular dependent part of this quantity only contributes to $1/F(p^2)$ beyond the leading order, and so here we can simply approximate $G(q^2)$ with $G(k^2)$.
 We can then carry out the angular integration in Eq.~(\ref{eq:ferfour}) after inserting the coefficients of the basis tensors, {\it i.e.}
 $\lambda_i$'s and $\tau_i$'s from Eqs.~(\ref{eq:longitcoeff},~\ref{eq:tauform})~:
\be
\frac{1}{F(p^2,\Lambda^2)}=1&+&\,\frac{\al_0 \,\xi}{4 \,\pi}\,\int_{p^2}^{\Lambda^2}\,\,\frac{dk^2}{k^2}\,
           \frac{F(k^2)}{F(p^2)}\nn\\
             &-& \frac{3\,\alpha_0}{8\,\pi}\,\,\int_{p^2}^{\Lambda^2}\,\,\frac{dk^2}{k^2}\,F(k^2)\,G(k^2)\,\lef[\frac{1}{2}\,\lef(\frac{1}{F(k^2)}-\,\frac{1}{F(p^2)}\rig)
                                                              +\lef({\overline{\tau}}^{anti}_f+ {\overline{\tau}}^{sym}_f\rig)\rig]\,,\nn\\
\label{eq:finalfersde}
\ee
where
\be
{\overline{\tau}}^{\,\,anti}_f &\equiv& \,
\,\beta_2\,\tau_2^{\,anti}+\lef(\beta_3-\gamma_3\rig)\,\tau_3^{\,anti}+\lef(\beta_6+\gamma_6\rig)\,\tau_6^{\,anti}-\beta_8\,\tau_8^{\,anti}\,,\nn\\
{\overline{\tau}}^{\,\,sym}_f &\equiv& \,
\,\delta_2\,\tau_2^{\,sym}+\lef(\delta_3-\varepsilon_3\rig)\,\tau_3^{\,sym}\,+\lef(\delta_6+\varepsilon_6\rig)\,\tau_6^{\,sym}\,-\delta_8\,\tau_8^{\,sym}\,.
\ee
\noindent
To evaluate this expression, we have to insert the
coefficients of the basis tensors, {\it i.e.} the $\tau_i^{\,anti,~sym}$ from Eqs.~(\ref{eq:nonperttauanti}, \ref{eq:nonperttausym}) into Eq.~(\ref{eq:finalfersde}).
%
$\Lambda$ is the ultraviolet cut-off for the momentum $k$ introduced in Eq.~(\ref{eq:finalfersde}) in accord with
 Eqs.~(\ref{eq:mrfermion},\,\ref{eq:mrphoton},\,\ref{eq:mrvertex},\,\ref{eq:Fseries},\,\ref{eq:Gseries}).
One observes from Eq.~(\ref{eq:finalfersde}) that there is no contribution to $1/F(p^2,\Lambda^2)$ from the $\lambda_1$ part of the longitudinal vertex, Eq.~(\ref{eq:longitudinal}), but only from
$\lambda_2$.
On laboriously integrating Eq.~(\ref{eq:finalfersde}) and using
Eqs.~(\ref{eq:Fseries},~\ref{eq:Gseries}) we arrive at~:
\be
&&\frac{1}{F(p^2,\Lambda^2)} = 1\,-\,\Bigg\{
\frac{\xi}{4\,\pi}\,\frac{1}{F(p^2)}\,\sum_{u=0}^{\infty} \alpha_0^{(u+1)} \,\frac{A_u}{(u+1)}\,\ln^{u+1}\pdl\nn\\[4mm]
&&\hspace*{5mm}+\frac{3}{16\,\pi}\,\lef[\frac{1}{F(p^2)}\sum_{u=0}^{\infty}\,\sum_{t=0}^{\infty}\, \alpha_0^{u+t+1} \,\frac{A_u\,B_t}{(u+t+1)}\,\ln^{u+t+1}\pdl
%
%
%
-\,\sum_{t=0}^{\infty}\, \alpha_0^{t+1} \,\frac{B_t}{(t+1)}\,\ln^{t+1}\pdl\rig]\nn\\[4mm]
&&\hspace*{5mm}-\frac{3}{8\,\pi}\sum_{u=0}^{\infty}\,\sum_{t=0}^{\infty} \,A_u\,B_t\,\sum_{m=1}^{\infty}\,\sum_{n=0}^{\infty}\,\sum_{r=0}^{\infty}
                                                      \alpha_0^{u+t+m+n+2r+1}\,\ln^{u+t+m+n+2r+1}\pdl  \nn\\[5mm]
&&\hspace*{10mm}                     \quad \times\,\,{\overline{\cal{A}}}^{\,f}_{mnrr}\,\, \lef[ \frac{1}{\lef(u+t+m+n+r+1\rig)} - \frac{1}{\lef(u+t+n+r+1\rig)}\rig]\nn\\ \nn\\
&&\hspace*{5mm}-\frac{3}{8\,\pi}\sum_{u=0}^{\infty}\,\sum_{t=0}^{\infty} \,A_u\,B_t\,\sum_{n=0}^{\infty}\,\sum_{m=0}^{\infty}\,\sum_{r=0}^{\infty}
                                                      \alpha_0^{u+t+m+n+2r+1}\,\ln^{u+t+m+n+2r+1}\pdl  \nn\\[0.4cm]
&&\hspace*{10mm} \quad \times\,\,{\overline{\cal{S}}}^{\,f}_{mnrr}\,\, \lef[ \frac{1}{\lef(u+t+m+n+r+1\rig)} + \frac{1}{\lef(u+t+n+r+1\rig)}\rig] \Bigg\}\,,
                      \label{eq:nonpertFresult}
\ee
where
\be
{\overline{\cal{A}}}^{\,\,f}_{mnrr}&\equiv&
\beta_2\,{\cal{A}}_{mnrr}^{2}+\lef(\beta_3-\gamma_3\rig)\,{\cal{A}}_{mnrr}^{3}+\lef(\beta_6+\gamma_6\rig)\,{\cal{A}}_{mnrr}^{6}-\beta_8\,{\cal{A}}_{mnrr}^{8} \,,\nn\\
{\overline{\cal{S}}}^{\,\,f}_{mnrr}&\equiv&
\delta_2\,{\cal{S}}_{mnrr}^{\,2}\,\,+\lef(\delta_3-\varepsilon_3\rig)\,{\cal{S}}_{mnrr}^{\,3}\,\,+\lef(\delta_6+\varepsilon_6\rig)\,{\cal{S}}_{mnrr}^{\,6}\,\,-\delta_8\,{\cal{S}}_{mnrr}^{\,8}\,.
\label{eq:AS}
\ee
In order to rearrange the infinite sums in  Eq.~(\ref{eq:nonpertFresult}) in terms of powers of $\alpha_0$,
we convert some of the infinite sums to finite sums~:

\be
\framebox{$
\begin{array}{rcl}
\displaystyle\frac{1}{F(p^2,\Lambda^2)}& = &1\,-\,\Bigg\{
\displaystyle\frac{\xi}{4\,\pi}\,\frac{1}{F(p^2)}\,\sum_{u=0}^{\infty}\,\frac{A_u}{(u+1)}\,\alpha_0^{u+1}\,\ln^{u+1}\pdl\nn\\[0.6cm]
&-&\displaystyle\frac{3}{16\,\pi}\,\sum_{u=1}^{\infty}\,{\alpha_0}^{u+1}\ln^{u+1} \pdl\,\Bigg\{\sum_{a=1}^{u}\,A_a\,B_{u-a}\frac{1}{(u+1)}\nn\\[0.6cm]
&&\hspace{5cm}+\displaystyle\sum_{b=1}^{u}\,\sum_{a=0}^{u-b}\,(-1)^b\,A_b\,A_a\,B_{u-b-a}\,\frac{1}{(u-b+1)}\,\Bigg\}\nn\\[0.6cm]
&+&\displaystyle\frac{3}{8\,\pi}\,\sum_{u=1}^{\infty}\,{\alpha_0}^{u+1}\ln^{u+1} \pdl\lef( H_u + {\overline H}_u\rig) \Bigg\}\,,
\end{array}$}\\
                      \label{eq:nonpertFresult2}
\ee

\noindent
where
\footnotesize
\be
H_u&=&\sum_{b=1}^{u}\,\sum_{c=1}^{b}\,\sum_{d=0}^{b-c}\sum_{a=1}^{c}\,A_d\,B_{b-c-d}\, R_{u-b}\,
\Bigg\{\frac{1}{\lef[\frac{1}{2}(u+b)+1\rig]} - \frac{1}{\lef[\frac{1}{2}(u+b)-a+1\rig]}\Bigg\}\, \, {\overline{\cal{A}}}^f_{a(c-a)\frac{u-b}{2}\frac{u-b}{2}}\,,\nn\\[-4mm]
\label{eq:Hfer}\\\nn\\[2mm]
{\overline H}_u&=&\sum_{b=0}^{u}\,\sum_{c=0}^{b}\,\sum_{d=0}^{b-c}\sum_{a=0}^{c}\,A_d\,B_{b-c-d}\, R_{u-b}\,
\Bigg\{\frac{1}{\lef[\frac{1}{2}(u+b)+1\rig]} + \frac{1}{\lef[\frac{1}{2}(u+b)-a+1\rig]}\Bigg\}\,\, {\overline{\cal{S}}}^f_{a(c-a)\frac{u-b}{2}\frac{u-b}{2}}\,,\quad\quad\nn\\[-4mm]
\label{eq:Hbarfer}
\ee
\normalsize
with
\be
&&
R_{j}=\left\{\begin{array}{ll}
                    1 &  \mbox{if}\quad j \quad \mbox {is even} \nn\\
                    0 &  \mbox{if} \quad j \quad \mbox {is odd}\quad.
 \end{array}\right.\\
 \label{eq:rij}
\ee

\noindent
The above expression for the fermion wavefunction renormalization, $1/F(p^2,\Lambda^2)$, is the exact non-pertubative calculation for the
massless fermions in a general covariant gauge at leading logarithmic order. In this equation the
${\cal{A}}^i_{mnrr}$'s and ${\cal{S}}^i_{mnrr}$'s are the constants to be constrained by multiplicative renormalization.
For the purpose of explaining how this works, we will first implement it order-by-order then we generalize. To do this,
we expand the fermion wavefunction renormalization, Eq.~(\ref{eq:nonpertFresult2}), in ${\cal{O}}(\alpha^4)$~:

\be
\framebox{$
\begin{array}{rcl}
\displaystyle\frac{1}{F(p^2,\Lambda^2)}&=&\displaystyle
1+\displaystyle \frac{1}{4\,\pi}\Bigg\{
-\,\mbox{\boldmath $ \al $}_{\bf{0}}\,\xi\,{\bf{ln}}{\bf{\pdl}}\nn\\[7mm]
&-&\displaystyle
{\mbox{\boldmath $ \al $}}_{\bf{0}}^{\bf{2}}\,{\bf{ln^2}}{\bf{\pdl}}
\,\lef[
\,-\lef(\frac{\xi}{2}+\frac{3}{8}\rig)\,A_1
\displaystyle
\,+\,\frac{3}{4}\,\lef({\overline{\cal{A}}}^{\,f}_{1000}\,-\,{\overline{\cal{S}}}^{\,f}_{1000}\rig)\rig]\nn\\[7mm]
&-&\displaystyle
{\mbox{\boldmath $ \al $}}_{\bf{0}}^{\bf{3}}\,{\bf{ln^3}}{\bf{\pdl}}\,\,\Bigg[
-\,\lef(\frac{\xi}{2}+\frac{3}{8}\rig)\,A_1^2
+\lef(\frac{4\,\xi}{3}\,+\,1\rig)\,A_2-\frac{A_1\,B_1}{8}\nn\\[6mm]
&& \hspace{20mm}
\displaystyle
+\,\frac{1}{4}\,(A_1+B_1)\,\lef({\overline{\cal{A}}}^{\,f}_{1000}\,-\,{\overline{\cal{S}}}^{\,f}_{1000}\,\rig)\,
+\,{\overline{\cal{A}}}^{\,f}_{2000}\,+\,\frac{1}{4}\,\,{\overline{\cal{A}}}^{\,f}_{1100}\nn\\[3mm]
&& \hspace{20mm}
\displaystyle
-\,\frac{3}{4}\,\,{\overline{\cal{S}}}^{\,f}_{2000}
+\,\frac{1}{4} \,{\overline{\cal{S}}}^{\,f}_{0200}
-\,\frac{1}{4}\,\,{\overline{\cal{S}}}^{\,f}_{0011}
%
%
\Bigg]
\nn\\[6mm]
&-&\displaystyle
{\mbox{\boldmath $ \al $}}_{\bf{0}}^{\bf{4}}\,{\bf{ln^4}}{\bf{\pdl}}\Bigg[
-A_3\,\lef(\frac{9}{16} + \frac{3}{4}\, \xi\rig)\,+\,A_1\,A_2\,\lef(\frac{1}{8}+ \frac{1}{6}\,\xi\rig)\nn\\[4mm]
&& \hspace{20mm}
\displaystyle
-\,\frac{1}{4}\, A_1^2\, B_1
+\,\frac{9}{16}\,A_2\,B_1
-\,\frac{1}{16}\,A_1\,B_2\nn\\[4mm]
&& \hspace{20mm}
\displaystyle
+\,\frac{1}{8}\,(A_2+B_2)\, \lef({\overline{\cal{A}}}^{\,f}_{1000}\,-\,{\overline{\cal{S}}}^{\,f}_{1000}\rig)\,
+\,\frac{1}{8}\,A_1\,B_1\lef({\overline{\cal{A}}}^{\,f}_{1000}\,-\,{\overline{\cal{S}}}^{\,f}_{1000}\rig)\nn\\[4mm]
%
%
&& \hspace{20mm}
\displaystyle
+\,(A_1\,+\,B_1)\,\Big(
+\,\frac{3}{8}\,{\overline{\cal{A}}}^{\,f}_{2000}\,+\,\frac{1}{8}\,{\overline{\cal{A}}}^{\,f}_{1100}\, \nn\\[4mm]
&& \hspace{47mm}
\displaystyle
-\,\frac{1}{4}\,{\overline{\cal{S}}}^{\,f}_{2000}
-\,\frac{1}{8}\,{\overline{\cal{S}}}^{\,f}_{0011}\,+\,\frac{1}{8}\,{\overline{\cal{S}}}^{\,f}_{0200}\Big)\nn\\[4mm]
&& \hspace{20mm}
\displaystyle
+\,\frac{1}{4}\,{\overline{\cal{A}}}^{\,f}_{1011}\,+\,\frac{1}{8}\,{\overline{\cal{A}}}^{\,f}_{1200}\,
+\,\frac{3}{8}\,{\overline{\cal{A}}}^{\,f}_{2100}\,+\,\frac{9}{8}\,{\overline{\cal{A}}}^{\,f}_{3000}\nn\\[4mm]
&& \hspace{20mm}
\displaystyle
+\,\frac{1}{8}\,{\overline{\cal{S}}}^{\,f}_{0111}\,+\,\frac{3}{8}\,{\overline{\cal{S}}}^{\,f}_{0300}
-\,\frac{1}{8}\,{\overline{\cal{S}}}^{\,f}_{1011}\,+\,\frac{1}{4}\,{\overline{\cal{S}}}^{\,f}_{1200}\,
-\,\frac{3}{4}\,{\overline{\cal{S}}}^{\,f}_{3000}\,\Bigg]\nn\\[4mm]
\displaystyle
&-&{\mbox{\boldmath $ {\cal O}(\al $}}_{\bf{0}}^{\bf{5}})\qquad\qquad\Bigg\}\,.
\end{array}$}\\
\label{eq:pertfermionresult}
\ee
\normalsize
\baselineskip=7.mm
\noindent
Eqs.~(\ref{eq:doubleprime}-\ref{eq:thirdorderS}) have been input to obtain this expression.
Eqs.~(\ref{eq:nonpertFresult},~\ref{eq:nonpertFresult2},~\ref{eq:pertfermionresult}) illustrate how
the fermion wavefunction renormalization depends on the explicit form
of the
full 3-point vertex. As we shall see in Sect. IV, the expansion to ${\cal O}(\alpha^4\,\ln^4)$ is the minimum
order at which we can recognize the pattern of constraints.
\newpage
\subsection{Photon propagator }
\label{subsec:unphoton}
\baselineskip=6.8mm
\noindent
Next we discuss the Schwinger-Dyson equation for the gauge boson.
This equation has some different features from the fermion SDE.
Now, the two fermion legs have to be treated equally.
We can ensure this symmetry property by dividing the external
momentum flow equally in the loop as shown in Fig.~\ref{fig:photon}~:
%
%
%
\begin{figure}[h]
\begin{center}
\refstepcounter{figure}
\addtocounter{figure}{-1}
~\epsfig{file=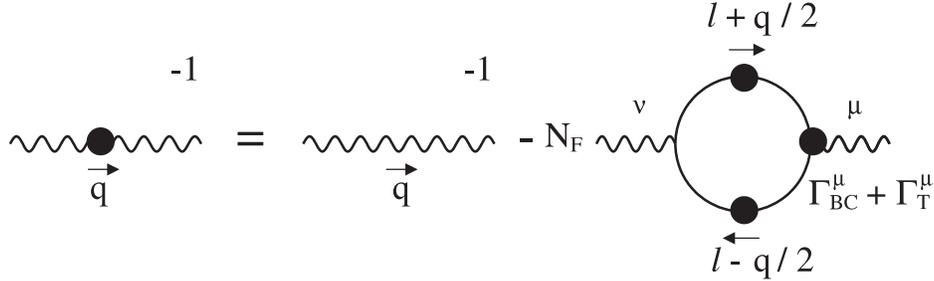,width=350pt}
\end{center}
\vspace{-1.8cm}
\hspace{4.5cm}
\caption{Unquenched Schwinger-Dyson equation for photon propagator.}
\label{fig:photon}
\end{figure}

\noindent
Using the Feynman rules, Fig.~\ref{fig:photon} can be expressed as~:
\be
-{\ii}\D_{\mu\nu}^{-1}(q)=-{\ii}{\D_{\mu\nu}^0}^{-1}(q)
-(-1)N_F\int_M \frac{d^4\e}{(2\pi)^4}
\Tr\lef[\left(-{\ii}e\Gamma^{\mu}({\ei},{\ea};q)\right){\ii}S_F(\ea)(-{\ii}e\g^{\nu}){\ii}S_F(\ei)\rig]\,,\nn\\
\label{eq:photonone},
\ee
which can be symbolically written as $\Delta^{-1}_{\mu\nu}(q)={\Delta_{\mu \nu}^{0}}^{-1}(q) \> +  \> \Pi_{\mu \nu}(q)$,
where $\Pi_{\mu\nu}$ is the photon self-energy and $
~~~\ea \equiv \at ~,~
\ei \equiv \eek.
$

\noindent
The definitions of the fermion and photon propagators~\footnotemark{{\footnotetext{Where appropriate, we  denote the fermion and photon
wavefunction renormalization functions as $F(p)$ or $F(p^2)$ and $G(p)$ or
$G(p^2)$, respectively. Where we wish to emphasize that the quantities are unrenormalized,
$\Lambda^2$ will be added to the list of arguments --- with similar conventions for
the renormalized quantities, for instance $F_R(p)$ and $G_R(p)$.}}}
 are given already in Sect.~\ref{sec:vertex},
\sbe
{\ii}S_F(\ea)&=&{\ii}\,F(\ea)/{\es_{+}} \quad,\\[5mm]
{\ii}\D^{\mu\nu}(q)&=&-\,\frac{{\ii}}{q^2}
\lef[G(q)\lef(g^{\mu\nu}-\frac{q^{\mu}q^{\nu}}{q^2}\rig)
+\,\xi\,\frac{q^{\mu}q^{\nu}}{q^2}\rig]\;.
\label{eq:photontwo}
\see
Eq.~(\ref{eq:photonone}) must satisfy the photon Ward identity, $q^\mu \Delta_{\mu \nu}^{-1}={q_\nu q^2}/{\xi}$,
which is, of course, fulfilled by the bare propagator in Eq.~(\ref{eq:photonone}). Consequently, the loop graph of Fig.~4 must be transverse.
Contracting Eq.~(\ref{eq:photonone}) with $q^\mu$ and using the Ward-Green-Takahashi identity of Eq.~(\ref{eq:WGT}), this transversality requires~:
\be
q_\mu\, \Pi^{\mu\nu} = \frac{i N_F e^2}{q^2 (2\pi)^4} \int_M d^4 \e \, \Tr\lef[ \gamma^\nu \lef\{ S_F(\ea) -S_F(\e_-)\rig\}\rig]\,=\,0\, .
\ee
If dimensional regularization is used, then this integral is automatically zero. However with  cut-off regularization, this is not the case. Then $\Pi_{\mu\nu}$ is not entirely transverse. To extract the correct component, we introduce the following tensor~\cite{Brown, Mike}~:
\be
P_{\mu\nu}=\frac{1}{3q^4}\,\lef(4q_{\mu}q_{\nu}-\,q^2 g_{\mu\nu}\rig)\quad.
\ee
\noindent
Projecting Eq.~(\ref{eq:photonone}) with $P_{\mu\nu}$ allows us to remove the potentially
quadratically divergent term in 4-dimensions, and project out the required
ultraviolet logarithmically divergent terms. It is easy to check that this leaves the correct leading logarithms.
We then have a scalar equation for the
photon wavefunction renormalization~:
\be \frac{1}{G(q^2,\Lambda^2)}&=&1+\,N_F\,\frac{{\ii}\,\al_0}{4\pi^3}\, \int_M
\frac{d^4\e}{\e_{+}^2\,\e_{-}^2}\,\,F(\e_{-})\,F(\e_{+})
\,P_{\mu\nu}\,
\Tr\lef[\,\Gamma^{\mu}_{F}\lef({\ei},{\ea},q)\right)\es_{+}\,\g^{\nu}\es_{-}\,\rig]\;.
\ee
Recalling the definition of the vertex of
Eqs.~(\ref{eq:longitudinal}-~\ref{eq:Ts}), we obtain~:
\vspace{3mm}
\be \frac{1}{G(q^2,\Lambda^2)}=1+\,N_F\,\frac{{\ii}\al_0}{4\pi^3}\,
\int_M\frac{d^4\e} {\ea^2\,\ei^2}\, F(\ei)\,F(\ea) &&\nn\\
\times\,
P_{\mu\nu}\,\Bigg\{\;\;\;\;\,\lambda_1^M(\ei^2,\ea^2,q^2)\,&& \Tr\lef(\g^{\mu}\es_{+}\,\g^{\nu}\es_{-}\rig)\nn\\
+\,\lambda_2^M(\ei^2,\ea^2,q^2)\,&& \Tr\lef(4\,\g^{\mu}\es_{+}\es\,\e^{\nu}\es_{-}\rig)\nn\\
+\hspace{27mm}&& \Tr\lef(\Gamma_T^{\mu}\,\es_{+}\,\g^{\nu}\,\es_{-}\rig)\;
\Bigg\}\;.
\label{eq:photonthree}
\ee

\noindent
Moving to Euclidean space, we  perform a Wick rotation. Substituting
\noindent
$d^4\e=2 \pi \,\e^2 \, d \e^2 \,d \Psi\,\sin^2 \Psi$ and the form of the transverse vertex from Eqs.~(9,~10), and then taking the traces leads to~:
\be \frac{1}{G(q^2,\Lambda^2)}=1&-&\frac{\al_0\,N_F}{6\pi^2\,q^2}\,\int_E
\frac{\e^2\,d\e^2}{\ea^2\,\ei^2} \int_0^\pi \sin^2\Psi\,d\Psi \, F(\ea)\,F(\ei)\nn\\[4mm]
&\times& \hspace*{3mm} \Bigg\{
\,\,\,\, 2 \, \lambda_1^E(\ei^2,\ea^2,q^2)\,
\lef\{\,16\,\frac{(\ld)^2}{q^2} - \,3\,q^2 -\,4\,\e^2\rig\}\nn\\[2mm]
&&\hspace*{5mm}
+\, 2 \,\lambda_2^E(\ei^2,\ea^2,q^2)\,
\lef\{-\lef(\,16\,\frac{\e^2}{q^2}-2\rig)\,(\ld)^2
         +\,4\,\e^4 + q^2\,\e^2\rig\}\nn\\[2mm]
&&\hspace*{5mm}
+\;\;\;\tau^E_2({\ei}^2,{\ea}^2,q^2)\,
\lef\{2\,\lef(4 \e^2+\ {q^2}\rig)\,\Delta^2 \rig\}\nn\\[2mm]
&&\hspace*{5mm}
+\;\;\; \tau^E_3({\ei}^2,{\ea}^2,q^2)\,
\lef\{-8\,\Delta^2- 3\,q^2\,\lef(4 \e^2- q^2\rig) \rig\}\nn\\[2mm]
&&\hspace*{5mm}
+\;\;\;\tau^E_6({\ei}^2,{\ea}^2,q^2)\,
\lef\{6\,\ld\,\lef(4\e^2-\ {q^2}\rig)\rig\}\nn\\[2mm]
&&\hspace*{5mm}
+\;\;\; \tau^E_8({\ei}^2,{\ea}^2,q^2)\,\lef\{8\,\Delta^2\rig\}\Bigg\}\;,
\label{eq:photonfour}
\ee
where $\Delta^2=(\ld)^2-\e^2q^2$ and the photon Schwinger-Dyson equation picks out loop momentum regions where
$ {\ea}^2  \sim {\ei}^2 \sim {\e}^2 \gg q^2 $. This allows us to carry out the angular integrals in Eq.~(\ref{eq:photonfour}) for the leading log terms
after inserting $\lambda_i$'s and $\tau_i$'s from Eqs.~(\ref{eq:longitcoeff},~\ref{eq:tauform}). This gives~:
\be
\frac{1}{G(q^2,\Lambda^2)} = 1 +
\frac{\alpha_0\,N_F}{3\,\pi}\,\int_{q^2}^{\Lambda^2}\,\frac{d\e^2}{\e^2}\,F^2(\e)\,\Bigg\{
\frac{1}{F(\e)}+\,\frac{3}{4}\,{\overline{\tau}}^{\,sym}_{\gamma} \Bigg\}\,,
\label{eq:finalsdeG} \ee
where
\be
{\overline{\tau}}^{sym}_{\gamma} \equiv\,\,(\delta_2+\varepsilon_2)\,\tau_2^{\,sym}-\,(\delta_3+\varepsilon_3)\,\tau_3^{\,sym}
                                  +\,(\delta_6+\varepsilon_6)\,\tau_6^{\,sym}-\,(\delta_8+\varepsilon_8)\,\tau_8^{\,sym}\,.
\ee
\noindent
This time the explicit longitudinal contribution comes from $\lambda_1$; $\lambda_2$ does not contribute to the leading log's.
Using Eq.~(\ref{eq:Fseries}) and performing the radial integration, Eq.~(\ref{eq:finalsdeG}) yields~:
\be
&&\frac{1}{G(q^2,\Lambda^2)} = 1 - \frac{N_F}{3\,\pi}\, \alpha_0\, \ln\qdl \nn\\[2mm]
&&\hspace*{10mm}+ \frac{N_F}{3\,\pi}\, \Bigg\{
   -\sum_{u=1}^{\infty}\alpha_0^{u+1}\,\frac{A_u}{(u+1)}\, \ln^{u+1}\qdl \nn\\[2mm]
&& \hspace*{10mm} -\frac{3}{2}\,\sum_{u=0}^{\infty}\,A'_u\,
\sum_{m=0}^{\infty}\,\sum_{n=0}^{\infty}\,\sum_{r=0}^{\infty}\,
       \alpha_0^{u+n+m+2r+1}\, \ln^{u+n+m+2r+1}\qdl\,\, \frac{{\overline{\cal{S}}}^{\,\,\gamma}_{mnrr}}{(u+n+2r+1)}\Bigg\}\,,\nn\\
\ee
\be
\mbox{where}\quad {\overline{\cal{S}}}^{\,\,\gamma}_{mnrr}&\equiv&
(\delta_2+\varepsilon_2)\,{\cal{S}}^2_{mnrr}-(\delta_3+\varepsilon_3)\,{\cal{S}}^3_{mnrr}+(\delta_6+\varepsilon_6)\,{\cal{S}}^6_{mnrr}-(\delta_8+\varepsilon_8)\,{\cal{S}}^8_{mnrr}\,,\qquad
\label{eq:Sphoton}\\[3mm]
A'_u&\equiv&\sum_{d=1}^u \frac{2d}{u}\,A_d\,A_{u-d}\,.
\ee
Evaluating the multiple sums using the symmetries and rearranging terms with respect to powers of $\alpha_0$ yields~:
\be
\framebox{$
\begin{array}{rcl}
\displaystyle \frac{1}{G(q^2,\Lambda^2)} = 1 &-& \displaystyle\frac{N_F}{3\,\pi}\, \alpha_0\, \ln\qdl \nn\\
&-& \displaystyle\frac{N_F}{3\,\pi}\, \sum_{u=1}^{\infty}\,\alpha_0^{u+1}\, \ln^{u+1}\qdl\Bigg\{
   \,\frac{A_u}{(u+1)}+\,\frac{3}{2}\,K_u \Bigg\} \,,
   \end{array}$}
   \label{eq:photonsde}
\ee
where
\be
K_u&=&\sum_{b=0}^u\,\frac{R_{u-b}}{(u+a-c+1)}\, \sum_{c=0}^b\,\sum_{a=0}^c\,
\lef(\sum_{d=1}^{(b-c)}\,\frac{2d}{(b-c)}\,A_d\,A_{(b-c-d)}\rig)\,\, {\overline{\cal{S}}}_{a(c-a)\frac{(u-b)}{2}\frac{(u-b)}{2}}\, ,
\label{eq:Ku}
\nn\\
K_0&=&0\quad ,
\ee
with $R_j$ defined by Eq.~(\ref{eq:rij}).
\noindent
Employing the expansion of the transverse vector coefficients introduced in
Eqs.~(\ref{eq:tauform}-\ref{eq:nonperttausym}),
we can then write $1/G(q^2)$ analogous to the fermion
result for $1/F(p^2)$ of Eq.~(\ref{eq:pertfermionresult}), after performing the many
integrals~:
\be
\framebox{$\begin{array}{rcl}
\nn\\
\displaystyle\frac{1}{G(q^2,\Lambda^2)}&=&\displaystyle1+\frac{N_F}{3\,\pi}
\Bigg\{ \;\;-\,\mbox{\boldmath $ \al $}_{\bf{0}}\,{\displaystyle \bf{ln}}{\bf{\qdl}}
\nn\\[5mm]
\displaystyle &-&\,{\mbox{\boldmath $ \al $}}_{\bf{0}}^{\bf{2}}\,{\displaystyle\bf{ln^2}}{\bf{\displaystyle\qdl}}
\displaystyle
\,\Bigg[
\;\frac{A_1}{2}\,-\,\frac{3}{4}\,\,{\overline{\cal{S}}}^{\,\,\gamma}_{1000}\,\Big]\nn\\[5mm]
\displaystyle &-&\,{\mbox{\boldmath $ \al $}}_{\bf{0}}^{\bf{3}}\,{\displaystyle\bf{ln^3}}{\displaystyle\bf{\qdl}}
\displaystyle
\,\Bigg[
\,\frac{A_2}{3}\,
-\,\frac{1}{2}\,A_1\,\,{\overline{\cal{S}}}^{\,\,\gamma}_{1000}
-\,\frac{1}{4}\,\,{\overline{\cal{S}}}^{\,\,\gamma}_{2000}\,
+\,\frac{3}{4}\,\,{\overline{\cal{S}}}^{\,\,\gamma}_{0200}\,
-\,\frac{1}{4}\,\,{\overline{\cal{S}}}^{\,\,\gamma}_{0011}\,
\Bigg]\nn\\[5mm]
\displaystyle &-&\,{\mbox{\boldmath $ \al $}}_{\bf{0}}^{\bf{4}}\,{\bf{ln^4}}{\displaystyle\bf{\qdl}}
\displaystyle
\,\Bigg[
\frac{A_3}{4}\,
-\,\frac{1}{4}\,A_1^2\,{\overline{\cal{S}}}^{\,\,\gamma}_{1000}\,
-\,\frac{1}{4}\,A_1\,{\overline{\cal{S}}}^{\,\,\gamma}_{0011}\,
+\,\frac{1}{2}\,A_1\,{\overline{\cal{S}}}^{\,\,\gamma}_{0200}\,
-\,\frac{1}{4}\,A_1\,{\overline{\cal{S}}}^{\,\,\gamma}_{2000}\nn\\[6mm]
&& \hspace*{20mm}
\displaystyle
+\,{\overline{\cal{S}}}^{\,\,\gamma}_{0300}\,
-\,\frac{1}{8}\,{\overline{\cal{S}}}^{\,\,\gamma}_{1011}
+\,\frac{1}{4}\,{\overline{\cal{S}}}^{\,\,\gamma}_{1200}\,
-\,\frac{1}{8}\,{\overline{\cal{S}}}^{\,\,\gamma}_{3000}\qquad
 \Bigg]\nn\\[3mm]
&&
\displaystyle
+{\mbox{\boldmath $ {\cal O}(\al $}}_{\bf{0}}^{\bf{5}})\qquad\qquad\Bigg\}\; .
\end{array}$}\\
\label{eq:photonresultpert}
\ee
\normalsize
%
%
\noindent
We have already made use of Eqs.(\ref{eq:doubleprime}-\ref{eq:thirdorderS}) in
above expression.
Now the transverse vertex must have the right structure,
{\it i.e.} the right coefficients
${\cal{A}}^i_{mnrr}$, ${\cal{S}}^i_{mnrr}$,  so that
the solution of the Schwinger-Dyson equations for $1/F(p^2)$ and $1/G(q^2)$,
Eqs.~(\ref{eq:nonpertFresult2},~\ref{eq:pertfermionresult},~\ref{eq:photonsde},~\ref{eq:photonresultpert}),
are multiplicatively renormalizable.
\baselineskip=8mm
\section{Multiplicatively Renormalizable $F(p^2)$ and $G(q^2)$}
\label{sec:MR}
\subsection{The Photon Propagator}
\label{subsec:mrphoton}
 \noindent
We shall first look for the most general form of the multiplicatively renormalizable
photon wavefunction renormalization. In order to do so, the renormalized $G{\sr}$ can be written in the following form
by using Eq.~(\ref{eq:mrphoton})~:
\be G_{\sr}(q^2,\mu^2)&=&{\it Z}_3^{-1}(\mu^2,\Lambda^2)\,G(q^2,\Lambda^2)\;. \ee
We define the most general leading logarithmic expansion of the unrenormalized
photon wavefunction renormalization by ~:
\be
G(q^2,\Lambda^2)&=&\sum_{u=0}^{\infty}\,\al_0^u\,B_u\,\ln^u\ql\nn\\
&=&1+\al_0\,B_1\,\ln\ql
+\,\al_0^2\,B_2\,\ln^2 \ql
+\al_0^3\,B_3\,\ln^3 \ql
+{\cal O}(\al_0^4)\;.
\label{eq:barepho}
\ee
We impose the renormalization
condition that $G_R(q^2=\mu^2)=1$. The coefficients $B_{i}$ $(i>2)$ are then
constrained by multiplicative renormalizability, {\it i.e.},
$B_2=B_1^{\,2}$, $B_n=(B_1)^{\,n}$ so that
the renormalized photon wavefunction renormalization can be written as~:
\be
G_{\sr}(q^2,\mu^2)&=&\sum_{u=0}^{\infty}\,\al_{\sr}^u\,(B_1)^u\,\ln^u\ql\nn\\
&=&
1+
\al_{\sr}\,B_1\ln\frac{q^2}{\mu^2} +\al_{\sr}^2\,B_1^2\,\ln^2\frac{q^2}{\mu^2}
+
\al_{\sr}^3\,B_1^3\,\ln^3\frac{q^2}{\mu^2}
+{\cal O}(\al^4_{\sr})\,.
\label{eq:renpho}
\ee
Then, as we shall use later, the inverse of $G$ and $G_R$ are~:
\vspace{3mm}
\begin{equation}
\framebox{\hspace*{1cm}
$\begin {array}{rcl}
\nn\\
\displaystyle \frac{1}{G_{\sr}(q^2,\mu^2)}=\displaystyle 1 &-&{\mbox{\boldmath $ \al $}}_{\sr} \displaystyle
B_1\,{\bf{ln}}\frac{\bf{q^2}}{\bf{\mu^2}}
\qquad , \qquad
\displaystyle \frac{1}{G(q^2,\Lambda^2)}=\displaystyle 1 -{\mbox{\boldmath $ \al $}}_{0} \displaystyle
B_1\,{\bf{ln}}\frac{\bf{q^2}}{\bf{\Lambda^2}}\,,
\nn\\\nn\\
\end{array}$\hspace*{1cm}}\\
\end{equation}
\be
\vspace*{-5cm}
\hspace*{2cm}
\label{eq:mrphoconst}
\ee
where, as is well-known, $1/G(q^2)$ in QED only has a leading logarithm
at one loop
order, just like $1/\alpha(q^2)$ and being related to this physical
 quantity is independent of the gauge.

\subsection{The Fermion Propagator}
\label{subsec:mrfermion}
\noindent
Analogously to the previous section, we deal with the fermion
wavefunction renormalization. We similarly  define the general
leading logarithmic expansion of the unrenormalized $F$ as~:
\be
F(p^2,\Lambda^2)&=&\sum_{u=0}^{\infty}\,\al_{0}^u\,A_u\,\ln^u\ql\nn\\
&=&
1+\al_0\,A_1\,\ln\pdl
+\al_0^2\,A_2\,\ln^2 \pdl
+\al_0^3\,A_3\,\ln^3 \pdl
+\,{\cal O}(\al_0^4)\;.
\label{eq:barefer}
\ee
Since not only the coupling, but the gauge parameter have to be renormalized, we need
to make the dependence of the $A_u$ on $\xi$ explicit. As
gauge dependence in the coefficients arises from photon propagators,
any $A_u$ cannot have a
higher power of $\xi$ than $\xi^u$. Consequently, $F(p^2, \Lambda^2)$ can be written as~:
\be
F(p^2,\Lambda^2)=
1&+&\al_0\,(a_1\xi+b_1)\,\ln\pdl\nn\\
&+&\al_0^2(a_2\,\xi^2+b_2\,\xi+c_2)\,\ln^2 \pdl\nn\\
&+&\al_0^3(a_3\,\xi^3+b_3\,\xi^2+c_3\,\xi+d_3)\,\ln^3 \pdl +\,{\cal O}(\al_0^4)\;,
\label{eq:mulbare} \ee
where $a_i, b_i, c_i, d_i$ are constants
related to the $A_u$ by comparing Eqs.~(\ref{eq:barefer}) and (\ref{eq:mulbare}).
Recalling Eqs.~(\ref{eq:mrgauge},~\ref{eq:mrcoupling}),~~ $\xi_0={\it Z}_3\,\xi_R \;$, $\al_0={\it
Z}_3^{-1}\,\al_R\quad$ we note that \be
\al_0\,\xi&=&\al_R\,\xi_R\,,\nn\\
\mbox{\rm{and}}\hspace{4cm} F_R(p^2,\mu^2)&=&{\it Z}_2^{-1}(\mu^2/\Lambda^2)\,F(p^2,\Lambda^2),\hspace{5cm} \ee with
the renormalization condition for the fermion wavefunction renormalization ~~~~~~~~~~
 $F_R(p^2=\mu^2)=1$. Eq.~(\ref{eq:mulbare}) can then be inserted in this equation to give

\normalsize
\be
F_R(p^2,\mu^2)=1&+&\al_R\,(a_1\xi_R+b_1)\,\ln\frac{p^2}{\mu^2}\nn\\
&+&\al_R^2(a_2\,\xi_R^2+b_2\,\xi_R+c_2)\,\ln^2 \frac{p^2}{\mu^2}\nn\\
&+&\al_R^3(a_3\,\xi_R^3+b_3\,\xi_R^2+c_3\,\xi_R+d_3)\,
\ln^3 \frac{p^2}{\mu^2}
+\,{\cal O}(\al_R^4)\;,
\ee

\noindent
Multiplicative renormalizability requires that the inverse unrenormalized fermion wavefunction
renormalization must have the following form keeping only the leading logarithms~:

\normalsize
\be
\framebox{$\begin {array}{rcl}
\nn\\
\displaystyle
\frac{1}{F(p^2,\Lambda^2)}&=&\displaystyle 1
+{\mbox{\boldmath $ \al $}}_{\bf{0}}\, \displaystyle{\bf{ln}}{\bf{\pdl}}
\displaystyle
\,\lef[-a_1\,\xi-b_1\rig]\,
\nn\\[5mm]
&+&{\mbox{\boldmath $ \al $}}_{\bf{0}}^{\bf{2}}\, \displaystyle{\bf{ln^2}}{\bf{\pdl}}
\displaystyle
\lef[\frac{a_1^2}{2}\,\xi^2+a_1\,b_1\,\xi
+\frac{b_1}{2}\,(b_1-B_{1})\rig]\nn\\[5mm]
&+&{\mbox{\boldmath $ \al $}}_{\bf{0}}^{\bf{3}}\, \displaystyle{\bf{ln^3}}{\bf{\pdl}}
\displaystyle
\lef[-\frac{a_1^3}{6}\,\xi^3-\frac{a_1^2\,b_1}{2}\,\xi^2
+\frac{a_1\,b_1}{2}\lef(-b_1+B_{1}\rig)\,\xi
-\frac{b_1^3}{6}-\frac{b_1\,B_{1}^2}{3}+\frac{b_1^2\,B_{1}}{2}\rig]
\nn\\[5mm]
&+&{\mbox{\boldmath $ \al $}}_{\bf{0}}^{\bf{4}}\, \displaystyle{\bf{ln^4}}{\bf{\pdl}}
\displaystyle
\Bigg[
\frac{a_1^4}{24}\,\xi^4
+\frac{a_1^3\,b_1}{6}\,\xi^3
+\frac{a_1^2\,b_1}{4}\,\lef(b_1 -\, B_1\rig)\,\xi^2
\nn\\
&& \hspace*{2cm}
\displaystyle
+\frac{a_1\, b_1}{2}\,\lef(\frac{b_1^2}{3} - b_1\, B_1 + \frac{2}{3}B_1^2\rig)\xi\nn\\
&& \hspace*{2cm}
\displaystyle
+ \frac{b_1}{4}\,\lef( \frac{b_1^3}{6} - b_1^2\, B_1 + \,\frac{11}{6}\, b_1\, B_1^2 - B_1^3\rig)\,\Bigg]+\,\,{\cal O}(\al_0^5)\\
\end{array}$}\\
\label{eq:mrferconstbare}
\ee
\normalsize
\noindent
The renormalized form of $1/F$ can be found by replacing $\alpha_0 \longrightarrow \alpha_R$, $\xi \longrightarrow \xi_R$ and
$\Lambda \longrightarrow \mu$ in the above expression.

\section{MR constraints on the vertex}
\label{sec:mrconsraint}
\noindent
In Sec.~{\ref{sec:unquenched}} we have shown exactly how the full vertex contributes in the fermion
and boson SDEs.
In principle, truncation of the
Schwinger-Dyson equations for the fermion
and boson propagators requires knowledge of the complete structure of the
vertex, all 12 independent components or here in massless QED all six.
While two are fixed by the
Ward-Green-Takahashi identity in terms of the fermion propagator,
the four transverse components appear to embody information about all
the higher point Green's functions.  Knowledge we do not have, unless we solve the theory
completely.  However, two simplifications have already occurred.
Firstly, the massless fermion and boson self energies involve just
two projections of the six independent vertex vectors, so we do not need to know
their complete
spin and Lorentz structure.  This is helpful, since even at ${\cal O}(\alpha_0)$
in perturbation theory, this is of daunting complexity \cite{Kizilersu:1995iz}.
The second simplification is that multiplicative
renormalizability (MR) is closely related to the ultraviolet behaviour of the loops
in Figs.~\ref{fig:fermion},~\ref{fig:photon}.
There not only is the structure of the vertex simpler,
but importantly for the present study the 2 graphs explore
the vertex in distinct kinematic regimes.  For the fermion self-energy, the
internal
fermion momentum $k$ and boson momentum $q$ are very much larger
than the external fermion momentum $p$, {\it i.e.}~$k^2 \simeq q^2 >> p^2$.
In contrast, for the boson self-energy, it is the fermion momenta that are
both large,
{\it i.e.} $k^2 \simeq p^2 >> q^2$.  We shall see that this distinction plays
a powerful role in our analysis.

\noindent
First, in this section we  combine the results of the previous two
sections to find the constraints
on the fermion-photon vertex imposed by multiplicative
renormalizability.
%
%
\vspace*{-0.4cm}
\subsection{MR constraints via fermion Schwinger-Dyson equation}
\label{subsec:mrconstfer}

\noindent
In this and the next section, we apply the above strategy first to the fermion
wavefunction renormalization in full massless QED. To do this, we start by comparing
order-by-order the results fixed by
multiplicatively renormalizable $F$, Eq.~(\ref{eq:mrferconstbare}), with those
found by solving the Schwinger-Dyson equation, Eq.~(\ref{eq:pertfermionresult}).
These comparisons will give what we refer to as the fermion conditions, labelled by $ FC1, FC2,$ {\it etc.}.

\noindent
\be
{\mbox{\boldmath $ \al $}}_{\bf{0}}\,{\bf{ln}}{\bf{\pdl}}
\;\mbox{{\bf{comparison:}}}\hspace{30mm}&&\hspace{85mm}\nn\\[2mm]
-A_1\equiv -(a_1\,\xi+b_1)&=&-\,\frac{\xi}{4\,\pi}\quad,\nn\\
&{{\Downarrow}}&\nn\\
a_1= \frac{1}{4\,\pi}\, , &&
b_1=\,0  \quad.
\label{eq:feral1log1}
\ee
\noindent
In this first order comparison MR fixes the value of $a_1$ and $b_1$ and by that all leading
order terms in $1/F$ or $F$, then Eq.~(\ref{eq:mrferconstbare}) requires
\begin{equation}
FC1~:\qquad \fbox{$\displaystyle
A_1=\,\frac{\xi}{4\pi}\, , \,A_2=\,\frac{A_1^2}{2}\,\,\,.$}
\label{eq:a01}
\end{equation}
%
%
\noindent
\be
{\mbox{\boldmath $ \al $}}^{\bf{2}}_{\bf{0}}\,{\bf{ln^2}}{\bf{\pdl}}\;
\mbox{{\bf{comparison:}}}\hspace{32mm}&&\hspace{85mm}\nn
\ee
\be
\frac{a_1^2}{2}\xi^2+a_1b_1\xi+\frac{b_1^2}{2}&=&
\frac{1}{4\,\pi}\,\lef[ \lef(\frac{\xi}{2}+\frac{3}{8}\rig)\,A_1-\frac{3}{4}\,{\overline{{\cal{A}}}}^{\,f}_{1000}+\frac{3}{4}\,{\overline{{\cal{S}}}}^{\,f}_{1000}\rig]\,,
\label{eq:ferorder1}
\ee
Making use of Eqs.~(\ref{eq:feral1log1},~\ref{eq:a01}) and keeping in mind that ${\overline{\cal{A}}}_{1000}^{\,f}$ and
${\overline{\cal{S}}}_{1000}^{\,f}$ can be at most proportional to $\xi$ or $N_F$
from Eqs.~(\ref{eq:perttauanti},~\ref{eq:perttausym}), we immediately see that
the $\xi^2$-term on both sides automatically matches and for the $\xi$-term we must have~:

\be
FC2~:\qquad
\fbox{$
\displaystyle \frac{A_1}{2}={\overline{\cal{A}}}^{\,f}_{1000}-{\overline{\cal{S}}}^{\,f}_{1000}\,\,\,.\,\,
$}
\label{eq:fercond1}
\ee
\noindent
\be
{\mbox{\boldmath $ \al $}}^{\bf{3}}_{\bf{0}}\,{\bf{ln^3}}{\bf{\pdl}}\;
\mbox{{\bf{comparison:}}}\hspace{32mm}&&\hspace{85mm}\nn
\ee
\be
-\frac{a_1^3}{6}\,\xi^3=-\frac{A_1^3}{3!}=&-&\frac{1}{4\,\pi}\,\Bigg\{
-\,\lef(\frac{\xi}{2}+\frac{3}{8}\rig)\,A_1^2
+\lef(\frac{4\,\xi}{3}\,+\,1\rig)\,A_2-\frac{A_1\,B_1}{8}\nn\\[6mm]
&& \hspace{12mm}
+\,\frac{1}{4}\,(A_1+B_1)\,\lef({\overline{\cal{A}}}^{\,f}_{1000}\,-\,{\overline{\cal{S}}}^{\,f}_{1000}\,\rig)\,
+\,{\overline{\cal{A}}}^{\,f}_{2000}\,+\,\frac{1}{4}\,\,{\overline{\cal{A}}}^{\,f}_{1100}\nn\\[3mm]
&& \hspace{12mm}
-\,\frac{3}{4}\,\,{\overline{\cal{S}}}^{\,f}_{2000}
+\,\frac{1}{4} \,{\overline{\cal{S}}}^{\,f}_{0200}
-\,\frac{1}{4}\,\,{\overline{\cal{S}}}^{\,f}_{0011}
\Bigg\}\,.
\label{eq:ferorder3}
\ee

\noindent
The leading terms in $\xi$ in
Eq.~(\ref{eq:ferorder3}) ({\it i.e.} $O(\xi^3)$)
automatically match on the left and right hand sides. Imposing Eq.~(\ref{eq:fercond1}), the $O(\xi^2)$ terms require the transverse part to be fixed so that

\be
FC3~:\qquad
\fbox{$
\displaystyle
\frac{A_1^2}{4}=\, -{\overline{\cal{A}}}^{\,f}_{2000}-\frac{1}{4}\, {\overline{\cal{A}}}^{\,f}_{1100}+\frac{3}{4}\,{\overline{\cal{S}}}^{\,f}_{2000}
                 -\frac{1}{4}\,{\overline{\cal{S}}}^{\,f}_{0200}+\,\frac{1}{4}\,{\overline{\cal{S}}}^{\,f}_{0011}\,.\,\,
                 $}
\label{eq:fercond2}
\ee

\noindent
As one can see the $B_1$ term in Eq.~(\ref{eq:ferorder3}) disappears from the above expression and this must repeat itself in every order, {\it i.e.} in
leading order terms the photon contribution will be cancelled out by the transverse vertex.
\newpage
%
\noindent
\be
{\mbox{\boldmath $ \al $}}^{\bf{4}}_{\bf{0}}\,{\bf{ln^4}}{\bf{\pdl}}\;
\mbox{{\bf{comparison:}}}\hspace{32mm}&&\hspace{85mm}\nn
\ee
\be
\frac{a_1^4}{24}\,\xi^4=\frac{A_1^4}{4!}=-\frac{1}{4\,\pi}\,\Bigg\{
&-&A_3\,\lef(\frac{9}{16} + \frac{3}{4}\, \xi\rig)\,+\,A_1\,A_2\,\lef(\frac{1}{8}+ \frac{1}{6}\,\xi\rig)
\nn\\[4mm]
&-&\frac{1}{4}\, A_1^2\, B_1
+\,\frac{9}{16}\,A_2\,B_1
-\,\frac{1}{16}\,A_1\,B_2\nn\\[4mm]
&+&\frac{1}{8}\,(A_2+B_2)\, \lef({\overline{\cal{A}}}^{\,f}_{1000}\,-\,{\overline{\cal{S}}}^{\,f}_{1000}\rig)\,
+\,\frac{1}{8}\,A_1\,B_1\lef({\overline{\cal{A}}}^{\,f}_{1000}\,-\,{\overline{\cal{S}}}^{\,f}_{1000}\rig)\nn\\[4mm]
&+&(A_1+B_1)\Big(
\,\frac{3}{8}\,{\overline{\cal{A}}}^{\,f}_{2000}\,+\,\frac{1}{8}\,{\overline{\cal{A}}}^{\,f}_{1100}
-\,\frac{1}{4}\,{\overline{\cal{S}}}^{\,f}_{2000}
-\,\frac{1}{8}\,{\overline{\cal{S}}}^{\,f}_{0011}\,+\,\frac{1}{8}\,{\overline{\cal{S}}}^{\,f}_{0200}\Big)\nn\\[4mm]
&+&\frac{1}{4}\,{\overline{\cal{A}}}^{\,f}_{1011}\,+\,\frac{1}{8}\,{\overline{\cal{A}}}^{\,f}_{1200}\,
+\,\frac{3}{8}\,{\overline{\cal{A}}}^{\,f}_{2100}\,+\,\frac{9}{8}\,{\overline{\cal{A}}}^{\,f}_{3000}\nn\\[4mm]
&+&\frac{1}{8}\,{\overline{\cal{S}}}^{\,f}_{0111}\,+\,\frac{3}{8}\,{\overline{\cal{S}}}^{\,f}_{0300}
-\,\frac{1}{8}\,{\overline{\cal{S}}}^{\,f}_{1011}\,+\,\frac{1}{4}\,{\overline{\cal{S}}}^{\,f}_{1200}\,
-\,\frac{3}{4}\,{\overline{\cal{S}}}^{\,f}_{3000}
\Bigg\}\,.
\label{eq:al4ln4}
\ee
\noindent
Once again in above expression the leading terms in $\xi$ (i.e. ${\cal{O}}(\xi^4)$) terms match on both sides.
After substituting the $ FC2$ and $ FC3$ conditions in Eq.~(\ref{eq:al4ln4}), we have the following combined constraints on
$\xi^3$ and $\xi^2\,N_F$ terms ~:
\be
FC4~:\qquad
\fbox{$\begin{array}{lll}\nn\\
\quad\displaystyle
\frac{A_1^3}{16}&=&\displaystyle\frac{(A_1+B_1)}{48}\Big\{
\,{\overline{\cal{A}}}^{\,f}_{1100}
-\,{\overline{\cal{S}}}^{\,f}_{0011}
+\,{\overline{\cal{S}}}^{\,f}_{2000}
+\,{\overline{\cal{S}}}^{\,f}_{0200}\Big\}\nn\\[4mm]
\displaystyle
&+&\displaystyle\frac{1}{6}\,{\overline{\cal{A}}}^{\,f}_{1011}\,+\,\frac{1}{12}\,{\overline{\cal{A}}}^{\,f}_{1200}\,
+\,\frac{1}{4}\,{\overline{\cal{A}}}^{\,f}_{2100}\,+\,\frac{3}{4}\,{\overline{\cal{A}}}^{\,f}_{3000}\nn\\[4mm]
\displaystyle
&+&\displaystyle\frac{1}{12}\,{\overline{\cal{S}}}^{\,f}_{0111}\,+\,\frac{1}{4}\,{\overline{\cal{S}}}^{\,f}_{0300}
-\,\frac{1}{12}\,{\overline{\cal{S}}}^{\,f}_{1011}\,+\,\frac{1}{6}\,{\overline{\cal{S}}}^{\,f}_{1200}\,
-\,\frac{1}{2}\,{\overline{\cal{S}}}^{\,f}_{3000}\quad.\nn\\[4mm]
\end{array}$}\\[-8mm]
\label{eq:fercond41}
\ee
\noindent
The Schwinger-Dyson Equation for the fermion propagator involves corrections from photon emission and absorption as
displayed in Fig.~\ref{fig:photon}. This requires the fermion renormalization function to depend on the photon renormalization
function, which in turn depends on the number of fermions $N_F$. Therefore in general
$\Big\{\,{\overline{\cal{A}}}^{\,f}_{1100}, {\overline{\cal{S}}}^{\,f}_{0011},{\overline{\cal{S}}}^{\,f}_{2000},{\overline{\cal{S}}}^{\,f}_{0200}\Big\}$
and
$\Big\{{\overline{\cal{A}}}^{\,f}_{1011},{\overline{\cal{A}}}^{\,f}_{1200},{\overline{\cal{A}}}^{\,f}_{2100},{\overline{\cal{A}}}^{\,f}_{3000},{\overline{\cal{S}}}^{\,f}_{0111},
{\overline{\cal{S}}}^{\,f}_{0300}, {\overline{\cal{S}}}^{\,f}_{1011},{\overline{\cal{S}}}^{\,f}_{1200}\,{\overline{\cal{S}}}^{\,f}_{3000}\Big\}$
 terms in Eq.~(\ref{eq:fercond41}) can be proportional to
$\lef(\xi^2~ \rm {or}~ N_F^2 ~{\rm or}~ \xi\, N_F \rig)$ and
$\lef( \xi^3 ~ \rm{or} ~ \xi^2\,N_F~\rm{or}~ \xi\, N_F^2~ \rm{or}~ N_F^3 \rig)$ respectively.
  Remarkably, the matching required by
multiplicatively renormalizability of these renormalization functions
 is automatically satisfied if the transverse fermion-boson vertex is independent of the photon renormalization function at leading logarithmic order.
 Therefore
 $\Big\{\,{\overline{\cal{A}}}^{\,f}_{1100}, {\overline{\cal{S}}}^{\,f}_{0011},{\overline{\cal{S}}}^{\,f}_{2000},{\overline{\cal{S}}}^{\,f}_{0200}\Big\}$
 and
$\Big\{{\overline{\cal{A}}}^{\,f}_{1011},{\overline{\cal{A}}}^{\,f}_{1200},{\overline{\cal{A}}}^{\,f}_{2100},{\overline{\cal{A}}}^{\,f}_{3000},{\overline{\cal{S}}}^{\,f}_{0111},
{\overline{\cal{S}}}^{\,f}_{0300}, {\overline{\cal{S}}}^{\,f}_{1011},{\overline{\cal{S}}}^{\,f}_{1200}\,{\overline{\cal{S}}}^{\,f}_{3000}\Big\}$ terms would be proportional to
only  $\xi^2$ and $ \xi^3$ terms respectively.
This will clearly constrain the non-perturbative forms of the transverse vertex that we wish to determine. In other words constraint $FC4$ of Eq.~(\ref{eq:fercond41}),
will divide
into two separate conditions for $\xi^2 N_F$ and $\xi^3$ comparisons~:
\be
FC41~:\qquad \,\,\,\,\,0&=&{\overline{\cal{A}}}^{\,f}_{1100}
-\,{\overline{\cal{S}}}^{\,f}_{0011}
+\,{\overline{\cal{S}}}^{\,f}_{2000}
+\,{\overline{\cal{S}}}^{\,f}_{0200}\,, \nn\\[4mm]
FC42~:\qquad \frac{A_1^3}{16}&=&
\frac{1}{6}\,{\overline{\cal{A}}}^{\,f}_{1011}\,+\,\frac{1}{12}\,{\overline{\cal{A}}}^{\,f}_{1200}\,
+\,\frac{1}{4}\,{\overline{\cal{A}}}^{\,f}_{2100}\,+\,\frac{3}{4}\,{\overline{\cal{A}}}^{\,f}_{3000}\,\nn\\[5mm]
&+&\frac{1}{12}\,{\overline{\cal{S}}}^{\,f}_{0111}\,+\,\frac{1}{4}\,{\overline{\cal{S}}}^{\,f}_{0300}
-\,\frac{1}{12}\,{\overline{\cal{S}}}^{\,f}_{1011}\,+\,\frac{1}{6}\,{\overline{\cal{S}}}^{\,f}_{1200}
-\,\frac{1}{2}\,{\overline{\cal{S}}}^{\,f}_{3000}\,.
\ee
The idea is then to find a non-perturbative structure for the transverse pieces that delivers
such relations.
This we do in the next section.  However, first we  determine the conditions
imposed by multiplicative renormalizability for the photon wavefunction renormalization.
%
\vspace*{-5mm}
\subsection{MR constraints via Photon Schwinger-Dyson equation}
\label{eq:mrconsphoton}
\noindent
 We now repeat the previous steps for the photon wavefunction renormalization.
Comparison takes place between  Eq.~(\ref{eq:photonresultpert})
and Eq.~(\ref{eq:mrphoconst})
order-by-order for $1/G$. Obviously, this time instead of looking at
the  terms depending on the gauge parameter $\xi$,
we compare  the dependence on  $N_F$, the number of flavours hidden in the
$B_i$ terms. These give what we refer to as the photon conditions labelled
$ PC1, PC2$, {\it etc.}. Then~:
\begin{eqnarray*}
{\mbox{\boldmath $ \al $}}_{\bf{0}}\,{\bf{ln}}{\bf{\pdl}}
\;\mbox{{\bf{comparison:}}}&&\hspace{115mm}
\end{eqnarray*}
\be
PC1~:\qquad
\fbox{$\displaystyle
B_1=\frac{N_F}{3\,\pi} \quad, \quad B_n=B_1^n=\lef(\frac{N_F}{3\,\pi}\rig)^n \,.\,\,$ }
\ee
First order comparison defines the value of $B_1$ in terms of $N_F$ and as given in Eq.~(\ref {eq:renpho}) fixes all the higher order terms.
\be
{\mbox{\boldmath $ \al $}}_{\bf{0}}^{\bf{2}}\,{\bf{ln^2}}{\bf{\pdl}}
\;\mbox{{\bf{comparison:}}}&&\hspace{20mm}
PC2~:\qquad
\fbox{$
\displaystyle
\,\,\,\frac{2}{3}\,A_1\,=\,\overline{S}^{\,\gamma}_{1000}\,.\,\,$}\hspace{40mm}
\label{eq:photoncond1}
\ee

\noindent
As we see above the second order comparison imposes this condition on the symmetric part of the transverse vertex.
\newpage
\begin{eqnarray*}
{\mbox{\boldmath $ \al $}}_{\bf{0}}^{\bf{3}}\,{\bf{ln^3}}{\bf{\pdl}}
\;\mbox{{\bf{comparison:}}}&&\hspace{110mm}
\end{eqnarray*}
\be
\frac{A_1^2}{6}\,=\,
\frac{A_1}{2}\,\overline{\cal{S}}^{\,\,\gamma}_{1000}
+\frac{1}{4}\,\overline{\cal{S}}^{\,\,\gamma}_{2000}
-\frac{3}{4}\,\overline{\cal{S}}^{\,\,\gamma}_{0200}
+\frac{1}{4}\,\overline{\cal{S}}^{\,\,\gamma}_{0011}\,.
\ee

\noindent
Substituting Eq.~(\ref{eq:photoncond1}) in above condition yields~:
\be
PC3~:\qquad
\fbox{$
\displaystyle
\frac{A_1^2}{3}\,=\,-\frac{1}{2}\,\,\overline{\cal{S}}^{\,\,\gamma}_{0011}+\frac{3}{2}\,\,\overline{\cal{S}}^{\,\,\gamma}_{0200}-\frac{1}{2}\,\,\overline{\cal{S}}^{\,\,\gamma}_{2000}\,,
\label{eq:photoncond2}
$}
\ee
where every term is proportional to $\xi^2$.
\begin{eqnarray*}
{\mbox{\boldmath $ \al $}}_{\bf{0}}^{\bf{4}}\,{\bf{ln^4}}{\bf{\pdl}}
\;\mbox{{\bf{comparison:}}}&&\hspace{110mm}
\end{eqnarray*}
\be
\frac{A_1^3}{24}= \frac{A_1}{4}\, \overline{\cal{S}}^{\,\,\gamma}_{0011}
                -\frac{A_1}{2}\, \overline{\cal{S}}^{\,\,\gamma}_{0200}
                +\frac{A_1^2}{4}\,\overline{\cal{S}}^{\,\,\gamma}_{1000}
                +\frac{A_1}{4}\,\overline{\cal{S}}^{\,\,\gamma}_{2000}
                +\frac{1}{8}\, \overline{\cal{S}}^{\,\,\gamma}_{3000}
                -\overline{\cal{S}}^{\,\,\gamma}_{0300}
                +\frac{1}{8}\,\overline{\cal{S}}^{\,\,\gamma}_{1011}
                -\frac{1}{4}\,\overline{\cal{S}}^{\,\,\gamma}_{1200}\,.\nn\\
\ee
Making use of Eqs.~(\ref{eq:photoncond1},~\ref{eq:photoncond2}), the above expression becomes~:
\be
PC4~:\qquad
\fbox{$
\displaystyle
\,\,\,\frac{A_1^3}{24}= \frac{A_1}{4} \,\,\overline{\cal{S}}^{\,\,\gamma}_{0200}
                 - \,\,\overline{\cal{S}}^{\,\,\gamma}_{0300}
                 +\,\frac{1}{8}\,\,\overline{\cal{S}}^{\,\,\gamma}_{1011}
                 -\,\frac{1}{4}\,\,\overline{\cal{S}}^{\,\,\gamma}_{1200}
                 +\,\frac{1}{8}\,\,\overline{\cal{S}}^{\,\,\gamma}_{3000}\,,\,\,
\label{eq:photoncond3}
$}
\ee
where every term is proportional to $\xi^3$.
So far we have expressed the general multiplicative
renormalizability constraints
on the 3-point vertex function in terms of the constants
${\overline{\cal{A}}}^{\,\,\gamma}_{mnrr}$ and ${\overline{\cal{S}}}^{\,\,\gamma}_{mnrr}$ up to ${\cal{O}}(\alpha^4)$.

\subsection{Generalized Fermion and Photon MR constraints }
\label{eq:mrconsphoton}
\noindent
Let us first look at the general picture. Firstly, $a_1$ and $b_1$ being fixed by Eq.~(\ref{eq:feral1log1}),
allows the expansion coefficients $A_u$ in Eq.~(\ref{eq:barefer}) to be fixed in all orders~:
\normalsize
\begin{equation}
\fbox{$ \vspace*{0.5cm}
\displaystyle
\,\,\,A_1=\,\frac{\xi}{4\pi}\, , \,A_2=\,\frac{A_1^{\,2}}{2!}\, ,A_3=\,\frac{A_1^{\,3}}{3!}\,,\cdots, \,A_u=\,\frac{A_1^{\,u}}{u!}\,,\,\,\,$}\\
\label{eq:a11}
\end{equation}

\noindent
and in turn the infinite leading log series of $F(p^2)$ in Eq.~(\ref{eq:barefer}) can be summed up as a power series~:
\be
F(p^2,\Lambda^2)\,=\,\sum_{u=0}^{\infty}\, \al_0^u\,\frac{A_1^{\,u}}{u!}\,\ln^u\pdl\,=\,\lef(\pdl\rig)^{\alpha\,A_1}\,.
\label{eq:summedF}
\ee
\noindent
This is the non-perturbative
expression for the unquenched (full) fermion wavefunction renormalization. Moreover, it has exactly the
same form as in the quenched QED \cite{Kizilersu:2001pd,Hawes:1994ce,Curtis:1990zs,Hawes:1996mw,Hawes:1996ig,Schreiber:1998ht,Curtis:1991fb,Gusynin:1998se}.
\noindent
Secondly, the relation between the photon coefficients are also found through $PC1$~:
\be
&&
\fbox{$\displaystyle
B_1\,=\,\frac{N_F}{3\,\pi} \quad, \quad B_n\,=\,B_1^{\,n}\,=\,\lef(\frac{N_F}{3\,\pi}\rig)^n \,,\,\,\,$ }
\label{eq:Bn}
\ee
hence the infinite series of $G(p^2,\Lambda^2)$, Eq.~(\ref{eq:barepho}) can also be summed up as~:
\be
G(q^2,\Lambda^2)&=&\sum_{u=0}^{\infty}\,\al_0^u\,B_1^u\,\displaystyle\ln^u\frac{q^2}{\Lambda^2}\,
=\,\frac{1}{1 -\, \al_{0}\,B_1\,\ln {q^2}/{\Lambda^2}}\,\,\,.
\label{eq:summedG}
\ee
\subsubsection{Generalized MR constraints from fermion SDE}
\noindent
Making use of Eqs.~(\ref{eq:a11}-~\ref{eq:summedG}), we can then rewrite the inverse fermion wavefunction
renormalization calculated from SDE, Eq.~(\ref{eq:nonpertFresult2}) as~:
\be
\frac{1}{F(p^2,\Lambda^2)} = 1\,&-&\,\Bigg\{
-\frac{1}{F(p^2,\Lambda^2)}+\,1\nn\\
&+& \frac{3}{8\,\pi}\,\sum_{u=1}^{\infty}\,{\alpha_0}^{u+1}\ln^{u+1} \pdl\,\Bigg[
-\frac{1}{2}\, \sum_{a=1}^{u}\,\frac{A_1^a}{a!}\,B_1^{u-a}\frac{1}{u+1}\nn\\
&&\hspace{4.5cm}
-\frac{1}{2}\,\sum_{b=1}^{u}\,\sum_{a=0}^{u-b}\,(-1)^b\,\frac{A_1^b}{b!}\,\frac{A_1^a}{a!}\,B_1^{u-b-a}\,\frac{1}{(u-b+1)}\,\Bigg]\nn\\
&+&\frac{3}{8\,\pi}\,\sum_{u=1}^{\infty}\,{\alpha_0}^{u+1}\ln^{u+1} \pdl\lef( H_u + {\overline H}_u\rig)\Bigg\}\,,
\label{eq:nonpertFresult3}
\ee
\noindent
and as a consequence of equating the multiplicatively renormalized $F$, Eq.~(\ref{eq:summedF}) to
Eq.~(\ref{eq:nonpertFresult3}) we can extract the generalized MR constraints to all orders,
which, of course, reproduces $FC1$ to $FC4$ ~:
\be
\fbox{$
\begin{array}{rcl}
\nn\\
\displaystyle 0&=&\displaystyle\sum_{u=1}^{\infty}\,{\alpha_0}^{u+1}\ln^{u+1} \pdl\lef( H_u + {\overline H}_u\rig) \nn\\
&+& \displaystyle \sum_{u=1}^{\infty}\,{\alpha_0}^{u+1}\ln^{u+1} \pdl\,\Bigg\{
-\frac{1}{2}\, \sum_{a=1}^{u}\,A_a\,B_{u-a}\,\frac{1}{(n+1)}\nn\\
&&\hspace{4.5cm}
-\displaystyle \frac{1}{2}\,\sum_{b=1}^{u}\,\sum_{a=0}^{u-b}\,(-1)^b\,A_b\,A_a\,B_{u-b-a}\,\frac{1}{(u-b+1)}\,\Bigg\}\,\,,\nn\\
\end{array}
$}\\
\ee
where in Eqs.~(\ref{eq:Hfer},~\ref{eq:Hbarfer}) for the $H_u$ and ${\overline H}_u$, one can now substitute for the $A_n$, $B_n$ from
Eqs.~(\ref{eq:a11}, \ref{eq:Bn}).
\subsubsection{Generalized MR constraints from photon SDE}
\noindent
Making use of Eqs.~(\ref{eq:a11}-\ref{eq:summedG}), we repeat the above procedure for photons,
which is analogous to the fermion case above, in order to rewrite the inverse photon wavefunction renormalization, Eq.~(\ref{eq:photonsde})~:
\be
\frac{1}{G(q^2,\Lambda^2)} = 1 &-& \frac{N_F}{3\,\pi}\, \alpha_0\, \ln\qdl \nn\\
&-& \frac{N_F}{3\,\pi}\, \sum_{u=1}^{\infty}\,\alpha_0^{u+1}\, \ln^{u+1}\qdl\Bigg\{
   \,\frac{A_1^{\,u}}{(u+1)!}+\,\frac{3}{2}\,K_u \Bigg\}\,\,,
   \label{eq:photonsde2}
\ee
where in the expression for $K_u$ of Eq.~(\ref{eq:Ku}) we can substitute the conditions for $A_n$ from Eq.~(\ref{eq:a11}).
\noindent
The generalized MR photon constraints can then be written as~:
%
%
\be
\fbox{$
\begin{array}{rcl}
\nn\\[-2mm]
&&\displaystyle \sum_{u=1}^{\infty}\,\alpha_0^{u+1}\, \ln^{u+1}\qdl\, \Bigg\{
\frac{A_1^{\,u}}{(u+1)!}\,+\, \frac{3}{2}\, K_u\Bigg\}\,=\,0\,\,,\nn\\[5mm]
\end{array}$}\\
\label{eq:nonpertphcond}
\ee
automatically satisfying $PC1$-$PC4$.

\subsection{Non-perturbative Fermion and Photon MR constraints }
\label{eq:nonpetmrcond}
\subsubsection{Non-perturbative MR constraints on transverse vertex from photon SDE}
\noindent
To understand the above conditions in full generality ({\it i.e.} beyond their expansion in  leading logarithms)
we first turn our attention to photon equation.
Starting from Eq.~(\ref{eq:mrphoconst}) for multiplicatively renormalizable $1/G(q^2,\Lambda^2)$, we see that
multiplicative renormalizability for  leading logs is given by just the $O(\alpha_0)$ term, $B_1$, which is gauge
independent.  Equating the photon Schwinger-Dyson equation at the leading logarithmic order,  Eq.~(\ref{eq:finalsdeG}), with the
multiplicatively renormalizable $G(q^2)$,  Eq.~(\ref{eq:mrphoconst})~:
\be
\frac{1}{G(q^2,\Lambda^2)}\, =\, 1 + \frac{\alpha_0\,N_F}{3\,\pi}\,\int_{q^2}^{\Lambda^2}\,\frac{d\e^2}{\e^2}\,F^2(\e)\,\Bigg\{
\frac{1}{F(\e)}+\,\frac{3}{4}\,{\overline{\tau}}^{sym}_{\gamma}
\Bigg\}\,=\,1 - \frac{\alpha_0\,N_F}{3\,\pi}\,\ln\left(\frac{q^2}{\Lambda^2}\right)\; .
\label{eq:finalsdeG1}
\ee

\noindent
We observe that the $\lambda_1$ term of the
Ball-Chiu longitudinal vertex generates this. However, importantly for the present purpose this is part of a whole
series:
\be
&& \frac{\alpha_0\,N_F}{3\,\pi}\,\int_{q^2}^{\Lambda^2}\,\frac{d\e^2}{\e^2}\,F(\e)=\frac{N_F}{3\,\pi}\,\alpha_0\,\ln\qdl \,\Bigg\{
        -1 \nn\\[3mm]
&& \hspace*{8mm}
-\lef[ \frac{1}{2}\, X + \frac{1}{6}\, X^2 + \frac{1}{24}\,X^3+\frac{1}{120}\,X^4 +\frac{1}{720}\, X^5 + \frac{1}{5040}\,X^6 + \frac{1}{40320}\,X^7
        + {\cal{O}}(\alpha_0^8)\rig]\,
        \Bigg\}\,, \nn\\
\label{eq:finalsdeG2}
\ee
where $X=\alpha_0\,A_1\,\ln\displaystyle\frac{q^2}{\Lambda^2}$. Beyond $O(\alpha_0)$, this series ({\it i.e.} terms inside the square bracket) has to be cancelled exactly by the
contribution from the vertex components.  Since the $\lambda_2$-term in the Ball-Chiu longitudinal component only
contributes at non-leading order, it is the symmetric part of the transverse vertex, ${\overline{\tau}}_{\gamma}^{sym}$, with its implicit gauge dependence that has to provide
this cancellation. $PC2$ to $PC4$ in Eqs.~(\ref{eq:photoncond1}-\ref{eq:photoncond3}) give the conditions for this cancellation to be achieved
at ${\cal{O}}(\alpha_0^2)$, ${\cal{O}}(\alpha_0^3)$ and  ${\cal{O}}(\alpha_0^4)$ and the general condition in Eqs.~(\ref{eq:nonpertphcond},~\ref{eq:finalsdeG1}) for all orders.
To go further, we
note that multiplicative renormalizability of the photon Schwinger-Dyson equation, Eq.~(\ref{eq:finalsdeG1}), picks out loop momentum regions where
$\e^2_+ \simeq \e^2_- \sim \e^2 \gg q^2$. The second term in Eq.~(\ref{eq:finalsdeG1}) must give the following result~:
\be
\frac{\alpha_0\,N_F}{4\,\pi}\,\int_{q^2}^{\Lambda^2}\,\,\frac{d\e^2}{\e^2}\,F^2(\e)\,{\overline{\tau}}^{sym}_{\gamma}
&=&\nn\\[3mm]
&& \hspace*{-55mm}
\frac{N_F}{3\,\pi}\,\alpha_0\,\ln\qdl \,\lef[ \frac{1}{2}\, X + \frac{1}{6}\, X^2 + \frac{1}{24}\,X^3+\frac{1}{120}\,X^4 +\frac{1}{720}\, X^5 + \frac{1}{5040}\,X^6 + \frac{1}{40320}\,X^7
        + {\cal{O}}(\alpha_0^8)\rig]\,.\nn\\
        \label{eq:finalsdeG3}
\ee
This surely determines the structure of
the ${\overline{\tau}}^{sym}_{\gamma}$'s for this to happen.  The dependence on the fermion wavefunction renormalization must be more complicated than $1/F$ times a kinematic factor.
It must be proportional to a function of a function of $F$'s so let us write~:
\be
{\overline{\tau}}^{sym}_{\gamma}\,\sim\,\frac{1}{F(q)}\; h(Y)\quad .
\label{eq:tauY}
\ee
In keeping with the ethos of this work, we assume that $Y$ is determined by
the fermion wavefunction renormalization. Since the renormalization of the $\tau_i$'s is replicated wholly by the factor of $1/F$,
$Y$ must be renormalization independent. As an example let us choose it to be:
\be
Y\;=\;\frac{F(q^2)}{F(\e^2)}\,-\,1 \quad ,
\label{eq:Y}
\ee
where the factor of $-1$ ensures that the leading logarithm expansion of $Y$ begins at ${\cal O}(\alpha_0\,\ln)$ as required by Eq.~(\ref{eq:finalsdeG3}).
Can we find what function $h(Y)$ is to satisfy Eqs.~(\ref{eq:finalsdeG1}, \ref{eq:finalsdeG3})? Let us assume we can expand $h(Y)$
as a power series in $Y$, and in turn expand this in leading logs of momenta. Then to produce the cancellation required, we deduce~:
\be
h(Y)\,=\,Y+\frac{1}{2}\,Y^2-\,\frac{1}{6}\,Y^3+\frac{1}{12} \,Y^4
           - \frac{1}{20}\,Y^5+ \frac{1}{30}\,Y^6 - \frac{1}{42}\,Y^7+\, \frac{1}{56}\,Y^8+ {\cal{O}}(Y^9) \; ,\nn\\
\ee
We recognise this as
\be
h(Y)&=& Y\left(1 - \sum_{n=1}^\infty\,\frac{(-Y)^n}{n(n+1)}\,\right)\,,\nn\\[2mm]
&=&(1+Y) \,\ln(1+Y)\,,
\ee
substituting $Y$ from Eq.~(\ref{eq:Y}), $h(Y)$ becomes~:
\be
h(Y)&=&\frac{F(q^2)}{F(\e^2)}\,\ln \frac{F(q^2)}{F(\e^2)}\quad .
\label{eq:fYform}
\ee

\noindent
Since a form like ${\overline{\tau}}^{sym}_{\gamma}\,\sim\,1/F(\ell^2)\;\ln(F(q^2)/F(\ell^2))$ in Eq.~(\ref{eq:tauY}) is the $k\to p = \ell$ limit of the evolving structure, this naturally generalises to
the $k \ne p$ configuration as~:
\be
{\overline{\tau}}^{sym}_{\gamma}(p^2,k^2,q^2)\,\sim\,
\frac{1}{2}\,\lef(\frac{1}{F(k^2)}+\frac{1}{F(p^2)}\rig)\,\ln\frac{F(q^2)}{2}\,\lef(\frac{1}{F(k^2)}+\frac{1}{F(p^2)}\rig)\quad .
\ee
\noindent
While the form in the photon limit is determined, the structure
in general momentum configurations is not unique and there are several possibilities differing only beyond leading logarithmic order.
Three of these are:
\be
{\cal{S}}^{(1)}&=&\frac{1}{2}\,\lef(\frac{1}{F(k^2)}+\frac{1}{F(p^2)}\rig)\,\ln\frac{F(q^2)}{2}\,\lef(\frac{1}{F(k^2)}+\frac{1}{F(p^2)}\rig)\,,\nn\\\nn\\
{\cal{S}}^{(2)}&=&\frac{1}{2}\,\frac{1}{(F(k^2)\,F(p^2))^{1/2}}\,\ln\,\frac{F(q^2)^2}{F(k^2)\,F(p^2)}\,,\nn\\\nn\\
{\cal{S}}^{(3)}&=&\frac{1}{4}\,\lef(\frac{1}{F(k^2)}+\frac{1}{F(p^2)}\rig)\,\ln\,\frac{F(q^2)^2}{F(k^2)\,F(p^2)} \qquad ,
\label{eq:Symxamples}
\ee
all of which give the same $h(Y)$ of Eq.~(\ref{eq:fYform}) in the photon limit of $k^2 \simeq p^2 \gg q^2$.
%

\subsubsection{Non-perturbative MR constraints on transverse vertex from fermion SDE}
\noindent
Similarly, for the multiplicatively renormalizable $1/F(q^2,\Lambda^2)$, the result at leading logarithmic order is given by the leading $\xi$
dependent piece, as required by the Landau-Khalatnikov-Fradkin transformation \cite{LK,Fradkin:Fradkin}. This leading term is provided by the
first term in the integrals of Eqs.~(\ref{eq:finalfersde},~\ref{eq:nonpertFresult},~\ref{eq:nonpertFresult2}).  Let us recall Eq.~(\ref{eq:finalfersde}) and
in this equation we perform both the radial and angular integration for the first term, but only the angular integration for the second term, then we find~:
\be
\frac{1}{F(p^2)}=1&+&\, \Bigg\{ \frac{1}{F(p^2)} -1 \nn\\
             &-& \frac{3\,\alpha_0}{8\,\pi}\,\,\int_{p^2}^{\Lambda^2}\,\,\frac{dk^2}{k^2}\,F(k^2)\,G(q^2)\,\lef[\frac{1}{2}\,\lef(\frac{1}{F(k^2)}-\,\frac{1}{F(p^2)}\rig)
                                                              +\lef({\overline{\tau}}_f^{anti}+ {\overline{\tau}}_f^{sym}\rig)\rig] \Bigg\}\,.\quad\quad
\label{eq:finalfersde1}
\ee

\noindent
 Imposing the MR fermion condition, Eq.~(\ref{eq:summedF}), on this expression yields the following constraint on the transverse vertex~:
\be
\framebox{$
\displaystyle
- \frac{3\,\alpha_0}{8\,\pi}\,\,\int_{p^2}^{\Lambda^2}\,\,\frac{dk^2}{k^2}\,F(k^2)\,G(q^2)\,\lef[\frac{1}{2}\,\lef(\frac{1}{F(k^2)}-\,\frac{1}{F(p^2)}\rig)
                                                           +\lef({\overline{\tau}}^{anti}_f+ {\overline{\tau}}^{sym}_f\rig)\rig]\,=\,0\,.\quad
$}
\label{eq:unqmrcondF}
\ee

\noindent
This cancellation involves both the longitudinal and transverse
pieces together. At leading logarithmic order the longitudinal contribution comes from just the $\lambda_2$-term in the Ball-Chiu
vertex.

\noindent
While antisymmetric forms do not contribute to the
leading logarithmic behaviour of the photon Schwinger-Dyson equation, this is not the case for the fermion equation.
Indeed, here the distinction between symmetric and antisymmetric disappears when $k^2 \simeq q^2 \gg p^2$.
Thus, a seemingly symmetric form like
\be
\ln\frac{F(q^2)}{2}\,\lef(\frac{1}{F(k^2)}+\frac{1}{F(p^2)}\rig)
\;\sim\ln \lef(\frac{F(k^2)}{F(p^2)}\rig)\;=\;\alpha_0\,A_1\,\ln \frac{k^2}{p^2} + O(\alpha_0^2)\,,
\ee

\noindent
is antisymmetric in $k$ and $p$ . Such a form contributes equally to the antisymmetric terms like
\be
\frac{1}{F(k^2)}\,-\,\frac{1}{F(p^2)}\;=\;-\alpha_0\,A_1\, \ln\frac{k^2}{p^2} + O(\alpha_0^2)\quad .
\ee

\noindent
The ${\cal{O}}(\alpha_0^2)$, ${\cal{O}}(\alpha_0^3)$ and ${\cal{O}}(\alpha_0^4)$ conditions of Eqs.~(\ref{eq:photoncond1}-\ref{eq:photoncond3}),
 which embody the gauge independence of
the photon wavefunction renormalization and the known gauge dependence of the fermion function, require the transverse
vertex to deliver a very particular gauge dependence itself. Our aim is to reproduce this by constructing the non-perturbative transverse
vertex from the fermion wavefunction renormalization.
This means from Eq.~(\ref{eq:unqmrcondF}) that
\be
\framebox{$
{\overline{\tau}}^{\,anti}_f\,+\,{\overline{\tau}}^{\,sym}_f\,=
\,-\displaystyle\frac{1}{2}\,\lef(\displaystyle \frac{1}{F(k^2)}-\,\frac{1}{F(p^2)}\rig)\,.
$}
\label{eq:nonpertfercond1}
\ee

\noindent
Hence this expression tells us that the total transverse vertex, {\it i.e.} combination of antisymmetric and symmetric parts,  must be proportional to
antisymmetric form in the limit $k^2 \simeq q^2 \gg p^2$.
These considerations suggest particular antisymmetric and symmetric vertex forms. In Table~I, we
give the specific coefficients ${\overline{\cal{A}}}_{mnrr}$ and ${\overline{\cal{S}}}_{mnrr}$ at ${\cal{O}}(\alpha_0^3)$ for these examples.

\small
\refstepcounter{table}
\addtocounter{table}{-1}
\begin{table}[htbp]
\begin{center}
\renewcommand{\arraystretch}{1.5}
\begin{tabular}{|c|c||l@{=}l l |}
\hline
            &                                  & $\displaystyle {\cal{A}}_{1000}$  & $-A_{1}$  &  \\
\,\,\,${\cal A}^{(1)}$\,  & $\frac{\displaystyle{1}}{\displaystyle F(k)}
              -\frac{\displaystyle{1}}{\displaystyle F(p)}$  & $\displaystyle {\cal{A}}_{2000}$  & $\displaystyle \frac{A_1^2}{2!}$,  &$\displaystyle {\cal{A}}_{1100}$=0 \\
            &                                  & $\displaystyle {\cal{A}}_{3000}$  & $-\displaystyle \frac{A_1^3}{3!}$,  &$\displaystyle {\cal{A}}_{2100}$=

                                                                                                                        $\displaystyle {\cal{A}}_{1011}$=
                                                                                                                        $\displaystyle {\cal{A}}_{1200}$=0 \\
           &                                  & $\displaystyle {\cal{A}}_{4000}$  & $-\displaystyle \frac{A_1^4}{4!}$,  &\\
           &                                  & $\displaystyle {\cal{A}}_{3100}$ & $\displaystyle {\cal{A}}_{2200}$=   &  $\displaystyle {\cal{A}}_{1300}$=
                                                                                                                           $\displaystyle {\cal{A}}_{2011}$=
                                                                                                                           $\displaystyle {\cal{A}}_{1111}$= 0\\
\hline\hline
            &                                  & $\displaystyle {\cal{S}}_{1000}$  & $-\displaystyle \frac{A_1}{2}$,  &$\displaystyle {\cal{S}}_{0100}$=$\displaystyle \frac{A_1}{2}$\\
\,\,\,${\cal S}^{(1)}$\,  & $\frac{\displaystyle 1}{\displaystyle 2}\,
                          \lef(\frac{\displaystyle 1}{\displaystyle F(k)}
                              +\frac{\displaystyle{1}}{\displaystyle F(p)}\rig)
                              \displaystyle{\ln}\frac{\displaystyle F(q)}{2}\,\lef(\frac{1}{\displaystyle F(k)}+\frac{1}{\displaystyle F(p)}\rig)$
                        & $\displaystyle {\cal{S}}_{2000}$  & $\displaystyle \frac{3}{8}\,A_1^2$, &                     $\displaystyle {\cal{S}}_{1100}$= $-\displaystyle \frac{A_1^2}{2}$,
                                                                                                                        $\displaystyle {\cal{S}}_{0011}$= $\displaystyle \frac{A_1^2}{8}$,
                                                                                                                        $\displaystyle {\cal{S}}_{0200}$=0\\
 &                      & $\displaystyle {\cal{S}}_{3000}$  & $\displaystyle -\frac{3}{16}A_1^3$,&                      $\displaystyle {\cal{S}}_{2100}$= $\displaystyle \frac{A_1^3}{4}$,
                                                                                                                        $\displaystyle {\cal{S}}_{1011}$= $\displaystyle -\frac{A_1^3}{16}$\\
 &                      &$\displaystyle {\cal{S}}_{0300}$   & $\displaystyle {\cal{S}}_{1200}$=  &                      $\displaystyle {\cal{S}}_{0111}$=0 \\
\hline\hline
            &                                  & $\displaystyle {\cal{S}}_{1000}$  & $-\displaystyle \frac{A_1}{2}$,  &$\displaystyle {\cal{S}}_{0100}$=$\displaystyle \frac{A_1}{2}$\\
\,\,\,${\cal S}^{(2)}$\,  & $\frac{\displaystyle 1}{\displaystyle 2}\,
                             \lef(\frac{\displaystyle 1}{\displaystyle F(k)\,F(p)}\rig)^{1/2}
                             \displaystyle{\ln}\frac{\displaystyle F(q)^2}{\displaystyle F(k)\,\displaystyle F(p)}$
                        & $\displaystyle {\cal{S}}_{2000}$  & $\displaystyle \frac{A_1^2}{4}$, &                     $\displaystyle {\cal{S}}_{1100}$= $-\displaystyle \frac{A_1^2}{2}$,
                                                                                                                        $\displaystyle {\cal{S}}_{0011}$= $\displaystyle \frac{A_1^2}{4}$
                                                                                                                        $\displaystyle {\cal{S}}_{0200}$=0\\
 &                      & $\displaystyle {\cal{S}}_{3000}$  & $\displaystyle -\frac{A_1^3}{16}$,&                      $\displaystyle {\cal{S}}_{2100}$= $\displaystyle \frac{A_1^3}{8}$,
                                                                                                                        $\displaystyle {\cal{S}}_{1011}$= $\displaystyle -\frac{3}{16}A_1^3$\\
 &                      &$\displaystyle {\cal{S}}_{0300}$   & $\displaystyle {\cal{S}}_{1200}$= 0 &                      $\displaystyle {\cal{S}}_{0111}$=$\displaystyle \frac{A_1^3}{8}$ \\

\hline\hline
            &                                  & $\displaystyle {\cal{S}}_{1000}$  & $-\displaystyle \frac{A_1}{2}$,  &$\displaystyle {\cal{S}}_{0100}$=$\displaystyle \frac{A_1}{2}$\\
\,\,\,${\cal S}^{(3)}$\,  & $\frac{\displaystyle 1}{\displaystyle 4}\,
                          \lef(\frac{\displaystyle{1}}{\displaystyle F(k)}
                              +\frac{\displaystyle{1}}{\displaystyle F(p)}\rig)
                              \displaystyle{\ln}\frac{\displaystyle F(q)^2}{\displaystyle F(k)\,\displaystyle F(p)}$
                        & $\displaystyle {\cal{S}}_{2000}$  & $\displaystyle \frac{A_1^2}{4}$, &                        $\displaystyle {\cal{S}}_{1100}$= $-\displaystyle \frac{A_1^2}{2}$,
                                                                                                                        $\displaystyle {\cal{S}}_{0011}$= $\displaystyle \frac{A_1^2}{4}$
                                                                                                                        $\displaystyle {\cal{S}}_{0200}$=0\\
 &                      & $\displaystyle {\cal{S}}_{3000}$  & $\displaystyle -\frac{A_1^3}{8}$,&                    $\displaystyle {\cal{S}}_{2100}$= $\displaystyle \frac{A_1^3}{4}$,
                                                                                                                        $\displaystyle {\cal{S}}_{1011}$= $\displaystyle -\frac{A_1^3}{8}$\\
 &                      & $\displaystyle {\cal{S}}_{0300}$  & $\displaystyle {\cal{S}}_{1200}$=&                    $\displaystyle {\cal{S}}_{0111}$=0 \\
\hline
\end{tabular}
\vspace{3mm}
\caption{Antisymmetric combinations of $F$ and $G$.}
\label{Tab:3}
\end{center}
\end{table}

\normalsize
\section{Application}
\label{sec:application}
\noindent
The next step is to make use of all the examples in Table~\ref{Tab:3} as inputs to
the multiplicative
renormalizability constraints. In order to satisfy these, we
have a  set of equations to solve. As a first step the coefficient
functions, $\tau_i$'s,
can in general be written as a sum of different non-perturbative forms of $F$ and $G$ using
 the above examples. Hence, an antisymmetric and symmetric combination of $F$ and $G$ in $\tau_i^{anti}$ and $\tau_i^{sym}$ respectively become~:
\be
\tau_i^{anti}&=&\lef(f^{(1)}\,{\cal A}^{(1)}\rig)_i
            +\,\lef(f^{(2)}\,{\cal A}^{(2)}\rig)_i +\, \cdots + \,\lef(f^{(n)}\,{\cal A}^{(n)}\rig)_i\nn\\
\tau_i^{sym}&=&\lef(\widetilde{f}^{(1)}\,{\cal S}^{(1)}\rig)_i
            +\,\lef(\widetilde{f}^{(2)}\,{\cal S}^{(2)}\rig)_i+\, \cdots + \,\lef(\widetilde{f}^{(n)}\,{\cal S}^{(n)}\rig)_i
\label{eq:taugen}
\ee
where ${\cal A}^{(n)}$ and ${\cal S}^{(n)}$ refer to the relevant expressions in the
left hand column of Table~\ref{Tab:3}. In general the number of constants needed to solve
these equations is proportional to the number, $n$, of various combinations of
the $F$ and $G$. These combinations will appear
in the ansatz for the non-perturbative transverse vertex.
We then try to solve these equations by choosing a minimal number
 of combinations, in order to find the simplest possible vertex
ansatz.

\noindent
From Eqs.~(\ref{eq:AS},~\ref{eq:Sphoton}), we see that the coefficients $\beta_i, \gamma_i,\delta_i, \epsilon_i$, defined in Eqs.~(\ref{eq:tauform}) appear in the fermion and photon conditions in rather
 specific combinations. To make this explicit and simplify the notation, it is useful to define~:
\be
  &&\beta_f\;\equiv\;(\beta_2+\beta_3+\beta_6-\beta_8)\quad ,\qquad\qquad\!\!\!\!\!\gamma_f\;\equiv\;(-\gamma_3+\gamma_6)\quad ,\nn\\
&&\delta_f\;\equiv\;(\delta_2+\delta_3+\delta_6-\delta_8)\quad ,\qquad\qquad\epsilon_f\;\equiv\;(-\epsilon_3-\epsilon_6)\quad ,\nn\\
&&\delta_\gamma\;\equiv\;(\delta_2-\delta_3+\delta_6-\delta_8)\quad , \qquad\qquad\epsilon_\gamma\;\equiv\;(\epsilon_2-\epsilon_3+\epsilon_6-\epsilon_8)\quad .
\label{eq:bdge}
\ee
Recall that antisymmetric forms for the $\tau_i$'s do not contribute to the photon renormalization at leading logarithmic order, and so we have no corresponding
combinations of $\beta_\gamma$ and $\gamma_\gamma$.

\subsection{Fermion constraints}
\noindent
We now wish to write down the fermion constraints $FC1-FC4$, Eqs.~(\ref{eq:fercond1},~\ref{eq:fercond41}), which we obtained in the previous section for the specific
choices for~ ${\overline{\tau}}^{\,anti}_f$ and ${\overline{\tau}}^{\,\,sym}_f$, namely
{ ${\cal{A}}^{(1)}$} as the antisymmetric form of the transverse vertex and {${\cal{S}}^{(1)}$} as the symmetric one in the Table~{\ref{Tab:3}}~:
\be
{\cal{A}}^{(1)}=\lef(\frac{1}{F(k^2)}-\frac{1}{F(p^2)}\rig)\,,\,
{\cal{S}}^{(1)}=\frac{1}{2}\,\lef(\frac{1}{F(k^2)}+\frac{1}{F(p^2)}\rig)\,\ln\frac{F(q^2)}{2}\,\lef(\frac{1}{F(k^2)}+\frac{1}{F(p^2)}\rig).\,\,\,
\label{eq:A1S1}
\ee

\noindent
After recalling the definition of ${\overline{\cal{A}}}_{mnrr}^{\,f}$ and ${\overline{\cal{S}}}_{mnrr}^{\,f}$ from Eq.~(\ref{eq:AS})
and reading off the specific values of ${\cal{A}}_{mnrr}$ and ${\cal{S}}_{mnrr}$'s from the Table~\ref{Tab:3},
the MR constraint $FC2$, Eq.~(\ref{eq:fercond1}), which comes from $\mathbf{\al_0^2\,\displaystyle \ln^2 p^2/\Lambda^2}$ order comparison together with
Eq.~(\ref{eq:bdge}) gives the following condition~:

\be
\fbox{$
\begin{array}{ccc}
-\displaystyle(\beta_f +\gamma_f)+\,\frac{1}{2}\lef(\delta_f+\varepsilon_f\rig)&=& \displaystyle\frac{1}{2}\,.
\end{array}$}
\label{eq:condfer}
\ee
The $\mathbf{\al_0^3\,\ln^3p^2/\Lambda^2}$ order constraint $FC3$, Eq.~(\ref{eq:fercond2}), splits the combined $\beta_f, \gamma_f, \delta_f, \varepsilon_f$ form of
previous constraint into two separate ones~:
\be
\fbox{$
\begin{array}{ccc}
\,\,\,\displaystyle\lef(\delta_f+\varepsilon_f\rig)&=& 0 \,,\,\,\\
(\beta_f +\gamma_f) & =&- \displaystyle\frac{1}{2}\,.
\end{array}$}
\label{eq:condfer2}
\ee
\noindent
The $\mathbf{\al_0^4\,\ln^4 p^2/\Lambda^2}$ order constraint $FC4$, Eq.~(\ref{eq:fercond41}) does not give
further new information, but again yields Eq.~(\ref{eq:condfer2}).
\vspace*{-5mm}
\subsection{Photon constraints}
\noindent
We repeat this procedure procedure for the photon constraints $PC2-PC4$ for the same choices of ${\cal{A}}^{(1)}$
and ${\cal{S}}^{(1)}$ in Table~\ref{Tab:3}.
All the MR constraints $PC2$ to $PC4$, Eqs.~(\ref{eq:photoncond1},~\ref{eq:photoncond2}), which follow from ${\al_0^2 \ln^2}$ to $\al_0^4 \ln^4$ comparisons
give the same condition and that is~:
\be
\fbox{$
\,\,\,\displaystyle \lef(\delta_{\gamma}+\varepsilon_{\gamma}\rig)= -\frac{4}{3}\,.\,\,
$}
\label{eq:condphoton2}
\ee
\noindent
Since this condition repeats itself at every order, this means we have the exact solutions! There are 14 constants to be fixed, and
Eqs.~(\ref{eq:condfer2},~(\ref{eq:condphoton2}) can only fix three of them in terms of the others, for instance~:
\be
\beta_2&=& -\frac{1}{2}-\beta_3-\beta_6+\beta_8+\gamma_3-\gamma_6\,,\nn\\
\delta_2&=&-\frac{2}{3}-\delta_6+\delta_8-\frac{\varepsilon_2}{2}+\varepsilon_3-\varepsilon_6+\frac{\varepsilon_8}{2}\,, \nn\\
\delta_3&=&\frac{2}{3}+\frac{\varepsilon_2}{2}-\frac{\varepsilon_8}{2}\,.
\ee
\noindent
Substituting these constants into the Eq.~(\ref{eq:tauform}) we
can write the non-perturbative coefficient functions $\tau_i$'s as ~:
\be
&&\tau_2^M(p^2,k^2,q^2)\,= \nn\\
&& \hspace*{1.3cm}
\displaystyle{\frac{2}{(k^4-p^4)}}
\,\,\lef[\lef(\frac{1}{2}-\beta_3-\beta_6+\beta_8+\gamma_3-\gamma_6 \rig)+\gamma_2\,\frac{2\, k\cdot p}{k^2+p^2}\rig]
                    \; \tau_2^{anti}(p^2, k^2,q^2) \nn\\[1.5mm]
&& \hspace*{1cm}
+\displaystyle{\frac{2}{(k^2+p^2)^2}}\lef[\lef( -\frac{2}{3}-\delta_6+\delta_8-\frac{\varepsilon_2}{2}+\varepsilon_3-\varepsilon_6+\frac{\varepsilon_8}{2}\rig)
+\epsilon_2 \frac{2\,k\cdot p}{k^2+p^2}\,\,\rig] \;\tau_2^{sym}(p^2, k^2, q^2) \,,\nn\\ [2mm]
&&\tau_3^M(p^2,k^2,q^2)=
\displaystyle{\frac{1}{(k^2-p^2)}}\,\lef[\beta_3+\gamma_3\,\frac{2\, k\cdot p}{k^2+p^2}\rig]\;
                  \tau_3^{anti}(p^2,k^2,q^2) \nn\\ [1.5mm]
&& \hspace*{2.4cm}
+\displaystyle{\frac{1}{(k^2+p^2)}}\,\lef[\lef( \frac{2}{3}+\frac{\varepsilon_2}{2}-\frac{\varepsilon_8}{2} \rig)+\epsilon_3\frac{2\, k\cdot p}{k^2+p^2}\,\,\rig] \;
                  \tau_3^{sym}(p^2, k^2, q^2) \,,\nn\\[2mm]
&&\tau_6^M(p^2,k^2,q^2)=
\displaystyle{\frac{1}{(k^2+p^2)}}\,\,\lef[\beta_6+\gamma_6\,\frac{2\, k\cdot p}{k^2+p^2}\rig] \;\tau_6^{anti}(p^2, k^2,q^2)\nn \\[1.5mm]
&& \hspace*{2.4cm}
+\displaystyle{\frac{(k^2-p^2)}{(k^2+p^2)^2}}\lef[\delta_6+\epsilon_6\frac{2\, k\cdot p}{k^2+p^2}\,\,\rig]\;
        \tau_6^{sym}(p^2, k^2,q^2) \,,\nn\\[2mm]
&&\tau_8^M(p^2,k^2,q^2)=
\displaystyle{\frac{1}{(k^2-p^2)}}\lef[\beta_8+\gamma_8\,\frac{2\, k\cdot p}{k^2+p^2}\rig]  \;
        \tau_8^{anti}(p^2, k^2,q^2) \nn\\[1.5 mm]
&& \hspace*{2.4cm}
+\displaystyle{\frac{1}{(k^2+p^2)}}\lef[\delta_8+\epsilon_8\frac{2\, k\cdot p}{k^2+p^2}\,\,\rig]\;
        \tau_8^{sym}(p^2, k^2, q^2)\,,
\label{eq:tauform2}
\ee
\normalsize
\noindent
with $\tau_i^{anti,~sym}$ having specific forms such as those determined in Sec.~V, Eqs.~(\ref{eq:Symxamples},~\ref{eq:nonpertfercond1}), examples of which are given in Table~\ref{Tab:3}.

\noindent
Multiplicative renormalizability relates the coefficients at order $(\alpha_0 \ln)^n$ to that at $n=1$. This lowest leading logarithm coefficient is fixed by
the longitudinal component of the fermion-boson vertex. Transverse components only enter at $n=2$. Remarkably, once the MR conditions at this first non-trivial
 order are satisfied, the conditions at all orders in leading logarithms for both the fermion and photon Schwinger-Dyson equations are fulfilled.

\noindent As far as the leading terms are concerned, the above
constraints ensure that both fermion and
photon propagators are multiplicatively renormalizable in massless unquenched
QED$_4$. These constraints impose conditions on the transverse
part of the vertex.
The 3-point vertex  calculated at ${\cal{O}}(\alpha_0)$ and the coefficient
constants, $\tau_i$'s, at one loop order \cite{Kizilersu:1995iz} will be very helpful in  fixing some of these constants.
\section{Perturbation Theory}

\noindent
The vertex coefficients $\tau_i$'s were calculated exactly in ${\cal{O}}(\alpha_0)$ for the
massive fermions in a general covariant gauge \cite{Kizilersu:1995iz} and for
our purpose their massless limits are given in Appendix~\ref{app:perttaus}.

\noindent
We observe in Eq.~(\ref{eq:perttau2}-\ref{eq:perttau8}) that all the four $\tau_i$'s ($i=2,3,6,8$), contains four different structures in general. The
first one is the $J_0$ dependent part, which contains Spence functions (or ~Dilogarithms) of momenta $p^2,k^2,q^2$ in Eq.~({\ref{eq:J0}). The second part is
proportional to $\ln{k^2/p^2}$  which is the perturbative expansion of the asymmetric combination of $F$ and $G$ in first order,
and the third part is proportional to $\ln{q^4/(k^2\,p^2)}$, which is the perturbative expansion of the symmetric combination of $F$ and $G$,
and the final one is the kinematical term dependent on $k^2,p^2,q^2$.

\noindent
In order to fix some of the individual constants $\beta_i,~ \gamma_i,~ \delta_i,~ \varepsilon_i$'s appearing in Eqs.~(\ref{eq:condfer2},~\ref{eq:condphoton2})
we need to make a comparison between
 perturbative transverse vertex coefficients $\tau_i^{pert}$ of Eq.~(\ref{eq:perttau2}-~\ref{eq:perttau8}), and the non-perturbative ones
we used in fermion and photon SDE, $\tau_i^{non-pert.}$ of Eqs.~(\ref{eq:nonperttauanti},~\ref{eq:nonperttausym}) in the previous sections. However
this comparison has to be made in a particular way in order to be meaningful. There are two points to be considered. The first  is
how these $\tau_i$ coefficients behave inside the fermion and photon SDEs, since these equations project out different parts of the vertex.
Recall, that with this in mind we started with a simplified ansatz for the explicit kinematic factors in the $\tau_i^{non-pert.}$, Eq.~(\ref{eq:tauform}),
and assumed their denominators did not depend on $k\cdot p$.
We therefore need to take the corresponding limits of both pure perturbative $\tau_i^{pert.}$'s, Eq.~(\ref{eq:perttau2}-~\ref{eq:perttau8}),  and
the $\tau_i^{non-pert.}$'s, Eqs.~(\ref{eq:nonperttauanti},~\ref{eq:nonperttausym}) which we inserted into SDE.
While for the fermion SDE the relevant limit would be
where either of the fermion momenta are large, {\it e.g.} $ k^2 \simeq q^2 \gg k \cdot\! p \gg p^2$, for the photon SDE the relevant one is where the both internal
fermion momenta are same and much greater than the photon momentum, e.g. $ k^2 \simeq p^2 \gg q^2$.

\noindent
The second point is that the real $\tau_i$ functions depend on
the angle between momenta $k$ and $p$.
This means that when we obtained MR constraints, Eqs.~(\ref{eq:condfer},~\ref{eq:condphoton2}),
on the vertex, {\it i.e.} on $\tau_i$ functions, their angular dependences were already integrated out. These angular averaged functions
we call {\underline{effective}} $\tau_i$'s~\cite{Bashir:1997qt}. It is these that we have to compare with perturbation theory.
\subsection{ $k^2 \simeq q^2 \gg p^2$: The Fermion Limit}
\noindent
Let us take the fermion limit   of the perturbative $\tau_i^{pert.}$'s, Eqs.~(\ref{eq:perttau2}-~\ref{eq:perttau8}) in Euclidean space.
In order to do this, $J_0$ of Eqs.~(\ref{eq:J0},~\ref{eq:J02}) has to be expanded up to
${\cal O}(1/k^7)$ to ensure we keep all  the terms of the required order.
As shown in Appendix~\ref{app:J0}, these results are~:
\normalsize
\be
(\tau_2^{E})^{pert.}_{Real}\,(p^2,k^2,q^2)&=&
\frac{\alpha_0\,\xi}{8\pi k^4}\,\,\ln\frac{k^2}{p^2}\,\,\Big\{\,\hspace*{5mm} \frac{4}{3} \, + \,2\,\,\,\,\,\frac{k \cdot p}{k^2}\,\,+\,\frac{14}{15}\,\frac{p^2}{k^2}\,\Big\}\,,\nn\\
(\tau_3^E)_{Real}^{pert.}\,(p^2,k^2,q^2)&=&\frac{\alpha_0\,\xi}{8\pi k^2}\,\,\ln\frac{k^2}{p^2}\,\,\Big\{ \,\hspace*{5mm}\frac{2}{3}\,  +\, \,\,\,\,\,\,\,\frac{k \cdot p}{k^2}\,\,
+\,\frac{2}{15}\,\frac{p^2}{k^2}\,\,\Big\}\,,\nn\\
(\tau_6^E)_{Real}^{pert.}\,(p^2,k^2,q^2)&=&\frac{\alpha_0\,\xi}{8\pi k^2}\,\,\ln\frac{k^2}{p^2}\,\,\Big\{\, -\frac{1}{3}\, -\, \frac{1}{3}\,\,\,\frac{k \cdot p}{k^2}\,\,
-\,\frac{1}{5}\,\,\,\frac{p^2}{k^2}\,\,\,\Big\}\,,
\nn\\[3mm]
(\tau_8^E)_{Real}^{pert.}\,(p^2,k^2,q^2)&=&0\,.
\label{eq:ferlimitperttaus}
\ee
In this limit one observes that both $J_0$ and $\ln\lef(q^4/{k^2\,p^2}\rig)$ behave like $\ln\lef(k^2/p^2\rig)$. Therefore all four coefficient
functions become proportional to $\ln\lef(k^2/p^2\rig)$ signaling that the structure of non-perturbative transverse vertex consists of purely asymmetric combination of
$F$ or $G$.
Next we expand the non-perturbative $\tau_i^{non-pert.}$'s, Eq.~(\ref{eq:tauform}), using Eqs.~(\ref{eq:perttauanti},~\ref{eq:perttausym}) at the order ${\cal{O}}(\alpha_0)$~:
\be
(\tau_2^E)^{non-pert.}\,(p^2,k^2,q^2)&={\,\,\displaystyle \frac{2}{k^4}}&
\lef(\beta_2+\gamma_2\,\frac{2\, k\cdot p}{k^2}\rig)
                    \; \lef[\,\,\,\,\alpha_0 \,{\cal{A}}_{1000}^2\,\ln\frac{k^2}{p^2}\rig]\nn\\
 \,\,&+{\displaystyle\frac{2}{k^4}}&\lef(\delta_2+\epsilon_2 \frac{2\,k\cdot p}{k^2}\,\,\rig) \;
          \lef[-\alpha_0\, {\cal{S}}_{1000}^2\,\ln\frac{k^2}{p^2}\,\,\rig]+{\cal{O}}(\alpha_0^2)\quad,
\nn\\[2mm]
(\tau_3^E)^{non-pert.}\,(p^2,k^2,q^2)&=
      {\,\,\displaystyle\frac{1}{k^2}}&
\lef(\beta_3+\gamma_3\,\frac{2\, k\cdot p}{k^2}\rig)\;
                  \lef[-\alpha_0 \,{\cal{A}}_{1000}^3\,\ln\frac{k^2}{p^2}\rig] \nn\\
 \,\,&+{\displaystyle\frac{1}{k^2}}&\lef(\delta_3+\epsilon_3\frac{2\, k\cdot p}{k^2}\,\,\rig) \;
                  \lef[\,\,\,\,\,\alpha_0 \,{\cal{S}}_{1000}^3\,\ln\frac{k^2}{p^2}\,\rig] +{\cal{O}}(\alpha_0^2)\quad,
\nn\\
(\tau_6^E)^{non-pert.}\,(p^2,k^2,q^2)&=
    {\,\,\displaystyle\frac{1}{k^2}}&
\lef(\beta_6+\gamma_6\,\frac{2\, k\cdot p}{k^2}\rig) \;
        \lef[-\alpha_0\, {\cal{A}}_{1000}^6\,\ln\frac{k^2}{p^2}\rig] \nn\\
\nn \,\,&+{\displaystyle\frac{1}{k^2}}&\lef(\delta_6+\epsilon_6\frac{2\, k\cdot p}{k^2}\,\,\rig)\;
        \lef[\,\,\,\,\alpha_0\, {\cal{S}}_{1000}^6\,\ln\frac{k^2}{p^2}\rig] +{\cal{O}}(\alpha_0^2)\quad,
\nn\\
(\tau_8^E)^{non-pert.}\,(p^2,k^2,q^2)&=
    {\,\,\displaystyle\frac{1}{k^2}}&
\lef(\beta_8+\gamma_8\,\frac{2\, k\cdot p}{k^2}\rig)  \;
        \lef[-\alpha_0\, {\cal{A}}_{1000}^8\,\ln\frac{k^2}{p^2}\rig]\nn\\
 \,\,&+\displaystyle{\displaystyle\frac{1}{k^2}}&\lef(\delta_8+\epsilon_8\frac{2\, k\cdot p}{k^2}\,\,\rig)\;
       \lef[\,\,\,\,\alpha_0 \,{\cal{S}}_{1000}^8\,\ln\frac{k^2}{p^2}\rig] +{\cal{O}}(\alpha_0^2)\quad.
\label{eq:ferlimitsdetaus}
\ee
\noindent
As we mentioned earlier, during the process of finding MR constraints in Eq.~(\ref{eq:condfer2}) from the fermion SDE we performed both radial and angular
integrations therefore these constraints on the vertex are for the $\tau_i$'s whose angular dependence has been integrated out,
{\it viz.} they are the effective $\tau_i^{non-pert.}$'s.
 To make consistent
 comparison between the Eqs.~(\ref{eq:ferlimitperttaus},~\ref{eq:ferlimitsdetaus}), we must
integrate out the angular dependence of both $\tau_i^{pert.}$ and $\tau_i^{non-pert.}$.
The details of this procedure can be found in Appendix~\ref{app:effectivetaus}. Following this, the
effective coefficient functions can be found from $\tau_{Real}^{pert.}$'s in Eq.~(\ref{eq:ferlimitperttaus})~:
\be
(\tau_{2}^E)_{eff}^{pert.}\,(p^2,k^2)&=&
\frac{\alpha_0\,\xi}{8\pi k^4}\,\,\,\ln\frac{k^2}{p^2}\,\,\,\lef(\,\frac{4}{3}\,\rig)\,,\nn\\
(\tau_{3}^E)_{eff}^{pert.}\,(p^2,k^2)&=&
\frac{\alpha_0\,\xi}{8\pi k^2}\,\,\,\ln\frac{k^2}{p^2}\,\,\,\lef(\,\frac{1}{6}\,\rig)\,,\nn\\
(\tau_{6}^E)_{eff}^{pert.}\,(p^2,k^2)&=&
\frac{\alpha_0\,\xi}{8\pi k^2}\,\,\,\ln\frac{k^2}{p^2}\,\,\,\lef(\,-\frac{1}{2}\,\rig)\,,\nn\\[4mm]
(\tau_{8}^E)_{eff}^{pert.}\,(p^2,k^2)&=&\hspace*{5mm}0\,.
\label{eq:effectiveperttaus}
\ee
\noindent
We repeat the same procedure for the first order expansion of the non-perturbative coefficients $\tau_{Real}^{non-pert.}$'s in Eq.~(\ref{eq:ferlimitsdetaus}) to give~:
\normalsize
\be
(\tau_2^E)_{eff}^{non-pert.}\,(p^2,k^2)&=&\frac{2}{k^4} \hspace*{7mm}
     \beta_2\hspace*{6.5mm}\lef(\,\,\,\,\,\alpha_0\,{\cal{A}}_{1000}^2\,\ln\frac{k^2}{p^2}\,\rig)\nn\\
 &+&{\frac{2}{k^4}}\hspace*{6mm}\delta_2 \hspace*{7mm}\lef(\,-\alpha_0\,{\cal{S}}_{1000}^2\,\ln\frac{k^2}{p^2}\rig)+{\cal{O}}(\alpha^2)\quad,
\nn\\ [4mm]
(\tau_3^E)_{eff}^{non-pert.}\,(p^2,k^2)&=&
    \frac{1}{k^2}\,
\lef(\beta_3-\gamma_3\rig)
                 \lef(-\alpha_0\,{\cal{A}}_{1000}^3\,\ln\frac{k^2}{p^2}\,\,\rig)\nn\\
&+&{\frac{1}{k^2}}\lef(\delta_3-\epsilon_3\rig)
                  \lef(\,\,\,\,\,\alpha_0 \,{\cal{S}}_{1000}^3\,\ln\frac{k^2}{p^2}\rig) +{\cal{O}}(\alpha^2)\quad,
\nn\\[4mm]
(\tau_6^E)_{eff}^{non-pert.}\,(p^2,k^2)&=&
    \frac{1}{k^2}\,
\lef(\beta_6+\gamma_6\rig)
       \lef(-\alpha_0 \,{\cal{A}}_{1000}^6\,\ln\frac{k^2}{p^2}\rig)\nn\\
&+&{\frac{1}{k^2}}\lef(\delta_6+\epsilon_6\rig)\;
        \lef(\alpha_0 {\cal{S}}_{1000}^6\ln\frac{k^2}{p^2}\rig) +{\cal{O}}(\alpha^2)\quad,
\nn\\[4mm]
(\tau_8^E)_{eff}^{non-pert.}\,(p^2,k^2)&=&
    \frac{1}{k^2}\,\hspace*{7mm}
\beta_8 \hspace*{7mm}
       \lef(-\alpha_0 {\cal{A}}_{1000}^8\ln\frac{k^2}{p^2}\rig)\nn\\
& +&{\frac{1}{k^2}}\hspace*{7mm}\delta_8 \hspace*{7mm}
       \lef(\alpha_0 {\cal{S}}_{1000}^8\ln\frac{k^2}{p^2}\rig) +{\cal{O}}(\alpha^2)\quad.
\label{eq:nonperttauseff}
\ee

\noindent
The constants $\beta_i$'s, $\delta_i$'s, $\gamma_i$'s and $\varepsilon_i$'s appearing in Eq.~(\ref{eq:nonperttauseff}) are the
ones which must satisfy the MR constraints, Eqs.~(\ref{eq:condfer2},~\ref{eq:condphoton2}).
Let us check we have obtained the correct result in three key situations.

\noindent
First we compare Eq.~(\ref{eq:effectiveperttaus}) with Eq.~(\ref{eq:nonperttauseff})
to read off the constraints on ${\cal{A}}^{\,i}_{1000}$ and ${\cal{S}}^{\,i}_{1000}$ for $i=2,3,6,8$~:
\be
\beta_2\,{\cal{A}}^2_{1000}-\delta_2\,{\cal{S}}^2_{1000}&=& \frac{A_1}{3}\,, \nn\\
(\beta_3-\gamma_3)\,{\cal{A}}^3_{1000}-(\delta_3-\varepsilon_3)\,{\cal{S}}^3_{1000}&=& -\frac{A_1}{12} \,,\nn\\
(\beta_6+\gamma_6)\,{\cal{A}}^6_{1000}-(\delta_6+\varepsilon_6)\,{\cal{S}}^6_{1000}&=& \frac{A_1}{4} \,,\nn\\
\beta_8\,{\cal{A}}^8_{1000}-\delta_8\,{\cal{S}}^8_{1000}&=& 0\,.
\label{eq:indivfertaus}
\ee

\noindent
\textbf{\underline{1a)~General Case at ${\cal{O}}(\alpha_0)$~:}}
Recall the definition of ${\overline{\cal{A}}}^{\,f}_{1000}$ and ${\overline{\cal{S}}}^{\,f}_{1000}$, Eq.~(\ref{eq:AS}), in order to
form the $FC2$ constraint in Eq.~(\ref{eq:fercond1}) using above expressions by adding them up appropriately ~:
\be
&&{\overline{\cal{A}}}^{\,f}_{1000}\,-\,{\overline{\cal{S}}}^{\,f}_{1000}\,=\,\lef( \frac{1}{3}-\frac{1}{12}+\frac{1}{4}\rig)\,A_1 \,=\, \frac{A_1}{2}\,.
\ee
\textbf{\underline{1b)~For the special vertex (${\cal{A}}^{(1)}$ and ${\cal{S}}^{(1)}$) at ${\cal{O}}(\alpha_0)$~:}}
Making use of Table~\ref{Tab:3} we can read off the value of ${\cal{A}}^i_{1000}$ and ${\cal{S}}^i_{1000}$ and insert them into Eq.(\ref{eq:indivfertaus})
to see whether we can satisfy the fermion MR constraint of Eq.~(\ref{eq:condfer}) by using Eq.~(\ref{eq:bdge})~:
\be
&&\lef[\lef(\beta_2+\beta_3+\beta_6-\beta_8\rig)+\lef(-\gamma_3+\gamma_6\rig)\rig]\,(-A_1)
-\lef[\lef(\delta_2+\delta_3+\delta_6-\delta_8\rig)+\lef(-\varepsilon_3+\varepsilon_6\rig)\rig]\,(-\frac{A_1}{2})\qquad\,,\nn\\
&&\qquad\qquad\qquad\qquad=\lef[-(\beta_f+\gamma_f)+\frac{1}{2}\,(\delta_f+\varepsilon_f)\rig]\,A_1\,,\nn\\
&&\qquad\qquad\qquad\qquad=\lef( \frac{1}{3}-\frac{1}{12}+\frac{1}{4}\rig)\,A_1\,,\nn\\
&&\qquad\qquad\qquad\qquad= \frac{A_1}{2}\,.
\ee

\noindent
As we see all effective $\tau^i_{eff}$'s, Eq.~(\ref{eq:effectiveperttaus}) add up to $A_1/2$, as required.

\noindent
\textbf{\underline{2)~Non-perturbative check~:}}
~If we trace back the MR constraint in fermion SDE equation, Eq.~(\ref {eq:unqmrcondF}), we have already observed that the $\xi$ dependent
part will give the right equality and the rest must be zero to give the
fermion MR condition. Hence this MR constraint for the effective $\tau_i$'s after the angular and before the radial integration was performed can be written as~:
\be
\frac{3\alpha_0}{8\pi}\;\int\;\frac{dk^2}{k^2}\; F(k^2)\;G(k^2)\;\lef[ \frac{1}{2}\,\lef(\frac{1}{F(k^2)}\,-\,\frac{1}{F(p^2)}\rig)\,+k^2\,\sum (\tau_i^E)_{eff}^{non-pert.}\,(p^2,k^2) \rig]=0\,,\nn\\
\label{eq:sumtau1}
\ee
where
\be
\sum (\tau_i^E)_{eff}^{non-pert.}\,(p^2,k^2) \;=\; \frac{1}{2} k^2 (\tau_2^E)_{eff}\,-\,(\tau_3^E)_{eff}\,-\,(\tau_6^E)_{eff}\,+\,(\tau_8^E)_{eff}\,.
\label{eq:sumtaus}
\ee

\noindent
At ${\cal{O}}(\alpha_0)$
\be
\frac{1}{2}\,\lef(\frac{1}{F(k^2)}\,-\,\frac{1}{F(p^2)}\rig)= -\frac{A_1}{2}\, \alpha_0\,\ln\frac{k^2}{p^2}+{\cal{O}}(\alpha_0^2)\,.
\label{eq:diffF}
\ee
Making use of $\tau_{eff}^i$'s in Eq.~(\ref{eq:effectiveperttaus}) to form Eq.~(\ref{eq:sumtaus}) gives~:
\be
k^2\ \sum (\tau_i^E)_{eff}\,(p^2,k^2) &=& \alpha_0\,\ln\frac{k^2}{p^2}\,\lef[\frac{2}{3}-\frac{1}{6}+\frac{1}{2}\rig]\frac{A_1}{2}\,,\nn\\
&=& \frac{A_1}{2}\,\alpha_0\,\ln\frac{k^2}{p^2}+{\cal{O}}(\alpha_0^2)\,.
\label{eq:sumtau2}
\ee
\noindent
Since Eq.~(\ref{eq:diffF}) cancels out in  Eq.~(\ref{eq:sumtau2}),  Eq.~(\ref{eq:sumtau1}) is satisfied.
\subsection{ $ k^2 \simeq p^2 \gg q^2$ ~: The Photon Limit}
\noindent
Let us turn our attention now to the photon limit of the perturbative $\tau_i^{pert.}$'s, Eqs.~(\ref{eq:perttau2}-~\ref{eq:perttau8}) in Euclidean space.
The technical details
of this limit can be found in Appendix~\ref{sec:photonlimit} and then we have~:
\be
(\tau_2^E)_{real}^{pert}\,(\e^2,q^2)&=&\frac{\alpha_0\,\xi}{12\pi \e^4}\,\ln\frac{\e^2}{q^2}+{\cal{O}}(\alpha_0^2)\,,\nn\\
(\tau_3^E)_{real}^{pert}\,(\e^2,q^2)&=&\frac{\alpha_0\,\xi}{12\pi \e^2}\,\ln\frac{\e^2}{q^2}+{\cal{O}}(\alpha_0^2)\,,\nn\\
(\tau_6^E)_{real}^{pert}\,(\e^2,q^2)&=&\qquad 0 \quad \quad \,\,\,\, +{\cal{O}}(\alpha_0^2)\,,\nn\\
(\tau_8^E)_{real}^{pert}\,(\e^2,q^2)&=&\qquad 0 \quad \quad \,\,\,\,+{\cal{O}}(\alpha_0^2)\,,
\label{eq:efftaus}
\ee
since in this limit $\ln\lef(k^2/p^2\rig)$ approaches $1$ and $\ln\lef(q^4/(k^2\,p^2)\rig)$ approaches $\ln\lef(q^4/\e^4\rig)$. Therefore all four coefficient
functions become proportional to $\ln\lef(q^2/\e^2\rig)$. This signals that the structure of the non-perturbative transverse vertex consists of purely symmetric combination of
$F$ or $G$.
We expand the non-perturbative $\tau_i^{non-pert.}$'s, Eq.~(\ref{eq:tauform}), using  Eqs.~(\ref{eq:perttauanti},~\ref{eq:perttausym}) at the order ${\cal{O}}(\alpha_0)$~:

\be
(\tau_2^E)_{real}^{non-pert.}\,(\e^2,q^2)&=
\displaystyle{\frac{1}{\e^4}}\,&\lef(\delta_2+\varepsilon_2 \rig) \; \alpha_0\, {\cal{S}}_{1000}^2\,\ln\frac{\e^2}{q^2}+{\cal{O}}(\alpha_0^2)\quad,
\nn\\ [4mm]
( \tau_3^E)_{real}^{non-pert.}\,(\e^2,q^2)&=
\displaystyle {-\frac{1}{\e^2}}\,&\lef(\delta_3+\varepsilon_3\rig)\;
                  \alpha_0 \,{\cal{S}}_{1000}^3\,\ln\frac{\e^2}{q^2} +{\cal{O}}(\alpha_0^2)\quad,
\nn\\[4mm]
(\tau_6^E)_{real}^{non-pert.}\,(\e^2,q^2)&=
\displaystyle{-\frac{\e\cdot q}{\e^2}}\,&\lef(\delta_6+\varepsilon_6\rig)\;
        \alpha_0\, {\cal{S}}_{1000}^6\,\ln\frac{\e^2}{q^2} +{\cal{O}}(\alpha_0^2)\quad,
\nn\\[4mm]
(\tau_8^E)_{real}^{non-pert.}\,(\e^2,q^2)&=
\displaystyle{-\frac{1}{\e^2}}\,&\lef(\delta_8+\varepsilon_8\rig)\;
       \alpha_0\, {\cal{S}}_{1000}^8\,\ln\frac{\e^2}{q^2} +{\cal{O}}(\alpha_0^2)\quad.
\label{eq:photonlimit1}
\ee
\noindent
Comparing Eq.~(\ref{eq:efftaus}) and  Eq.~(\ref{eq:photonlimit1}) one can read off the symmetric coefficients as~:
\be
(\delta_2+\varepsilon_2)\,{\cal{S}}^2_{1000}&=& \frac{A_1}{3}\,,\nn\\
(\delta_3+\varepsilon_3)\,{\cal{S}}^3_{1000}&=& -\frac{A_1}{3}\,,\nn\\
(\delta_6+\varepsilon_6)\,{\cal{S}}^6_{1000}&=& 0\,,\nn\\
(\delta_8+\varepsilon_8)\,{\cal{S}}^8_{1000}&=& 0\,.
\label{eq:indivphotaus}
\ee
Analogously to the fermion case, we now perform similar checks for the photon constraints in the same three situations~:

\noindent
\textbf{\underline{1a)~General Case at ${\cal{O}}(\alpha_0)$~:}}

\noindent
Recalling Eq.~(\ref{eq:Sphoton}) let us check whether the photon MR constraint $PC2$, Eq.~(\ref{eq:photoncond1}), at ${\cal{O}}(\alpha_0)$ is satisfied by Eq.~(\ref{eq:indivphotaus})
after adding them appropriately~:
\be
(\delta_2+\varepsilon_2)\,{\cal{S}}^2_{1000}-(\delta_3+\varepsilon_3)\,{\cal{S}}^3_{1000}+(\delta_6+\varepsilon_6)\,{\cal{S}}^6_{1000}-(\delta_8+\varepsilon_8)\,{\cal{S}}^8_{1000}
&=&\lef(\frac{A_1}{3}-\lef(-\frac{A_1}{3}\rig)\rig)\,,\nn\\[4mm]
i.e. \hspace*{7.5cm} {\overline{\cal{S}}}_{1000}^{\,\gamma}&=&\frac{2}{3}\,A_1\,.
\label{eq:check1}
\ee
\textbf{\underline{1b)~For the special vertex (${\cal{A}}^{(1)}$ and ${\cal{S}}^{(1)}$) at ${\cal{O}}(\alpha_0)$~:}}

\noindent
We also check if the photon MR constraint, Eq.~(\ref{eq:condphoton2} ) at ${\cal{O}}(\alpha_0)$ is satisfied for this special choice of the vertex~:
\be
\lef((\delta_2+\varepsilon_2)\,-(\delta_3+\varepsilon_3)\,+(\delta_6+\varepsilon_6)\,-(\delta_8+\varepsilon_8)\rig)\,\lef(\frac{-A_1}{2} \rig) &=&\frac{2}{3}\,A_1\,,\nn\\[4mm]
i.e. \hspace*{7.5cm} \delta_{\gamma}+\varepsilon_{\gamma}&=&-\frac{4}{3}\,.
\label{eq:check2}
\ee
\noindent
As we can see from both results, Eqs.~(\ref{eq:check1},~\ref{eq:check2}), the effective $\tau^i_{eff}$'s satisfy the photon MR constraint.

\noindent
\textbf{\underline{2)~Non-perturbative check~:}}

\noindent
Recalling Eq.~(\ref {eq:finalsdeG1}) and after extracting the non-perturbative MR constraints, we can usefully rewrite this as~:
\be
 \frac{\alpha\,N_F}{3\,\pi}\,\int_{q^2}^{\Lambda^2}\,\frac{d\e^2}{\e^2}\,\Bigg\{
\lef[F(\e)-1\rig]\,+\,\frac{3}{2}\,\e^2\,F^2(\e)\,\sum (\tau_i^E)^{non-pert.}_{eff}\,(\e^2,q^2)
\Bigg\}\,=\,0\; ,
\label{eq:sumtausphoton}
\ee
where
\be
\sum (\tau_i^E)_{eff}^{non-pert.}\,(\e^2,q^2) \;=\; \e^2 \,(\tau_2^E)_{eff}\,+\,(\tau_3^E)_{eff}\,+\,(\tau_8^E)_{eff}\,.
\label{eq:sumpho}
\ee
\noindent
At ${\cal{O}}(\alpha_0)$
\be
\int_{q^2}^{\Lambda^2}\,\frac{d\e^2}{\e^2}\,\lef[F(\e^2)\,-\,1\rig]\,=\,-\frac{A_1}{2}\,\alpha_0\,\ln^2 \frac{q^2}{\Lambda^2}+{\cal{O}}(\alpha_0^2)\,.
\label{eq:diffFpho}
\ee
\noindent
Making use of Eq.~(\ref{eq:efftaus}) to form Eq.~(\ref{eq:sumpho}) we obtain ~:
\be
\int_{q^2}^{\Lambda^2}\,\frac{d\e^2}{\e^2}\,\lef[\frac{3}{2}\,\e^2\,F^2(\e)\, \sum (\tau_i^E)_{eff}\,(\e^2,q^2)\rig] &=& \frac{A_1}{2}\,\alpha_0\,\ln^2\frac{q^2}{\Lambda^2}+{\cal{O}}(\alpha_0^2)\,.
\label{eq:sumtau2pho}
\ee
\noindent
We see  Eq.~(\ref{eq:diffFpho}) cancels Eq.~(\ref{eq:sumtau2pho}) and so Eq.~(\ref{eq:sumtausphoton}) is satisfied.
\subsection{Individual coefficients}
\noindent
With guidance from perturbation theory, we can now find further relations between the
constants, Eq.~(\ref{eq:indivfertaus}) and Eq.~(\ref{eq:indivphotaus}). These eight equations
fix eight of the 14 unknown constants ($\delta_2$, $\delta_3$, $\delta_6$, $\delta_8$, .....). In general these are~:

\be
\fbox{$
\begin{array}{rcl}
\displaystyle \delta_2\,{\cal{S}}_{1000}^2 &=&\displaystyle \frac{A_1}{3}-\,\varepsilon_2\,{\cal{S}}_{1000}^2\,
\qquad,\qquad
\displaystyle\qquad\,\,\,\,\,\,\beta_2\,{\cal{A}}^2_{1000}=\displaystyle \frac{2}{3}\,A_1\,-\varepsilon_2\,{\cal{S}}_{1000}^2\,,\nn\\
\displaystyle\delta_3\,{\cal{S}}_{1000}^3 &=& \displaystyle-\frac{A_1}{3}-\,\varepsilon_3\,{\cal{S}}_{1000}^3\,
\quad\,,\qquad
\displaystyle(\beta_3-\gamma_3)\,{\cal{A}}^3_{1000}= \displaystyle-\frac{5}{12}\,A_1\,-2\,\varepsilon_3\,{\cal{S}}_{1000}^3\,,\nn\\
\displaystyle\delta_6\,{\cal{S}}_{1000}^6 &=&\displaystyle -\varepsilon_6\,{\cal{S}}_{1000}^6\,
\qquad\qquad,\qquad
\displaystyle(\beta_6+\gamma_6)\,{\cal{A}}^6_{1000}=\displaystyle \frac{1}{4}\,A_1\,,\nn\\
\displaystyle\delta_8\,{\cal{S}}_{1000}^8 &=&\displaystyle -\varepsilon_8\,{\cal{S}}_{1000}^8\,
\qquad\qquad,\qquad
\displaystyle\qquad\quad\beta_8\,{\cal{A}}^8_{1000}=-\displaystyle\varepsilon_8\,{\cal{S}}_{1000}^8\,.
\end{array} $}\\
\label{eq:finalconst}
\ee
%
%
%
\noindent
For the specific choices of  antisymmetric, ${\cal{A}}^{(1)}$ and symmetric ${\cal{S}}^{(1)}$ transverse vertex forms
given in  Table~\ref{Tab:3}, Eq.~(\ref{eq:finalconst}) becomes~:
\begin{alignat}{2}
         \delta_2 &=-\frac{2}{3}+2\,\varepsilon_3-\varepsilon_8  \quad, \qquad\qquad           \beta_2  = -\frac{2}{3}+\,\varepsilon_3-\frac{1}{2}\,\varepsilon_8  \,, \nn\\
         \delta_3 &=\,\,\,\,\,\frac{2}{3}-\varepsilon_3    \qquad\qquad, \qquad\qquad                  \beta_3  = \,\,\, \frac{5}{12}+\gamma_3-\,\varepsilon_3   \,,  \nn\\
         \delta_6 &=\,\,\,\,\,\,\,\,\,\,\,-\varepsilon_6               \qquad\qquad, \qquad\qquad      \beta_6  = -\,\frac{1}{4}-\gamma_6      \,,                \nn\\
         \delta_8 &=\,\,\,\,\,\,\,\,\,\,\,-\varepsilon_8               \qquad\qquad, \qquad\qquad      \beta_8  = -\,\,\frac{1}{2}\,\varepsilon_8\,,\nn\\
         \varepsilon_2 &=-2\,\varepsilon_3+\varepsilon_8\,.
           \label{eq:finalfix}
\end{alignat}
\noindent
As we can see the unknown constraints in $\tau_i$'s, Eqs.~(\ref{eq:tauform},~\ref{eq:tauform2}), have now been fixed to match with perturbation theory.
If we insert these constants in Eqs.~(\ref{eq:tauform},~\ref{eq:tauform2}), we can write the coefficient functions, $\tau_i$'s, in Euclidean space to obtain our
final non-perturbative result~:
\newpage
\be
\tau_2^E(p^2,k^2,q^2)&=\displaystyle{\frac{2}{(k^4-p^4)}}&
\lef[\lef(-\frac{2}{3}+\,\varepsilon_3-\frac{\varepsilon_8}{2}  \rig)+\gamma_2\,\frac{2\, k\cdot p}{k^2+p^2}\rig]
                    \; \tau_2^{\,anti}\nn\\[2mm]
&+\displaystyle{\frac{2}{(k^2+p^2)^2}}&\lef[-\frac{2}{3}+(2\,\varepsilon_3-\varepsilon_8) \frac{q^2}{k^2+p^2}\,\rig]
                       \tau_2^{\,sym},
\nn\\ [4mm]
\nn \tau_3^E(p^2,k^2,q^2)&=
    \displaystyle{-\frac{1}{(k^2-p^2)}}&
\lef[\lef(\frac{5}{12}-\,\varepsilon_3\rig)+\gamma_3\,\frac{(k+ p)^2}{k^2+p^2}\rig]\;
                  \tau_3^{\,anti} \\ [2mm]
\nn &-\displaystyle{\frac{1}{(k^2+p^2)}}&\lef[\frac{2}{3}-\varepsilon_3\frac{q^2}{k^2+p^2}\,\,\rig] \;
                \tau_3^{\,sym} \quad,
\\[4mm]
\nn
\tau_6^E(p^2,k^2,q^2)&=
    \displaystyle{-\frac{1}{(k^2+p^2)}}&
\lef[-\frac{1}{4}-\gamma_6\,\frac{q^2}{k^2+p^2}\rig] \;
        \tau_6^{\,anti}  \\[2mm]
\nn &-\displaystyle{\frac{(k^2-p^2)}{(k^2+p^2)^2}}&\lef[-\varepsilon_6\,\frac{q^2}{k^2+p^2}\,\,\rig]\;
                     \tau_6^{\,sym} \quad,
\\[4mm]
\nn
\tau_8^E(p^2,k^2,q^2)&=
    \displaystyle{-\frac{1}{(k^2-p^2)}}&
\lef[-\frac{1}{2}\,\varepsilon_8+\gamma_8\,\frac{2\, k\cdot p}{k^2+p^2}\rig]  \;
        \tau_8^{\,anti} \\[2mm]
\nn &-\displaystyle{\frac{1}{(k^2+p^2)}}&\lef[-\varepsilon_8\,\frac{q^2}{k^2+p^2}\,\,\rig]\;
        \tau_8^{\,sym}\quad,
\ee
where
\be
 \tau_i^{\,anti}&=&\lef(\frac{1}{F(k^2)}\,-\,\frac{1}{F(p^2)}\rig)\,,\nn\\[3mm]
\mbox{and}\quad \tau_i^{\,sym} &=& \frac{1}{4}\,\lef(\frac{1}{F(k^2)}\,+\,\frac{1}{F(p^2)}\rig)\,\ln\lef( \frac{F(q^2)^2}{F(k^2) F(p^2)}\rig)\nn\\
 \mbox{OR}\qquad\quad\,\,&&\nn\\
\tau_i^{\,sym} &=&\frac{1}{2}\lef(\frac{1}{F(k^2)}+\frac{1}{F(p^2)}\rig)\ln\lef[\frac{1}{2}\lef(\frac{F(q^2)}{F(k^2)}+\frac{F(q^2)}{ F(p^2)}\rig)\rig]\,.
\label{eq:finaltaus}
\ee
\normalsize
\noindent
The fermion and photon SDE's at leading log order do not fix the constants $\gamma_i, \varepsilon_i$, Eq.~(\ref{eq:finaltaus}). As the simplest
example for later exploration we choose $\gamma_i=\varepsilon_i=0$ in the above expressions and insert the second form of $\tau^{\,sym}_i$ in Eq.~(\ref{eq:finaltaus}),
we  then have~:
\newpage
\be
\fbox{$
\begin{array}{lll}
\nn\\[-5mm]
\displaystyle \tau_2^E(p^2,k^2,q^2)&=\displaystyle{\frac{1}{(k^4-p^4)}}&
\displaystyle \lef(-\frac{4}{3}\rig)
 \displaystyle                   \; \lef(\frac{1}{F(k^2)}\,-\,\frac{1}{F(p^2)}\rig)\nn\\[3mm]
&+\displaystyle{\frac{1}{(k^2+p^2)^2}}&\displaystyle\lef(-\frac{2}{3}\rig)
  \displaystyle                     \lef(\frac{1}{F(k^2)}+\frac{1}{F(p^2)}\rig)\ln\lef[\frac{1}{2}\lef(\frac{F(q^2)}{F(k^2)}+\frac{F(q^2)}{ F(p^2)}\rig)\rig]\,,
\nn\\ [3mm]
\nn \displaystyle \tau_3^E(p^2,k^2,q^2)&=
    \displaystyle{-\frac{1}{(k^2-p^2)}}&
\displaystyle \lef(\frac{5}{12}\rig)\;
                  \lef(\frac{1}{F(k^2)}\,-\,\frac{1}{F(p^2)}\rig) \\ [3mm]
\nn &-\displaystyle{\frac{1}{(k^2+p^2)}}&\displaystyle\lef(\frac{1}{3}\,\,\rig) \;
  \displaystyle              \lef(\frac{1}{F(k^2)}+\frac{1}{F(p^2)}\rig)\ln\lef[\frac{1}{2}\lef(\frac{F(q^2)}{F(k^2)}+\frac{F(q^2)}{ F(p^2)}\rig)\rig] \,,
\\[3mm]
\nn
\displaystyle \tau_6^E(p^2,k^2,q^2)&=
    \displaystyle{-\frac{1}{(k^2+p^2)}}&
\displaystyle \lef(-\frac{1}{4}\rig) \;
        \lef(\frac{1}{F(k^2)}\,-\,\frac{1}{F(p^2)}\rig)\,,
\\[3mm]
\tau_8^E(p^2,k^2,q^2)&= 0\,.& \nn\\[-1mm]
\end{array} $}\\
\label{eq:finaltaus2}
\ee

\vspace*{-5.5mm}
\noindent
This is our simplest expression for the transverse part.
We can then construct the full vertex from this using

\vspace*{-10.5mm}
\be
\Gamma^{\mu}(p,k;q)\;=\;\sum_{i=1}^4\,\lambda_i(p^2,k^2,q^2)\,L^{\mu}_i(p,k;q)\;+\;\sum_{j=2,3,6,8}\,\tau_i(p^2,k^2,q^2)\,T_i^{\mu}(p,k;q)\,,
\label{eq:finalansatz}
\ee

\vspace*{-3mm}
\noindent
from Eqs.~(\ref{eq:longitudinal}-~\ref{eq:Ts}). This is our final result. Phenomenological studies of strong coupling QED
with this vertex ansatz are presently underway~\cite{Richard:thesis,Kizilersu:inprep}.
\section{Conclusions}
\noindent The Schwinger-Dyson equations constitute the field equations of
a theory.  Being an infinite set of nested integral equations, they are in
general intractable without some form of truncation.  To date, the only known
consistent truncation procedure is perturbation theory.  This satisfies
gauge invariance and multiplicative renormalizability order-by-order,
and the meaning of any truncation is well-defined.  In the case of
non-perturbative truncations, like the rainbow approximation, one
has always been unsure as to how much physics has been encoded
and how much lost. The calculation of dynamical mass generation
nicely illustrates this.
\noindent   The properties of gauge invariance and multiplicative
renormalizability
are fundamental to our ability to calculate consistently in a gauge
theory.  It is thus natural that any truncation should respect these
properties.  They ensure not only the elimination of overlapping divergences
that plague Schwinger-Dyson calculations, but allow all
 ultraviolet divergences to
be handled appropriately.
\noindent   Here we have considered the fermion and boson propagators in 4-dimensional
massless QED.  To be able to study these
requires an {\it ansatz} for the full fermion-boson vertex.
This interaction involves 11 non-zero components, three of which are fixed
by the Ward-Green-Takahashi identity in terms of the fermion
propagator functions.  The other eight (transverse) components in principle
require knowledge of the four, five, six ... point functions.  However, very
specific projections of this vertex appear in the fermion and boson
self-energies.
We have seen that these projections are strongly constrained by
the multiplicative renormalizability of the fermion and boson propagators.
At its simplest, multiplicative renormalzability is closely related
to the ultra-violet behaviour of loop integrals. This probes
distinct limits for the fermion-boson vertex : one in the fermion
equation and the other in the boson.  In these two limits, the vertex has
quite different structures.  Such behaviour ensures the multiplicative renormalizability
of leading logarithms and shows that the 2-point Green's
functions for both fermion and photon are wholly determined
by the fermion wavefunction renormalization.
This has enabled us to unravel for the first time the non-perturbative structure of the full vertex, Eqs.~(\ref{eq:finaltaus2},~\ref{eq:finalansatz} ),
at least as far as concerns the fermion and photon Schwinger-Dyson equations.

\noindent
While the form of the 3-point vertex is determined in three kinematic limits,
when $k^2, p^2 \gg q^2$, when $k^2,q^2 \gg p^2$ and when $p^2,q^2 \gg k^2$, its form at general
momenta when all six vector structures of massless QED contribute involves free parameters.  Imposing the known
perturbative ${\cal O}(\alpha)$ result for the individual vertex components fixes these.  This marks a significant step in the
development of non-perturbative Feynman rules
needed for realistic calculations in strong QED.
   There are many steps to go :
\begin{itemize}
\item to solve the extended constraints beyond leading logarithmic order and include masses~\cite{Kizilersu:Ayse-Mike},
\item to compute the Lamb shift of hydrogen and calculate the  properties of positronium to asses how well our vertex ansatz automatically sums higher orders in $\alpha$,
\item to explore strong physics with such a complete, unquenched vertex --- extending the existing studies
using bare, Ball-Chiu and CP
vertices~\cite{Bloch:1994if,Hawes:1991qr,Hawes:1996ig,Hawes:1996mw,Hawes:1996pe,Kizilersu:2001,Kizilersu:2001pd,Curtis:1993py,Schreiber:1998ht,Roberts:1994dr,Atkinson:1993mz}.
Such calculations are under way and will be reported elsewhere~\cite{Kizilersu:inprep}
\end{itemize}
\noindent
Eventually an extension to QCD will be our target.
\vspace*{-1cm}
\begin{acknowledgments}
\noindent
The authors are grateful to the IPPP in Durham and CSSM in Adelaide for
providing ideal working environments and for their kind hospitality. We also acknowledge partial support from the
Australian Research Council Linkage International grant "LX0776452" that allowed this collaboration to continue.
\end{acknowledgments}
\appendix
\section{Perturbative $\tau$'s}
\label{app:perttaus}
\noindent
The vertex coefficients $\tau_i$'s given below are the massless limit of the exact ${\cal{O}}(\alpha)$ calculation for the
massive fermions in general covariant gauge \cite{Kizilersu:1995iz}.
\be
\tau_2^M(p^2,k^2,q^2)&=&\frac{\alpha_0}{8\pi\D^2}\Bigg\{
{\it J}_0\Bigg[\lef(\frac{k^2+p^2}{2}+\frac{3}{4\D^2}\,p^2k^2q^2\rig)
\lef(\xi-2\rig)
+k\cdot p \Bigg]\nn\\[1mm]
&& \hspace{10mm}
+\ln\frac{k^2}{p^2}\Bigg[\lef(\frac{(k+p)^2}{2\,(p^2-k^2)}
+\frac{3}{4\D^2}\,k\cdot p\,(p^2-k^2)\rig)\lef(\xi-2\rig)
+\frac{(p+k)^2}{(p^2-k^2)}\Bigg]\nn\\[1mm]
&& \hspace{10mm}
+\ln\frac{q^4}{k^2p^2}\Bigg[\lef(\frac{3}{4\D^2}\,k\cdot p\, q^2
+1\rig)\lef(\xi-2\rig)
+1\Bigg]\nn\\[1mm]
&& \hspace{10mm}
+\lef(\xi-2\rig)\Bigg\}\,,
\label{eq:perttau2}\\\nn\\
\tau_3^M(p^2,k^2,q^2)&=& \frac{\alpha_0}{8\pi\D^2}\Bigg\{
{\it J}_0\Bigg[
\lef(\frac{(k^2+p^2)^2}{8}
-\frac{3}{8\D^2}(\mul)^2(k^2-p^2)^2\rig)\lef(\xi-2\rig)-\D^2\Bigg]\nn\\[1mm]
&& \hspace{10mm}
+\ln\frac{k^2}{p^2}\Bigg[\frac{(k^2-p^2)}{4}
\lef(-1+\frac{3}{2\D^2} \mul\, (k+p)^2\rig)\lef(\xi-2\rig)\Bigg]\nn\\[1mm]
&& \hspace{10mm}
+\ln\frac{q^4}{k^2p^2}\Bigg[
\frac{\mul}{2}\lef(1-\frac{3}{4\D^2}(k^2-p^2)^2\rig)\lef(\xi-2\rig)\Bigg]
\nn\\[1mm]
&& \hspace{10mm}
-\frac{(k+p)^2}{2}\lef(\xi-2\rig)\Bigg\}\,,
\label{eq:perttau3}\\\nn\\
\tau_6^M(p^2,k^2,q^2)&=&\frac{\alpha_0}{8\pi\D^2}\frac{(p^2-k^2)}{2}\Bigg\{
{\it J}_0\Bigg[
\lef(-\frac{q^2}{4}+\frac{3}{4\D^2}q^2 (\mul)^2\rig)\lef(\xi-2\rig)\Bigg]
\nn\\[3mm]
&& \hspace{30mm}
+\ln\frac{k^2}{p^2}\Bigg[\lef(\frac{3}{4\D^2}\mul\, (p^2-k^2)
-\frac{(p+k)^2}{2(p^2-k^2)}\rig)\lef(\xi-2\rig)\Bigg]\nn\\[1mm]
&& \hspace{30mm}
+\ln\frac{q^4}{k^2p^2}\Bigg[\frac{3}{4\D^2}\mul\, q^2 \lef(\xi-2\rig)\Bigg]
\nn\\[1mm]
&& \hspace{30mm}
+\lef(\xi-2\rig)\Bigg\}\,,
\label{eq:perttau6}\\ \nn\\
\tau_8^M(p^2,k^2,q^2)&=&\frac{\alpha_0}{8\pi\D^2}\Bigg\{q^2\lef[\mul\,  {\it J}_0
+\ln\frac{q^4}{k^2p^2}\rig]+(p^2-k^2) \ln\lef(\frac{k^2}{p^2}\rig)\Bigg\}\,,
\label{eq:perttau8}
\ee
where
\be
J_0=\frac{2}{\Delta}\,\lef[
{\it{f}}\,\lef(\frac{k\cdot p-\Delta}{p^2} \rig)-\,{\it{f}}\,\lef(\frac{k\cdot p+\Delta}{p^2} \rig)
+\frac{1}{2}\,\ln\lef(\frac{q^2}{p^2}\rig)\,\ln\lef(\frac{k\cdot p-\Delta}{k\cdot p+\Delta} \rig)
\rig]\,,
\label{eq:J0}
\ee
and
\be
{\it{f}}(x)={\it{Sp}}(1-x)=-\int_x^1\,dy\,\frac{\ln y}{1-y}\,.
\ee

\noindent
\section{Limits of $\tau_i$'s}
\label{app:J0}

\subsection{Fermion Limit}
\label{sec:fermionlimit}
\noindent
In order to take the $k^2 \simeq q^2 \gg p^2$ limit of the perturbative transverse vertex coefficients namely the $\tau_i$ functions, Eq.~(\ref{eq:perttau2}-\ref{eq:perttau8})
we need to expand $J_0$ function, Eqs.~(\ref{eq:J0},~\ref{eq:J02}), up to ${\cal{O}}(1/k^7)$~:
\be
J_0= \frac{2}{k^2}\,\Bigg\{
1& +& \frac{1}{k^2}\,\lef( k \cdot p - \frac{p^2}{3}\rig)
  + \frac{1}{k^4}\,\lef(\frac{4}{3} (k \cdot p)^2 - (k  \cdot p)\, p^2 +\frac{1}{5}\, p^4\rig)\nn\\
  &+& \frac{1}{k^6}\,\lef(2\,(k \cdot p)^3 - \,\frac{12}{5}\, (k \cdot p)^2\, p^2 + (k \cdot p)\, p^4 - \,\frac{1}{7}\,p^6\rig)\nn\\
  &+& \frac{1}{k^8}\,\lef(\frac{16}{5}\,(k \cdot p)^4 - \frac{16}{3}\, (k \cdot p)^3\, p^2 + \frac{24}{7}\, (k \cdot p)^2 \,p^4 - (k \cdot p)\, p^6 +\frac{1}{9} p^8\rig)\nn\\
  &+& \frac{1}{k^{10}}\,\lef(\frac{16}{3}\,(k \cdot p)^5 - \frac{80}{7}( k \cdot p)^4 p^2 +10 (k \cdot p)^3 p^4 - \frac{40}{9}(k \cdot p)^2 p^6 + (k \cdot p) p^8
   - \frac{p^{10}}{11}\rig)\nn\\
   &+& {\cal{O}}(1/k^{7})
\Bigg\}\,\,
\ln\lef( \frac{k^2}{p^2}\rig)\,.
\label{eq:J02}
\ee

\subsection{Photon Limit}
\label{sec:photonlimit}
In the photon limit, $k^2 \simeq p^2 \gg q^2 $, $J_0$ behaves like~:
\be
J_0 = \frac{2}{(p^2-k^2)}\,\lef[ \frac{2\,(p^2-k^2)}{p^2}+ \frac{(p^2-k^2)^2}{p^4}+\,\frac{13}{18}\,\frac{(p^2-k^2)^3}{p^6}+ \,\cdots\rig]\,.
\label{eq:photonJ0}
\ee

\section{Effective $\tau$'s}
\label{app:effectivetaus}

\noindent
The connection between the effective and real $\tau_i$ functions are given below and the detail of this procedure can be found elsewhere~\cite{Bashir:1997qt}~:
\be
(\tau^E_2)_{eff}(p^2 ,k^2) & = & \frac{1}{ f(k^2,p^2)}\,\int_0^{\pi}\,d\Psi\,\frac{ sin^2 \Psi}{q^2}\,(\tau^E_2)_{Real}(p^2 ,k^2 ,q^2)\,\lef\{-\Delta^2\rig\}\,,\nn\\
(\tau^E_3)_{eff}(p^2 ,k^2) & = & \frac{1}{ f(k^2,p^2)}\,\int_0^{\pi}\,d\Psi\,\frac{ sin^2 \Psi}{q^2}\,(\tau^E_3)_{Real}(p^2 ,k^2 ,q^2)\,\lef\{-\Delta^2-\frac{3}{2}\,q^2\,k \cdot p \rig\}\,,\nn\\
(\tau^E_6)_{eff}(p^2 ,k^2) & = & \frac{1}{ f_6(k^2,p^2)}\,\int_0^{\pi}\,d\Psi\,\frac{ sin^2 \Psi}{q^2}\,(\tau^E_6)_{Real}(p^2 ,k^2 ,q^2)\, \lef\{k \cdot p \rig\}\,,\nn\\
(\tau^E_8)_{eff}(p^2 ,k^2) & = & \frac{1}{ f(k^2,p^2)}\,\int_0^{\pi}\,d\Psi\,\frac{ sin^2 \Psi}{q^2}\,(\tau^E_8)_{Real}(p^2 ,k^2 ,q^2)\,\lef\{-\Delta^2\rig\}\,,
\ee
where
\be
f(k^2,p^2)&=& \frac{\pi}{8}\,\frac{p^2}{k^2}\,(3\,k^2-p^2)\,,\nn\\
f_6(k^2,p^2)&=&\frac{\pi}{4}\,\frac{k^2}{k^2}\,.
\ee
\bibliography{C:/AYSE/Library/myLibrary}
\end{document}